\documentclass[aps,10pt,prx,twocolumn]{revtex4-2}

\usepackage{graphicx} 
\usepackage{comment}

\usepackage{graphicx}  
\usepackage{epstopdf}
\usepackage{dcolumn}   
\usepackage{amsmath,amssymb,dsfont}
\usepackage{wrapfig}
\usepackage{array}
\usepackage[caption=false]{subfig}
\usepackage[bookmarks=false,linkcolor=blue,urlcolor=blue,colorlinks,citecolor=blue]{hyperref}
\usepackage{exscale,relsize}
\usepackage[usenames, dvipsnames]{color}
\usepackage{bbold}
\usepackage{xcolor}

\usepackage{cleveref}
\Crefformat{appendix}{Appendix~#2#1#3}

\usepackage{microtype}

\usepackage{siunitx}
\DeclareSIUnit{\ueV}{\micro\electronvolt}
\DeclareSIUnit{\um}{\micro\meter}
\DeclareSIUnit{\nm}{\nano\meter}
\usepackage[]{cleveref}
\Crefname{section}{Sec.}{Secs.}
\Crefname{subsection}{Sec.}{Secs.}
\Crefname{appendix}{\IfAppendix{Sec.}{Sec.}}{\IfAppendix{Secs.}{Secs.}}
\Crefname{subappendix}{\IfAppendix{Sec.}{Sec.}}{\IfAppendix{Secs.}{Secs.}}
\Crefname{equation}{Eq.}{Eqs.}
\Crefname{figure}{Fig.}{Figs.}
\Crefname{tabular}{Tab.}{Tabs.}

\DeclareMathOperator{\tr}{tr}

\newcommand{\assignerr}{{\sf{err}_a}}
\newcommand{\biaserr}{{\sf{err}_b}}

\newcommand{\bra}[1]{\langle #1|}
\newcommand{\ket}[1]{| #1\rangle}
\newcommand{\proj}[1]{{\ket{#1}\!\bra{#1}}}

\newcommand{\instr}[1]{{\mathcal{M}^{#1}}}
\newcommand{\sop}[1]{{\mathcal{#1}}}
\newcommand{\mop}[2]{{\mathcal{M}^{#1}_{#2}}}
\newcommand{\mst}[2]{{{\Pi}^{#1}_{#2}}}

\newcommand{\Tlife}{T_\text{life}}
\newcommand{\decay}[1]{\lambda_{#1}}
\newcommand{\eres}{\varepsilon_\text{res}}
\newcommand{\emst}{\varepsilon_\text{mst}}

\newcommand{\logical}[1]{\overline{#1}}
\newcommand{\repcode}[1]{\widetilde{#1}}

\begin{document}

\title{Roadmap to fault tolerant quantum computation using topological qubit arrays}
\author{Microsoft Quantum$^\dagger$}
\noaffiliation
\date{\today}

\begin{abstract}
We describe a concrete device roadmap towards a fault-tolerant quantum computing architecture based on noise-resilient, topologically protected Majorana-based qubits. Our roadmap encompasses
four generations of devices: a single-qubit device that enables a measurement-based qubit benchmarking protocol; a two-qubit device that uses measurement-based braiding to perform single-qubit Clifford operations; an eight-qubit device that can be used to show an improvement of a two-qubit operation when performed on logical qubits rather than directly on physical qubits; and a topological qubit array supporting lattice surgery demonstrations on two logical qubits. 
Devices that enable this path require a superconductor-semiconductor heterostructure that supports a topological phase, quantum dots and coupling between those quantum dots that can create the appropriate loops for interferometric measurements, and a microwave readout system that can perform fast, low-error single-shot measurements.
We describe the key design components of these qubit devices, along with the associated protocols for demonstrations of single-qubit benchmarking, Clifford gate execution, quantum error detection, and quantum error correction, which differ greatly from those in more conventional qubits. Finally, we comment on implications and advantages of this architecture for utility-scale quantum computation.
\end{abstract}

\maketitle

\tableofcontents

\section{Introduction: fault-tolerant quantum computation using tetrons}\label{sec:intro}

Practical schemes for quantum error correction overwhelmingly rely on the ability to perform multi-qubit Pauli measurements~\cite{Nielsen00,Preskill22}.
Here, \emph{logical} qubits are encoded in a multitude of underlying \emph{physical qubits}, and a Pauli measurement sequence is carefully designed to allow the identification of any errors without affecting the encoded quantum information. 
The number of physical qubits required per logical qubit as well as the depth of the measurement sequence---both commonly referred to as \emph{overhead}---depend on the performance of the underlying physical measurements, making this a crucial performance metric for a large-scale fault-tolerant quantum computer.

\begin{figure*}
    \centering
    \includegraphics[width=2\columnwidth]{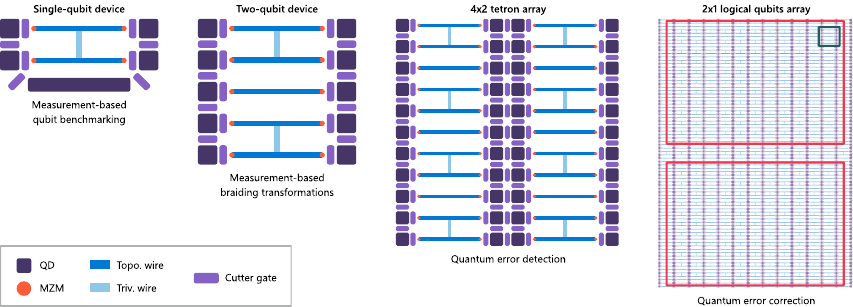}
    \caption{
    {\it Roadmap to fault-tolerant quantum computation with tetrons.}  The first panel shows a single-qubit device.  The tetron is formed through two parallel topological wires (blue) with a MZM at each end (orange dot) connected by a perpendicular trivial superconducting wire (light blue).  The regions adjacent to the MZMs have quantum dots (QDs) shown in dark purple with cutter gates (light purple) controlling the tunnel coupling between the quantum dots and MZMs.  The device supports measurement-based qubit benchmarking.  The next panel shows a two-qubit device that supports measurement-based braiding transformations.  The third panel shows a $4\times2$ array of tetrons supporting a quantum error detection demonstration on two logical qubits.  These demonstrations build towards quantum error correction, such as on the device shown in the right panel (a $27\times 13$ tetron array). The red boxes highlight the two fault-distance 7 logical qubits; the device is compatible with lattice surgery demonstrations.  The blue-box highlights a two-qubit tile identical to the device in the second panel. Additional demonstrations may be inserted between the third and fourth panels to gradually scale the system. 
    }
    \label{fig:roadmap}
\end{figure*}

Measurement-based topological qubits built upon Majorana zero modes (MZMs) stand out in this context as an architecture in which multi-qubit Pauli measurements are the native instruction set. 
This is unlike most conventional platforms, where these measurements are implemented by sequences of multi-qubit Clifford gates and single-qubit measurements. 
A specific MZM-based qubit architecture that enables such measurements was proposed in Refs.~\onlinecite{Karzig17,Plugge17},
building on the ideas in
Refs.~\onlinecite{Fidkowski11b, Plugge16, Vijay15} (see also Ref.~\onlinecite{Schrade18}).
The qubits in this architecture are {\it tetrons}, which store the qubit state in four MZMs. 
The MZMs are located at the endpoints of two topological superconducting wires~\cite{Kitaev01} formed from proximitized semiconductor nanowires~\cite{Lutchyn10,Oreg10}.
In this paper, we focus on two-sided tetrons, in which two parallel topological wires are joined by a trivial superconducting backbone to form a single island with charging energy~\cite{Karzig17}. 
In the remainder of this paper, we will drop the modifier and simply use {\it tetrons} to refer to two-sided tetrons. 
The measurements are performed by forming interferometric loops between qubit islands and nearby quantum dots, leading to a shift in the quantum capacitance of the dots that can be detected using standard microwave techniques.
Such an interferometric parity measurement was first demonstrated in Ref.~\onlinecite{Aghaee24} (using a linear tetron), and more recently a hardware realization of the two-sided tetron along with single-shot measurements of both loops was presented in Ref.~\onlinecite{Aghaee25}.

Tetrons are predicted to suppress idle and measurement errors exponentially in three dimensionless quantities: (i) the ratio of the topological gap to the temperature, (ii) the ratio of the length of the device to the topological superconducting coherence length, and (iii) the signal-to-noise ratio of the measurement system~\cite{Kitaev97, Nayak08,Cheng09,Knapp18a,mishmash2020dephasing}. 
The resulting low error rates help enable scalable fault-tolerant quantum computation using recently introduced topological codes specifically tailored towards measurement-based qubits; see, {\it e.g.,} Refs.~\onlinecite{Plugge16,Litinski18,Hastings21, Gidney23,Grans-Samuelsson23}. 
For a more detailed discussion of the practical advantages offered by this architecture for building a utility-scale fault-tolerant quantum computer, see the discussion in \Cref{sec:outlook}.

We present a roadmap, summarized in \Cref{fig:roadmap}, building towards fault tolerant quantum computing with tetrons.
The roadmap incrementally scales the system while building the capabilities required for scalable quantum error correction.
Previous works discussing demonstrations for Majorana-based qubits have either focused on introducing the particular qubit platform~\cite{Karzig17,Plugge17,Hassler11,Heck12,Hyart13}, detailing pre-qubit demonstrations for Majorana nanowires~\cite{Aasen16,Alicea11,Vijay16b}, or physical implementations of quantum error correction with Majorana-based qubits~\cite{Vijay15,Vijay16a,Plugge16,Litinski18,Paetznick23,Grans-Samuelsson23}; in contrast the roadmap here focuses on tetrons realized in gate-defined 2DEG realizations of Majorana nanowires and describes milestone demonstrations building towards implementing the pair-wise Pauli-measurement-based quantum error correcting codes of Refs.~\onlinecite{Hastings21,Paetznick23,Grans-Samuelsson23}.
We present a gate schematic of the minimal device supporting each demonstration, and describe the key design elements, operating principles, and dominant error sources.

We support this roadmap by several new technical results.
In \Cref{sec:MV11}, we introduce a specific protocol and associated performance metrics for measurement-based qubit benchmarking (MBQB). This protocol can both be used as a first demonstration of tetron operations as well as for benchmarking qubits in a scalable tetron-based system.
\Cref{sec:MV21} reviews the concept of measurement-based braiding transformations and its role in the context of fault-tolerant quantum computing.
In addition to discussing details of a two-qubit device that supports such an initial braiding demonstration, we simulate the specific measurement sequences for that device in a tetron-specific noise model.
In \Cref{sec:MV42}, we propose a specific quantum error detection scheme for tetrons that utilizes the same measurements as required for the syndrome extraction circuit of the Floquet codes in Refs.~\onlinecite{Hastings21,Paetznick23}, only a subset of which are needed for the pair-wise measurement-based surface code of Ref.~\onlinecite{Grans-Samuelsson23}.
We simulate the expected performance of this demonstration on an eight-qubit device, identifying a region of two-qubit measurement logical improvement.
Finally, in \Cref{sec:outlook} we discuss the outlook to the above-mentioned scalable quantum error correcting codes with the tetron architecture. 
An example device is the $2\times 1$ logical qubit array shown in the right-most panel of \Cref{fig:roadmap}; each logical qubit (red box) supports the surface code with fault distance $7$.
Additional details of the roadmap may be found in the appendices.

\section{Single-qubit tetron device}
\label{sec:MV11} 

\subsection{Measurement-based benchmarking of a topological qubit}

For most conventional qubit platforms, coherent rotations are the elementary single-qubit operations, along with measurements in a distinguished ``computational basis."  For measurement-based topological qubits, such rotations are only used to implement non-Clifford gates and, in a fault-tolerant setting, do not generally limit the system's performance, see discussion in \Cref{app:T-gates}.  Instead, topologically protected operations are performed through single- and two-qubit Pauli measurements.  Notably, measurements, including two-qubit measurements, can be performed in several different bases without the need for unitary rotations~\footnote{The measurement bases are labeled by the corresponding Pauli operators. Even a two qubit measurement will still yield only two possible outcomes, corresponding to the different eigenvalues of the corresponding (two-qubit) Pauli operator. Equivalently, we sometimes refer to these measurements as measuring only the parity of the operators, since outcomes on individual qubits are not revealed, only the product of the corresponding single qubit $\pm1$ eigenvalues of each tensor factor.}.  Therefore, the most natural way to validate such a qubit is through these Pauli measurements.  In the single-qubit setting, this amounts to testing the ability to measure the qubit state in more than one Pauli basis. It is worth noting that more conventional quality metrics, such as qubit lifetimes $T_1$ and $T_2^*$, along with protocols to measure them, can be defined for tetron qubits~\cite{mishmash2020dephasing,Sau24}; however, since these timescales are not directly relevant to measurement-based operations, which this paper is focused on, we defer their discussion to \Cref{app:coh-times}.

\begin{figure}
    \centering
    \includegraphics[width=\columnwidth]{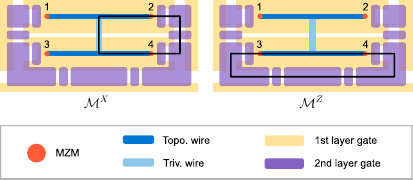}
    \caption{{\it Gate schematic of the single-qubit device with measurement loops indicated.}  The single-qubit device shown functionally in the left-most panel of \Cref{fig:roadmap} can be enabled by the gate schematic shown here. The tetron is formed by two parallel topological wires connected by a trivial superconducting backbone.  MZMs form at each end of the topological wires and are labeled $1-4$ as shown.  The chemical potential in the wires is controlled by first layer gates that deplete the 2DEG away from the superconductor.  Quantum dot plungers and cutter gates between quantum dots and the MZMs are second layer gates.  Depletion gates surround the device. 
    The top and bottom panels show $X$ and $Z$ measurements formed by tuning the gates along the black lines and performing dispersive gate sensing on one of the quantum dot plunger gates involved in the loop, see discussion in \Cref{sec:MV11-ops}.
    }
    \label{fig:MV11}
\end{figure}

It is worth noting that such non-commuting measurements have played an important role in the early development of quantum mechanics. The Stern-Gerlach experiment~\cite{Gerlach1922}, which uses a magnetic field gradient to split a beam of spinful particles into two beams of well-defined spin orientation, was one of the first demonstrations of the discrete nature of quantum measurement outcomes. By combining two such Stern-Gerlach apparatuses with different orientation, one can probe the non-commutative nature of different spin observables. Here, we will describe a formally similar experiment, albeit in a completely different physical system, where the spin orientation is replaced by measuring different Majorana bilinears.

For an initial demonstration of a topological qubit, only two Pauli measurements, $X$ and $Z$ are needed. In the ideal case, these are projective measurements that anticommute exactly, $\lbrace X, Z \rbrace = 0$.  Therefore, for two subsequent $X$ measurements, the outcome of the second should be the same as the first, and likewise for two subsequent $Z$ measurements. In the Stern-Gerlach analogy, this corresponds to two sequential apparatuses having the same orientation, where the second does not split the individual beams any further. If, conversely, an $X$ measurement is followed by a $Z$ measurement (or vice versa $Z$ followed by $X$), the outcome of the second measurement should be equally distributed regardless of any previous measurement outcome. In the Stern-Gerlach analogy, here the second apparatus has a different orientation and is able to split each output beam of the first one in half again. Performing a sequence of $X$ and $Z$ measurements and statistically analyzing the outcomes allows us to infer whether they indeed correspond to two anticommuting, projective measurements, within some well-defined approximation.  We propose a number of such statistical tests, which we refer to collectively as measurement-based qubit benchmarking (MBQB)~\footnote{The idea of verifying the existence of a qubit through two anticommuting measurements has been introduced before---Chao {\it et al}.~\cite{chao2017overlapping} \emph{define} a qubit in this way, and discuss the relation to more conventional notions of a qubit. Similarly, stronger guarantees about the action of the measurements can be obtained through {\em self-testing} protocols~\cite{Irfan20}, although this is beyond the scope of this roadmap.}. Similar tests were proposed and experimentally performed on nuclear spin qubits in Ref.~\onlinecite{Huie23}. Importantly, these protocols test not only the accuracy of the classical outcomes reported by the measurements, but also the quality of the post-measurement quantum state~\footnote{Note that the same protocols could be applied to logical qubits.}.

To quantify deviations from the ideal case, we introduce two operational error metrics: the {\it operational assignment error} $\assignerr$~\footnote{We emphasize the distinction between the usual definition of assignment error, which is the probability that a measurement classifier will assign an outcome to the observed measurement response that is different from the (inaccessible) true measurement outcome. Unlike the usual definition, the operational assignment error is directly measurable, but also accounts for state transitions during the measurement, not simply the classification errors.} and the {\it operational bias} $\biaserr$.  The former is the deviation of the experimental $X$ and $Z$ measurements from their ideal projective behavior, quantified by the total variational distance between the observed and expected outcome statistics maximized over different preparations and bases.
More explicitly, writing $\mop{P}{r}$ to be the superoperator corresponding to the (generally imperfect) implementation of the measurement of the Pauli operator $P$ with outcome $r\in \{+1, -1\}$, then~\footnote{\Cref{eqn:pA,eqn:pM} can be obtained by noting $ \Pr\left(\mop{P}{-1}\Big|\mop{Q}{s};\frac{1}{2}\tilde{\openone}\right)=1- \Pr\left(\mop{P}{+1}\Big|\mop{Q}{s};\frac{1}{2}\tilde{\openone}\right)$. }
\begin{align} 
\assignerr & =  \max_{s, P} \frac{1}{2}\sum_r \left| \Pr\left(\mop{P}{r} \Big| \mop{P}{s};\frac{1}{2}\tilde{\openone}\right) - \delta_{r,s} \right|\\
&= \max_{s, P} \left| \Pr\left(\mop{P}{+1} \Big| \mop{P}{s};\frac{1}{2}\tilde{\openone}\right) - \delta_{+1,s} \right|,\label{eqn:pA}
\end{align}
where $\frac{1}{2}\tilde{\openone}$ is an experimental approximation to the maximally mixed state.  We have written the conditional probability of getting outcome $r$ immediately after outcome $s$ having prepared the initial state $\frac{1}{2}\tilde{\openone}$ as $\Pr\left(\mop{X}{r} \Big| \mop{X}{s};\frac{1}{2}\tilde{\openone}\right) $.

The operational bias error quantifies the deviation from the {\em mutually unbiased bases} behavior of $X$ and $Z$ measurements~\cite{Schwinger60,Wooters89}.  For a single-qubit, this means that an $X$ measurement after a $Z$ measurement should yield either outcome with probability $1/2$.  As such, the operational bias corresponds to maximizing the total variation distance over sequences of pairs of nominally distinct measurements:
\begin{align}
\biaserr & = 
\max_{s,P\neq Q}
\frac{1}{2}\sum_r \left| \Pr\left(\mop{P}{r}\Big|\mop{Q}{s};\frac{1}{2}\tilde{\openone}\right) - \frac{1}{2}
\right|\\
& = 
\max_{s, P\neq Q} \left|  \Pr\left(\mop{P}{+1} \Big|\mop{Q}{s};\frac{1}{2}\tilde{\openone}\right) - \frac{1}{2}
\right|.\label{eqn:pM}
\end{align}
Together, $\assignerr$ and $\biaserr$ inform not only how distinguishable the experiments are from the ideal behavior, but also what aspect of the experiment deviates the most.  Further details of these operational error metrics are discussed in \Cref{app:MBQB-details}, including how the experiments used to estimate them can be used to reconstruct the full action of the measurements on a quantum state.

It is worth considering some concrete examples of non-ideal behavior. If the measurement operations yield uniformly random outcomes, such as in the case where switching the measurement on and off completely randomizes the state of the qubit, $\assignerr=1/2$ and $\biaserr=0$. 
 On the other hand, if the quantum device is able to implement the measurement projector perfectly but noise in the readout chain, {\it e.g.,} due to the amplifiers, flips the recorded measurement with probability $p_f$,
 one would find $\assignerr = 2p_f(1-p_f)$ and $\biaserr=0$.
Finally, if both measurements are actually identical instead of anticommuting, but otherwise perfect, then $\assignerr=0$ and $\biaserr=\frac{1}{2}$. 

The experiments implied by the definition of the $\assignerr$ and $\biaserr$ metrics assume the ability to prepare $\frac{1}{2}\tilde{\openone}$, an approximation to the maximally mixed state. In practice, the maximally mixed state can be approximated by applying a sequence of approximately non-commuting measurements without conditioning on the outcomes.  Estimating $\assignerr$ and $\biaserr$ then requires estimating the probability of subsequences 
\begin{align} \label{eqn:rqm-seqs}
    \mop{P}{r} \mop{Q}{s} & \mop{R}{*} \mop{S}{*}, & \mop{P}{r} \mop{P}{s} & \mop{R}{*} \mop{S}{*},
\end{align}
where $P\neq Q, R\neq S \in \{X,Z\}$ and the $*$ indicates that we do not care about that particular measurement outcome.  
One can then imagine running an experiment consisting of a long random or pseudorandom sequence of measurements and counting the separate occurrence frequencies of \Cref{eqn:rqm-seqs} as subsequences, see \Cref{app:MBQB-details}.

While the above definitions could be generalized to include $Y$ measurements, initial demonstrations of topological qubits will be restricted to use $X$ and $Z$ measurements for simplicity of device control, as shown in \Cref{fig:MV11}. Under these restrictions, only the real part of the states and operators acting on the states can be fully reconstructed, but that suffices for the reconstruction of the action of $X$ and $Z$ measurements and for the definition of the natural error metrics introduced above (see \Cref{app:MBQB-details} for more details).
MBQB can be generalized in other ways. For instance, the measurement sequence could include experimentally distinct implementations of the $X$ and $Z$ measurements ({\it e.g.}, sometimes measuring $X$ using $i\gamma_1\gamma_3$ and sometimes measuring $X$ using $i\gamma_2\gamma_4$, which is equivalent in the computational subspace). It is also possible to include two-qubit measurements in the sequence.

MBQB allows us to extract operational error metrics for a qubit assuming only the ability to perform sequences comprising two non-commuting Pauli measurements. However, at least one unitary outside the Clifford group is needed to perform universal quantum computation~\cite{Barenco95}. A $\pi/4$ rotation around $Z$, known as the $T$ gate, is often chosen as the operation to complete a universal set of gates, but this is not a topologically-protected operation in Majorana-based qubits. However, a noisy implementation of the $T$ gate together with low-noise Clifford operations can be used to distill low noise resource states from which fault-tolerant $T$-gates can be implemented~\cite{Knill04,Bravyi05}. We review implementing a $T$-gate in tetrons in \Cref{app:T-gates}; the required Clifford operations can be performed through topologically protected single- and two-qubit measurements, as described in \Cref{sec:braiding-protocols}.

\subsection{Device design}

\subsubsection{Design elements}

\Cref{fig:MV11} shows a gate schematic of the two-sided tetron (for a specific realization and measurements of such a device, see Ref.~\onlinecite{Aghaee25}).
The principal elements of the design are: 
(1) two parallel topological superconducting nanowires (blue), with MZMs at their ends (orange dots); 
(2) a trivial superconducting nanowire that connects the midpoints of the topological wires (light blue),
so that the three wires together form a (sideways) `H';
(3) five quantum dots (with chemical potential controlled by second-layer plunger gates); 
(4) tunnel junctions (controlled by second-layer cutter gates) coupling these quantum dots to each other as well as to the MZMs.
As in the case of the linear tetron device described in Ref.~\onlinecite{Aghaee24} (as well as the devices of Refs.~\onlinecite{Nichele17,Suominen17,Aghaee23}), the nanowires are gate-defined and can be realized in a material stack featuring a two-dimensional electron gas (2DEG) formed in an InAs quantum well that is proximitized by an epitaxially-grown $s$-wave superconductor such as Al.
The most crucial differences between the two-sided design and the linear tetron device described in Ref.~\onlinecite{Aghaee24} are that the topological superconducting segments are no longer colinear and the trivial superconducting segment, which is perpendicular to them, remains far from the MZMs. The placement of MZMs at the four corners of the H is advantageous for multi-qubit layouts with 2D connectivity. As the qubit islands are floating, a complete theoretical treatment of their ground state degeneracy requires a number-conserving formalism, as discussed in Refs.~\onlinecite{Fidkowski11b,Sau11b,Knapp18a,Knapp20}.

The ``horizontal'' topological superconducting wires can be gate-defined as follows. 
Each ``horizontal" superconducting strip can be covered by a ``plunger" gate that is used to tune the density of electrons in the semiconductor under the corresponding superconducting wire. 
The wires are narrow enough that the semiconductor can be tuned into the single-subband regime~\cite{Aghaee23}, and their length is chosen to minimize residual coupling between the Majorana zero modes while still supporting sufficient charging energy. For material stacks similar to the one used in Ref.~\onlinecite{Aghaee24}, a length of $\SIrange{3}{5}{\micro\meter}$ meets these criteria.  
Under normal operating conditions, there is an in-plane magnetic field of several Teslas
in the ``horizontal'' direction, {\it i.e.,} along the topological wires. 
The material stack and device can be designed~\cite{Antipov18, Winkler19} so that there is a range of in-plane magnetic fields and plunger gate voltages (within the lowest
sub-band) over which both of the horizontal wires are in the topological phase with MZMs at the ends, provided disorder is sufficiently low~\cite{Motrunich01, Brouwer11a, Brouwer11b, Akhmerov11, Stanescu11, Kells12, Prada12, DeGottardi13, Adagideli14,Reeg18b, Vuik19, Woods21, Pan20b, DasSarma21, DasSarma23, DasSarma23b, Pan24, Antipov25}.
The ``vertical" trivial superconducting wire connects the topological segments (the cross-bar of the H) and needs to be engineered in such a way that it
does not introduce low-energy subgap states.
The distance between the horizontal wires is approximately $\SI{1}{\micro\meter}$, dictated by requirements on the quantum dots discussed below.

There is a quantum dot adjacent to each endpoint of the horizontal wires and, thus, to the MZM that resides there. There is also a longer quantum dot parallel to the bottom wire that enables connections between the left and the right side of the device. These quantum dots have a form and function similar to those of the device used in Ref.~\onlinecite{Aghaee24}, with the extended dot mirroring the dimensions of the extended dots on the linear tetron device, and the dots adjacent to the bottom wire extended somewhat to connect with the long quantum dot. All the dots can be designed so that their charging energies and level spacings are large compared to the electronic temperature.
Unlike in the linear tetron device, the junctions connecting these quantum dots to the topological superconducting wires are at the ends of the wires.

Related methods for parity measurement
were discussed in
Refs.~\onlinecite{Akhmerov09,Fu09a,Fu10,Burnell13,Houzet13, Cheng15, Plugge16, Plugge17, Vijay15, Vijay16a, Vijay16b, Thamm21, Drechsler24,Sau24}.
Analogous device architectures
might apply to other physical systems
that have been proposed for MZMs,
including quasi-one-dimensional systems composed of chains of magnetic atoms on the surface of a superconductor~\cite{Nadj-Perge13, Klinovaja13, Braunecker13, Pientka13, Nadj-Perge14, Zhang16b}; in nanowires that are completely encircled by a superconducting shell in which the order parameter winds around the wire due to the orbital effect of the magnetic field~\cite{Cook11, Hosur11, Vaitieknas20}; in the vortex cores of three-dimensional superconductors
\cite{Wang18, Kong19}; in
two-dimensional $p + ip$ superconductors~\cite{Read00}; at the surface of a topological insulator~\cite{Fu08, Fu09, Hasan10}; in ferromagnetic insulator-semiconductor-superconductor heterostructures~\cite{Sau10a, Alicea10, Sau10b, Chung11, Duckheim11, Potter12, Lutchyn18, Liu19a, Laubscher24}; in $s$-wave superfluids of ultra-cold fermionic atoms~\cite{Sato09b, Zhang08}; and
potential non-Abelian fractional quantum Hall
states \cite{Willett87,Pan99b,Kumar10,Zibrov17}.
In the latter system, interferometric transport
measurements of fermion parity were proposed
in Refs.~\onlinecite{Chamon97, Fradkin98, Bonderson06a, Stern06},
and intriguing results have been observed in
experiments \cite{Willett10,Nakamura20}.

\subsubsection{Operating principles and dominant error sources}\label{sec:MV11-ops}

The quantum state of a tetron is encoded in four Majorana zero modes (MZMs), which can be numbered as shown in \Cref{fig:MV11}. Assuming a fixed MZM parity $-\gamma_1\gamma_2\gamma_3\gamma_4=+1$, we can define Pauli operators~\footnote{When the overall island parity can change, the Pauli operators depend on the overall MZM parity.}. 
\begin{align}
Z &= i \gamma_1 \gamma_2 = i \gamma_3 \gamma_4 \label{eqn:Z}\\
X &= i \gamma_1 \gamma_3 = -i \gamma_2 \gamma_4 \label{eqn:X}.
\end{align}
Since topological qubits have near-degenerate ground states, there is no preferred ``computational basis``, and the assignment of Pauli operators above is purely a matter of convenience. With the choice made above, the basis states of the qubit can be chosen as the eigenstates of the $Z$ operator, i.e. $Z\ket{0}=+\ket{0}$ and $Z\ket{1}=-\ket{1}$. 

The primary operations supported by the single-qubit device are measurements of $X$ and $Z$, performed by forming interferometric loops as indicated in \Cref{fig:MV11}. By enabling single-electron tunneling between the Majorana zero modes and an adjacent chain of quantum dots, a state-dependent shift of the energy spectrum of the quantum dots is induced. This shift can be measured as a change in the capacitance of a nearby gate to the quantum dot (referred to as \emph{quantum capacitance}), which in turn is detected as a frequency shift of a microwave resonator that is coupled to that gate. This approach is explained in more detail in Refs.~\onlinecite{karzig2017scalable} and has been experimentally demonstrated in Ref.~\onlinecite{Aghaee24}. Qubit states are prepared using such measurements, which need to be followed either by post-selection, a Clifford operation to correct the outcome, or classical tracking of the Pauli reference frame; in practice, the latter approach is usually favorable (see \Cref{sec:braiding-protocols}). Qubit states can be transformed into each other using the approaches discussed in \Cref{sec:braiding-protocols}.

While topological qubits are affected by errors due to imperfect control, coupling to the environment, and inherent physical limitations, their topological protection suppresses these errors exponentially in the dimensionless quantities discussed in \Cref{sec:intro}.  
These errors can be separated into classical errors that flip the measurement outcomes, and quantum errors that couple to the MZMs.  
The former are limited by the signal-to-noise ratio (SNR), and occur with probability~\footnote{\Cref{eqn:pa} assumes that measurement corresponds to distinguishing two Gaussians of equal width $\sigma$ for an SNR definition of the distance between Gaussians divided by $2\sigma$.} 
\begin{align}\label{eqn:pa}
    p_a &= \left[ 1-\text{erf} \left( \frac{\text{SNR}(\tau_\text{meas})}{\sqrt 2} \right) \right]/2.
\end{align}
Typically, $\text{SNR}(\tau_\text{meas})\sim \sqrt{\tau_\text{meas}}$ for sufficiently long measurement times, and $\lim_{x\to\infty} \left[1-\text{erf}(x)\right]\sim e^{-x^2}/x$, thus the probability decays exponentially with increasing $\tau_\text{meas}$. The interferometric parity measurement proposed here was demonstrated in Ref.~\onlinecite{Aghaee24},
where an SNR of $0.52$ was achieved for a measurement time of $\SI{1}{\micro\second}$. We comment on future improvements to the SNR in \Cref{sec:outlook}.

Errors on the quantum state, on the other hand, become more likely when the measurement time is increased because the qubit state may change during the measurement. The typical timescale for undesired changes of the quantum state is denoted by $\Tlife$.  The resulting error probability during a measurement is expected to scale as 
\begin{align} \label{eqn:p1}
    p_1 &\sim  1-e^{-\tau_\text{meas}/\Tlife} .
\end{align}  
While in conventional qubits, $\Tlife$ is governed by relaxation into the lower-energy state and typically denoted by $T_1$, in topological qubits, there is no strongly preferred basis and $\Tlife$ will have contributions from processes other than relaxation.  The processes predominantly limiting $\Tlife$ are due to (i) thermal fluctuations exciting a fermion from the MZMs to an above-gap quasiparticle state (scaling as $e^{-\Delta/k_B T}$)~\footnote{Additionally, non-equilibrium quasiparticles may contribute to the same error process with different scaling than thermally excited quasiparticles, however for the expected qubit volumes these are estimated to be subleading. This is in contrast to Ref.~\onlinecite{Aghaee24}, where a lifetime of $\SI{2}{\milli\second}$ was observed in a configuration where all but one segment of the qubit device are tuned into a trivial superconducting regime. As such a configuration does not support a qubit, many of the error processes discussed here do not apply, and non-equilibrium quasiparticles remain as the dominant source of parity flips.} and (ii) state flips due to the residual coupling between a MZM involved in a measurement and a MZM not involved in a measurement.
Both errors can be modeled phenomenologically as Pauli errors, see Appendix~\ref{app:noise-model-details} for details~\footnote{In the MBQB demonstration, it is never necessary for qubits to idle.  Idle qubits are additionally subject to coherent rotations along the axis of the residual MZM coupling; these rotations result in an error since the measurement-based approach cannot be formulated in a rotating frame.}.

To perform an $X$ (or $Z$) measurement,
the quantum dots that lie on the
$X$ ($Z$) loop must be
coupled to the qubit island
while those on the $Z$ ($X$) loop
must be decoupled from the island.
In an MBQB demonstration, the
device must switch between
these two configurations.
This can be done by coupling/decoupling
quantum dots from the qubit island
in one of two ways.

The ``detuning-based" approach decouples
a dot from the qubit island and
other dots by setting its chemical potential to lie in a Coulomb valley.
We couple a dot to the qubit
island by tuning it near its charge
degeneracy point (which also
couples it to other dots that are
near their charge degeneracy points).
More explicitly, a detuning-based $X$ measurement pulse sequence begins from an idle qubit configuration (all quantum dots tuned to Coulomb valleys), then tunes the quantum dots touched by the loop shown in the left panel of \Cref{fig:MV11} near the charge degeneracy point, stays in this configuration for the measurement time, then tunes the same quantum dots back to Coulomb valleys to end back in the idle configuration.
For this pulse sequence, the quantum dots can be tuned using any gate with significant lever arm to the quantum dot of interest and small lever arm to other regions, {\it e.g.,} the quantum dot plunger gates shown as larger second-layer gates in \Cref{fig:MV11}.

Alternatively, the ``cutter-based''
approach additionally uses a cutter gate (the small rectangular second-layer gates shown in \Cref{fig:MV11})
to open or close a junction, thereby decoupling a dot on one side of
the junction from the
qubit island (or dot) on the other
side of the junction.
A cutter-based pulse sequence again begins from an idle qubit configuration, which in this case corresponds to all quantum dots tuned to Coulomb valleys and all cutter gates pinched off.  
In the first step of the sequence, cutter gates intersected by the desired measurement loop are opened to allow tunnel-coupling between the quantum dots and MZMs.  
Then, the quantum dots intersecting the desired measurement loop are tuned near charge degeneracy.
This measurement configuration is maintained for the length of the measurement time, then the process is reversed (first detuning the quantum dots, then closing the cutter gates) to return the qubit to its idle configuration.

The detuning-based approach uses small amplitude pulses on the plunger gates (or equivalently on any gate with significant lever arm to the quantum dot of interest and small lever arm to other regions).
However, in this approach, the coupling through the $X$ ($Z$) loop is only suppressed by a factor $\sim t^2/E_C^{\text{QD}}$ during the $Z$ ($X$) measurement, where $t$ is the tunnel coupling between dots or dot and wire and $E_C^{\text{QD}}$ is the QD charging energy. This residual coupling
limits $\Tlife$ (and makes the measurement-bases not-strictly orthogonal) because the
unwanted coupling fluctuates
as a result of charge noise, thereby
decohering the qubit.
With cutter-based control,
the residual coupling can be made
exponentially small in the width of the tunnel barrier~\footnote{In practice, the band gap $E_g$ of the semiconductor sets a limit $\mathcal{O}(e^{-E_g/k_B T})$.}, so
qubit coherence is limited instead by the residual coupling of the MZMs through the qubit island. This scales as $e^{-L/\xi}$ where $L$ is the length of the topological wires and $\xi$ is the superconducting coherence length in the topological phase. Note that a $Y$ measurement in the device shown in \Cref{fig:MV11}, {\it e.g.,} $i\gamma_1\gamma_4$, requires cutter-based control as the $Z$ and $X$ loops are only closed off by the cutter adjacent to MZM $\gamma_3$.

\section{Measurement-based braiding transformations}
\label{sec:MV21}

Unitary gate operations can be performed via measurements with the aid of auxiliary degrees of freedom~\cite{Nielsen2003,Perdrix2005}.
Such measurement-based operations on topological qubits, pioneered in Ref.~\onlinecite{Bonderson08a}, preserve topological protection, while removing the need to physically move or adiabatically couple anyons~\cite{Alicea11,Sau11a,Heck12,Hyart13,Aasen16}. (Braiding schemes that rely on adiabatic braiding protocols are subject to diabatic errors that in general are not exponentially suppressed in macroscopic parameter ratios of the systems~\cite{Cheng09,Scheurer13,Hell16,Knapp16}.)
In the case of tetrons, Clifford operations are topologically protected and can be enacted entirely through single-qubit and two-qubit (2-MZM and 4-MZM) measurements involving auxiliary tetrons.
For example, to perform a full set of single-qubit Clifford operations, a system of two qubits is required, where one serves as the computational qubit and one as an auxiliary qubit.
In this section, we propose the corresponding measurement sequences for such measurement-based single-qubit Clifford operations and introduce a design for a device that is able to execute them.
We simulate the fidelity of one such measurement sequence subject to a tetron-motivated noise model.

It is worth noting that performing Clifford gates directly on physical qubits is not necessary for scalable quantum computation in the architecture described in this paper. Instead, Clifford gates are performed directly on logical qubits through so-called lattice surgery operations (see also the discussion in \Cref{sec:MV42}), and the only required physical operations are the measurements needed to perform error correction and the $T$-gate to achieve universality. 
Nevertheless, performing such Clifford gates via measurement-based braiding transformations is an important step on this roadmap as it implies a powerful demonstration of the non-Abelian nature of MZMs~\footnote{It is worth noting that braiding statistics have been experimentally emulated in qubit systems~\cite{Song2018,Satzinger2021,Stenger2021,Google2023,Xu2024,Iqbal2023a,Iqbal2023b}. However, these experiments differ from topologically protected braiding transformations as discussed in this paper in that the topological state is not prepared as the ground state of the system's Hamiltonian, but through a quantum circuit without active error correction. Therefore, there is no topological protection (such as that endowed by the topological gap) of these operations.}.
While these Clifford gates allow for a comparison to conventional qubit platforms, a more natural comparison is the fidelity of operations used in a quantum error correction syndrome extraction circuit, which in our architecture are the Pauli measurements themselves, and in other platforms include the Clifford gates.

As the measurements involved are topological, the operations generated by these measurement sequences directly reflect the fusion and braiding statistics of the corresponding topological phase~\cite{Bonderson08a}. 
However, the charging energy of a superconducting island fixes the total fermion parity of the four MZMs on each tetron, so we can only perform operations that preserve this parity~\cite{Karzig17,Knapp20}.
Equivalently, all measurements must involve an even number of Majorana operators on each island. This is sufficient to generate a complete set of Clifford gates.

We note also that this approach can be extended to multi-qubit Clifford operations; for a more complete discussion, see Refs.~\onlinecite{Karzig17,Tran20}.
On the other hand, going beyond Clifford operations to obtain a computationally universal gate set requires additional, non-topologically-protected operations. 
As such operations do not limit the performance of a fault-tolerant quantum computer based on tetrons~\cite{beverland2022assessing}, they are not the focus of this paper, but their implementation in tetrons is reviewed in \Cref{app:T-gates}. 

\subsection{Protocols}
\label{sec:braiding-protocols}

\begin{figure}
    \centering
    \includegraphics[width=\columnwidth]{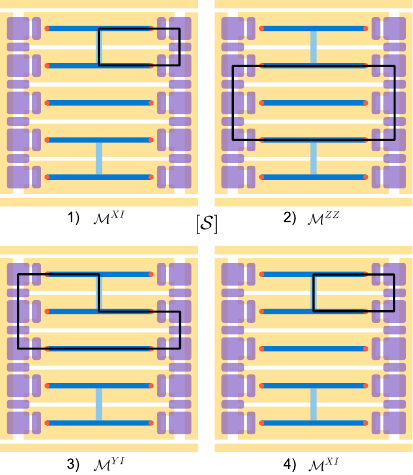}
    \caption{
    Measurement sequence for implementing $[\sop{S}]$ in \Cref{eqn:S} on the two-qubit device (same legend as in \Cref{fig:MV11}). The sequence is equivalent to applying a braid to MZMs $1$ and $2$ on the computational (bottom) qubit, see \Cref{app:braiding-details} for the explicit connection.}
    \label{fig:braiding-loops}
\end{figure}

An ideal measurement of two-qubit Pauli $PQ$ with outcome $s \in \{\pm 1\}$ corresponds to the projector  
\begin{align}
    \mst{PQ}{s} &= \frac{\openone + s \cdot PQ}{2}.
\end{align}
Using this, one can verify that the phase gate $S$ applied to the computational qubit can be generated, up to a Pauli correction~\footnote{Although these Pauli operations are commonly referred to as corrections, they are very different in nature from error corrections inferred by a decoder associated with an error correction code. In this context, they are simply operations that depend on previous non-deterministic measurement outcomes, like the corrections that arise in quantum teleportation.}, by the sequences of four measurements shown in \Cref{fig:braiding-loops}: $\instr{XI} \instr{YI} \instr{ZZ} \instr{XI}$.
(Here, the sequence of measurements is to be read right to left, like operators applied to a state.)
In particular, we find
\begin{align}
\mst{XI}{s_3} \mst{YI}{s_2} \mst{ZZ}{s_1} \mst{XI}{s_0} &= \mathcal{O}^{s_0}_{s_3} \otimes Z^{\frac{1+s_0 s_1 s_2}{2}} S
,
    \label{eqn:Sexplicit}
\end{align}
where the operator $\mathcal{O}^{s_0}_{s_3}$ on the auxiliary qubit is a shorthand for $\ket{X_{s_3}}\bra{X_{s_0}}$.
(The equivalence in Eq.~\eqref{eqn:Sexplicit} is up to overall constants, which are unimportant for the effective operations generated.)
The diagrammatic representation of this measurement sequence, shown in Appendix~\ref{app:braiding-details}, illuminates the relation to the corresponding braiding transformation.
We note that this measurement sequence requires only a single two-qubit measurement. 
Furthermore, the first and last measurement in the sequence initialize and reset the auxiliary qubit in the $X$ basis, and therefore one of them could potentially be omitted in a series of gate operations, though this comes with the potential risk of propagation of errors between subsequent Clifford operations.
Finally, this measurement sequence is not unique; in general, many different sequences can be used to generate a given Clifford gate.
The optimal choice of sequence will depend on the details of the underlying hardware implementation.
Here, we have chosen sequences that are well-suited to the qubit design described below in \Cref{sec:two-qubit-design}; for a broader discussion of this optimization, see Ref.~\onlinecite{Tran20}.

We note that the Pauli correction arising from the measurement sequence in \Cref{eqn:Sexplicit} (or from similar operator-generating measurement sequences) is not deterministic, since the measurement outcomes are probabilistic.
In principle, these outcomes can be made deterministic by employing a repeat-until-success protocol called ``forced measurement'' \cite{Bonderson08a,Tran20}, where measurements in the sequence are replaced by (longer) adaptive measurement sequences.
However, this is not necessary: instead, the Pauli corrections can be tracked in software, where they lead to a classical correction of the final measurement outcome or potentially a modification of subsequent non-Clifford operations.
Therefore, it is convenient to work directly with the Pauli equivalence class $[C]$ of a Clifford gate $C$, defined as $[C] = \lbrace C' \in \mathfrak{C} | \exists P : PC'=C \rbrace$, where $\mathfrak{C}$ is the Clifford group of unitaries acting on a single qubit, and $P$ is a Pauli operator. Any given element of the equivalence class can be accessed through changes to the Pauli frame~\cite{Knill05}, without any additional operations on the qubits themselves.

While the single-qubit Clifford group has 24 elements, there are only 6 Pauli equivalence classes, which we denote $[\openone]$, $[H]$, $[S]$, $[HSH]$, $[HS]$, and $[SH]$~\footnote{Here $H$ denotes the Hadamard gate, while $S$ denotes the phase gate.}.
For the two-qubit device described in the next section, examples of optimized measurement sequences for the nontrivial Pauli equivalence classes of single-qubit Clifford gates are
\begin{align}
    [\sop{H}] &= \instr{XI}\instr{YI}\instr{ZY} \instr{XI}, \label{eqn:H}
    \\ [\sop{S}] &= \instr{XI} \instr{YI} \instr{ZZ} \instr{XI}, \label{eqn:S}
    \\ [\sop{HSH}] &= \instr{XI} \instr{ZY} \instr{ZZ} \instr{XI}, \label{eqn:HSH}
    \\ [\sop{SH}] &= \instr{XI} \instr{YI} \instr{ZY} \instr{ZZ} \instr{XI}, \label{eqn:SH}
    \\ [\sop{HS}] &= \instr{XI} \instr{YI} \instr{ZZ} \instr{ZY} \instr{XI}. \label{eqn:HS}
\end{align}
Here we use $\sop{C}$ to represent the superoperator corresponding to the unitary $C\in \{H,S,HSH,SH,HS\}$, and use $\instr{P}$ to represent the instrument used to measure the observable $P$.
The corresponding Pauli corrections are elided for brevity, but can be efficiently computed (see Appendix~\ref{app:braiding-details}).

To experimentally validate successful execution of such measurement-based protocols, we can tomographically reconstruct the action of each of the measurement sequences on the computational qubit. The impact of state preparation and measurement (SPAM) errors can be separated and accounted for through reference experiments, and the most robust way to do so is through gate-set tomography~\cite{Nielsen21}. To this end, we prepare the computational qubit in the eigenstate of a Pauli operator via measurement, and characterize the final state in the Pauli basis, leading to the nine inequivalent sequences~\footnote{Similar to how partial gate-set tomography is performed for X and Z measurements, the sequences must be prefixed with a randomized state preparation sequence, which we elide here for brevity. See \Cref{app:MBQB-details} for details.}
\begin{subequations}
\begin{align}
\instr{IX}[\mathcal{C}]\instr{IX}&, & \instr{IY}[\mathcal{C}]\instr{IX}&, & \instr{IZ}[\mathcal{C}]\instr{IX}&,\\  
\instr{IX}[\mathcal{C}]\instr{IY}&, & \instr{IY}[\mathcal{C}]\instr{IY}&, & \instr{IZ}[\mathcal{C}]\instr{IY}&,\\  
\instr{IX}[\mathcal{C}]\instr{IZ}&, & \instr{IY}[\mathcal{C}]\instr{IZ}&, & \instr{IZ}[\mathcal{C}]\instr{IZ}&.
\end{align}
\end{subequations}
The reference experiments correspond to similar experiments replacing $[\mathcal{C}]$ with a trivial operation, resulting in the additional nine sequences
\begin{subequations}
\begin{align}
\instr{IX}\instr{IX}&, & \instr{IY}\instr{IX}&, & \instr{IZ}\instr{IX}&,\\  
\instr{IX}\instr{IY}&, & \instr{IY}\instr{IY}&, & \instr{IZ}\instr{IY}&,\\  
\instr{IX}\instr{IZ}&, & \instr{IY}\instr{IZ}&, & \instr{IZ}\instr{IZ}&.
\end{align}
\end{subequations}
In its simplest form, the gate set tomography procedure solves a linear problem to obtain a superoperator representation $\sop{C}$, including imperfections due to noise, implementation imperfections, and 
finite statistics. Error metrics such as gate fidelity can be directly extracted by comparing $\sop{C}$ to the ideal superoperator. It is worth noting that such metrics for the execution of unitary gates allow a direct comparison to more conventional qubit platforms.

\subsection{Two-qubit device design}
\label{sec:two-qubit-design}

\begin{figure}
    \centering
    \includegraphics[width=\columnwidth]{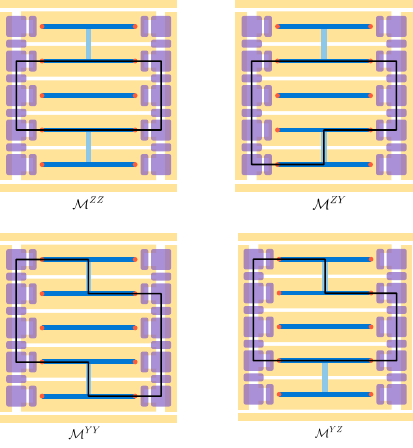}
    \caption{Two-qubit measurement loops supported on the two-qubit device, with the same legend as in \Cref{fig:MV11}.  }
    \label{fig:MV21}
\end{figure}

The two-qubit device supporting a measurement-based braiding demonstration builds off the single-qubit device described in the previous section.  
Two tetrons are stacked vertically, such that the device supports all single-qubit Pauli measurements and the two-qubit measurements: $\instr{ZZ}, \instr{YY}, \instr{YZ}, \instr{ZY}$. 
A device schematic with these two-qubit loops is shown in \Cref{fig:MV21}. 

A key design change is that the longer quantum dot of the single-qubit device is replaced by a {\it coherent link} (floating topological wire), which is situated between the two tetrons.
This component is used to facilitate measurements involving MZMs on opposite sides of a tetron ({\it i.e.}, single qubit $Z$ and $Y$ measurements).
For the long quantum dot, the relevant energy scale to low-lying excitations is the level spacing, which decreases with length; for the coherent link, on the other hand, the relevant scale is the topological gap. Therefore, switching to a coherent link facilitates using longer topological wires and thus achieving smaller MZM energy splitting through the topological wires of the qubits, which helps increase qubit lifetime.

\begin{figure}
    \centering
    \includegraphics[width=\columnwidth]{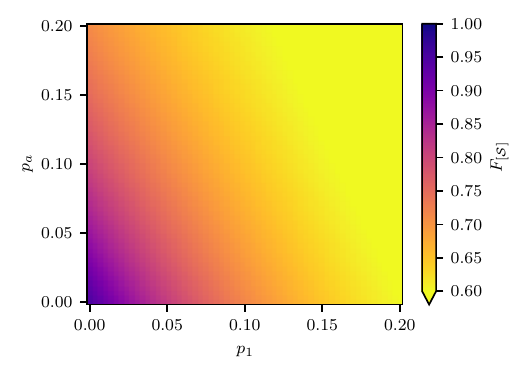}
    \caption{Fidelity of $[\sop{S}]$ (\Cref{eqn:Fid-def}) plotted as a function of single-qubit Pauli error probability $p_1$ and classical assignment error probability $p_a$ for two-qubit Pauli error probability $p_2=0.1$. The estimates are obtained through numerical simulation of process tomography.}
    \label{fig:braiding_sim}
\end{figure}

Two-qubit measurements introduce the potential for correlated two-qubit errors.  Such errors can occur from, {\it e.g.}, tetrons exchanging an electron through the MZMs during the measurement, such that the final charge states of the qubit islands have changed before and after the measurement.  These two-qubit state errors can also be modeled as Pauli errors, see \Cref{app:noise-model-details} for details~\footnote{Residual MZM coupling through the wire can also result in coherent rotations when the measurement basis commutes with the MZM coupling basis.  For example, MZM overlap along the same wire results in a single-qubit $Z$ rotation during a $ZZ$ measurement, which does not affect that measurement outcome but can affect a subsequent non-commuting measurement.}.

We simulate the measurement sequence to implement $[\sop{S}]$ (\Cref{eqn:S}) for the effective noise model outlined above and described in more detail in \Cref{app:noise-model-details}.
Using process tomography to reconstruct the noisy circuit while accounting for the Pauli correction associated with the different measurement outcomes (\Cref{eqn:Sexplicit}), we find its associated noisy channel $\tilde{\sop{S}}$~\footnote{Process tomography and gateset tomography accomplish similar tasks (the reconstruction of a noisy quantum evolution), but process tomography as usually described~\cite{Nielsen00} assumes knowledge of the state preparation and measurements used to reconstruct the evolution. As we are describing numerical simulations, we have knowledge of preparation and measurement details, so we can rely on the simpler process tomography procedure instead of gateset tomography.}.
Comparing this channel to the ideal Clifford group unitary $S$, we define the associated average fidelity between the noisy channel and the ideal unitary as the output state fidelity averaged uniformly over all initial pure states~\cite{Nielsen2002}
\begin{align}\label{eqn:Fid-def}
    F_{[\sop{S}]} &\equiv \int d\psi~\tr\left[\tilde{\sop{S}}(\proj{\psi})~S\proj{\psi}S^\dagger\right]\\
    & = \int d\psi~\bra{\psi} S^\dagger~\tilde{\sop{S}}(\proj{\psi})~S\ket{\psi}.
\end{align}
The dependence of $F_{[\sop{S}]}$ on the single-qubit Pauli error probability $p_1$ and classical assignment error probability $p_a$ are illustrated in \Cref{fig:braiding_sim} for fixed two-qubit Pauli error probability $p_2=0.1$.  The gate fidelity is only weakly dependent on $p_2$ as there is only one two-qubit measurement in the sequence (see \Cref{app:braiding-details} for additional details).

\section{Quantum error detection in a $4 \times 2$ array of tetrons}
\label{sec:MV42}

\begin{figure}
    \centering
    \includegraphics[width=\columnwidth]{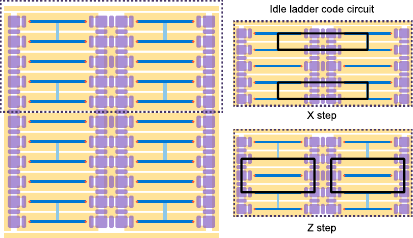}
    \caption{ {\it Left panel.}  Gate schematic for a $4\times 2$ qubit array following the same legend as in \Cref{fig:MV11}.  The two columns of qubits are connected by a double-rail of quantum dots separated by depletion gates and a dot cutter.  {\it Right panel.} Syndrome extraction circuit for the idle ladder code: the $X$ step is alternated with the $Z$ step as shown; after each step the instantaneous stabilizer group corresponds to the two-qubit measurements that were just performed, as well as the product $YYYY$ on the four qubits, inferred from the measurement outcomes of the previous two steps.}
    \label{fig:mv42_and_idle}
\end{figure}

\subsection{Overview}

Fault-tolerant quantum error correction~\cite{shor96,Steane1996,Dorit1997,Knill1998,kitaev2003fault,Terhal2005,aliferisgottesmanpreskill05} is a crucial ingredient for any scalable quantum computing platform.
Recently, the field of quantum error correction has advanced drastically from demonstrations with encoded states (and no repeated error correction)~\cite{willsch_2018,vuillot_2018,linke17edexp,cane2021,hong24cat4onh2,bluvsteinharvard23neutralatoms} to successful demonstrations of repeated error correction across several platforms. 
In ion traps, both repeated error detection~\cite{self22icebergcode,yamamoto23phaseestimattionerrordetection}, error correction~\cite{honeywell21steane,postler23steaneec,silva24microsoft12qubitcode}, and error corrected computation~\cite{reichardtmicrosoft24tesseract} have been performed---in some notable cases demonstrating logical error rates below the corresponding physical error rates~\cite{silva24microsoft12qubitcode,reichardtmicrosoft24tesseract}. 
In superconducting qubits, recent progress on repeated error detection and correction~\cite{harper19edd,krinner21repeatedsurfaceec,google23surface,amazon24catec,ofek2016extending,sivak2023real,ni2023beating} has culminated in a demonstration of sub-threshold error correction in the surface code~\cite{google24surfacecode,eickbusch2024demonstrating} and the color code (including logical operations)~\cite{lacroix2024scaling}.

For tetron qubits, a very natural class of codes are the Hastings-Haah Floquet codes~\cite{Hastings21,Haah2022,Gidney2021b,Gidney2022,Paetznick23}.  These codes are closely related to the surface code, but rely entirely on one- and two-body measurements to extract error syndromes, making them very naturally well suited to the capabilities of the tetron platform. Within the class of Floquet codes, the ladder code is an intriguing first demonstration, as it uses few qubits but builds on the same set of two-qubit measurements as the scalable Hastings-Haah Floquet codes and thus falls naturally on a roadmap towards scalable quantum computation.

To demonstrate a non-trivial improvement of system performance, it is important to go beyond idle operation of such a code to demonstrate a logical operation between two qubits. A promising way to perform such logical operations at scale is \emph{lattice surgery}~\cite{Horsman12}: here, two logical qubit patches are stitched together to form a single, larger patch, which effectively performs a two-logical-qubit measurement. In a later step, the enlarged patch is split back into two separate logical qubits. Leveraging these two-qubit measurements, logical Clifford gates can be performed in a way that is conceptually very similar to the measurement-based braiding transformations discussed in \Cref{sec:braiding-protocols}. Surgery is generally accomplished by turning on additional stabilizer measurements connecting the boundaries of the two code patches, with details depending strongly on the specific code and circuit realization being used. As we discuss below, such a surgery operation is particularly simple to implement in the ladder code being used here.
The logical two-qubit measurement performance obtained through this scheme can be compared to the physical two-qubit measurement performance by extracting the logical error per round from a repetition code decay experiment.
Demonstrating logical improvement for a two-qubit measurement is a key advance on the path to scalable fault-tolerant quantum computation with Majorana-based qubits, as it directly compares the native operation of the physical qubits with the corresponding operation of the logical qubits.

\subsection{Ladder code}

The ladder code can be defined on an array of $N \times 2$ tetrons ($N>1$), and relies on $XX$ measurements between the columns as well as $YY$ and $ZZ$ measurements between rows of qubits. Similar to the repetition code, its logical operators are highly asymmetric in weight; therefore, for sufficiently large $N$, it is able to correct errors in one basis but not the other~\footnote{For the scheme described below, the ladder code can only correct $X$ errors.}. For $N=2$, the ladder code is equivalent to the $d=2$ Bacon-Shor error detecting code~\cite{bacon05operator,shor1995scheme}. It is important to note that due to the structure of the circuits used to perform error detection, some elementary faults, such as correlated two-qubit errors occurring during a single two-qubit measurement, will remain undetected. 

The idle ladder code circuit on a $2\times2$ array of qubits has two steps shown in the right panel of \Cref{fig:mv42_and_idle}: in the first step an $XX$ measurement is performed on the two pairs of horizontal nearest neighbor qubits, while in the second step a $ZZ$ measurement is performed between the two pairs of vertical nearest neighbor qubits.   Let $\logical{P}$ denote the logical Pauli $P$ of the ladder code.
The logical group generators are $\logical{Z}=ZZ$ between horizontal nearest neighbor qubits and $\logical{X}=XX$ between vertical nearest neighbor qubits.  The instantaneous stabilizer group after a given step corresponds to the product of $Y$ on all four qubits as well as the two-qubit measurements performed at that step; any change in the stabilizer outcome indicates an error has occurred. 

\begin{figure}
    \centering
    \includegraphics[width=\columnwidth]{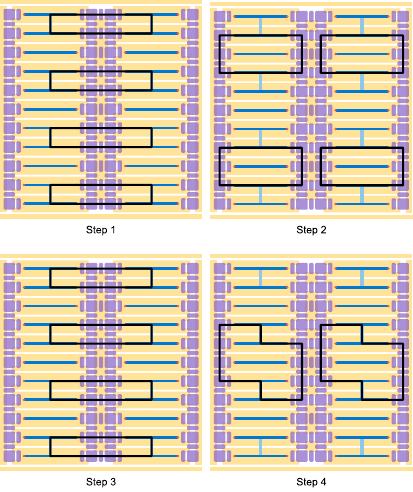}
    \caption{ The four steps of the $\logical{ZZ}$ measurement circuit for the ladder code (same legend as in \Cref{fig:MV11}).  The $\logical{ZZ}$ measurement outcome is inferred after steps 4 and 1 (assuming at least one previous round of the circuit). }
    \label{fig:protological_circuit}
\end{figure}

The eight-qubit $4 \times 2$ device required for the lattice surgery demonstration has two of the logical qubits described above stacked vertically. 
A $\logical{ZZ}$ measurement on the two logical qubits is performed by the circuit shown in Fig.~\ref{fig:protological_circuit}.  The $X$ and $Z$ steps of the idle circuit are performed, followed by an additional $X$ step and then a new $Y$ step entangling the two logical qubits.  
The instantaneous stabilizer group after a given step is given by the two-qubit measurements performed in that step, and additionally each logical qubit's $YYYY$ stabilizer after steps 2 and 3, and $\logical{ZZ}$ after steps 1 and 4. 
Again, any change in the stabilizer outcomes indicates an error has occurred.

\subsection{Decay experiment and simulations}

To compare the $\logical{ZZ}$ measurement on the logical qubits to a $ZZ$ measurement on a pair of physical qubits, we compare the repetition code run directly on the physical qubits with the repetition code concatenated with the ladder code. 
Specifically, we envision the following decay experiment: (1) initialize the qubits (logical or physical) in a logical state of the repetition code; (2) perform $N$ repetitions of the (logical or physical) $ZZ$ measurement circuit, corresponding to error detection cycles for the repetition code; (3) characterize the final state to identify undetected faults.
For both the logical and physical qubit case, post-select on runs for which no errors are detected, {\it i.e.,} for which the stabilizer outcomes are unchanged. 
The decay experiment is repeated for different initial logical states, which can be used to characterize the decay of the repetition code logical operators $\repcode{X}=XX$ and $\repcode{Z}={ZI}$.
Due to the inherent asymmetry of the repetition code, the associated decay rates can behave qualitatively differently. Details of this are discussed in \Cref{app:protological-details}. Defining the decay rate of Pauli $P$ as $\decay{P}$, we characterize the performance through the average logical improvement $\Lambda$
\begin{align}\label{eqn:log-imp}
    \Lambda &= \frac{\decay{XX} + \decay{ZI} }{\decay{\overline{XX}} + \decay{\overline{ZI}}},
\end{align}
{\it i.e.,} the ratio of the average decay rate for the physical circuit to the average decay rate for the logical circuit.

\begin{figure}
    \centering
    \includegraphics[width=\columnwidth]{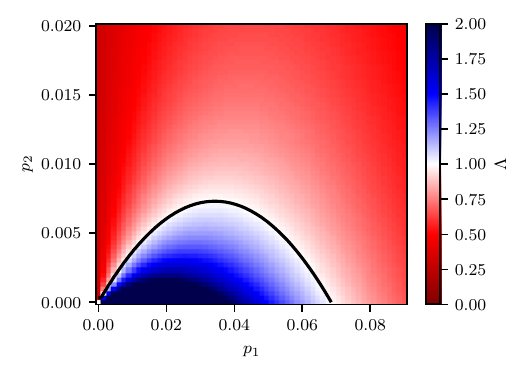}
    \caption{
    Average logical improvement $\Lambda$ (\Cref{eqn:log-imp}) for the $ZZ$ measurement, plotted as a function of single-qubit Pauli error probability $p_1$ and two-qubit Pauli error probability $p_2$ for assignment error $p_a=0.01$.  The blue region bounded by the black curves shows parameter values for which logical improvement is expected.}
    \label{fig:MV42-results}
\end{figure}

The decay rates (equivalently, logical errors per round) are extracted by varying the number of rounds $N$ and fitting the results to an exponential for both the logical and physical circuit.  We simulate this decay experiment for an effective noise model in which single-qubit Pauli errors occur with probability $p_1$, two-qubit Pauli errors occur with probability $p_2$ and classical assignment errors occur with probability $p_a$ (see \Cref{app:noise-model-details} for additional details).
Defining the average logical improvement according to \Cref{eqn:log-imp}, we extract the parameter space for which quantum error detection shows logical improvement of the $ZZ$ measurement, shown by the black curve in \Cref{fig:MV42-results}. Note there is an optimal single-qubit Pauli error probability that maximizes the allowed two-qubit Pauli error probability for which there is still logical improvement. Furthermore, there is generally no improvement if $p_1=p_2=p$, which is to be expected since these circuits are not fault-tolerant for circuit-level noise. However, in practice we expect $p_2 < p_1$ for tetron qubits, see discussion in Appendix~\ref{app:noise-model-details}.

\section{Outlook: scalable quantum error correction schemes}
\label{sec:outlook}

The roadmap outlined above builds towards scalable quantum error correction with tetrons.
Most importantly, the basic qubit design and functionality is unchanged after the two-qubit device, up to optimizations of the component dimensions. 
The operations required for error correction in the Hastings-Haah Floquet codes~\cite{Hastings21,Paetznick23} or the pair-wise measurement-based surface code~\cite{Grans-Samuelsson23} are a subset of the measurements used in the three demonstrations described above; no additional qubit capabilities are required.  In particular, the syndrome extraction circuit for the Hastings-Haah Floquet codes use exactly the same set of two-qubit measurements as for the quantum error detection demonstration.
Finally, each successive milestone described in the above roadmap requires improved physical qubit performance while increasing the scale of the system, the same two axes along which we must progress to achieve scalable quantum error correction.  Demonstrating successive milestones both de-risks our understanding of the errors affecting tetrons and indicates the path forward.

\begin{figure}
    \centering
    \includegraphics[width=\columnwidth]{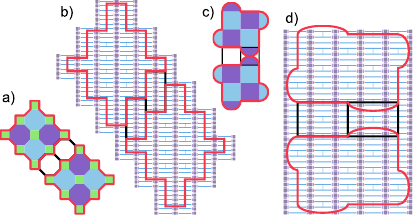}
    \caption{
    A device with two logical qubits with fault distance $3$ encoded in the 4.8.8 Hastings-Haah code~\cite{Hastings21,Paetznick23} (a, b) and the 3 aux surface code (c, d)~\cite{Grans-Samuelsson23}.  
    The colored faces indicate how the syndrome extraction circuits are constructed, and how the corresponding plaquettes map to a tetron array.  Throughout, a thick, red line highlights the region representing a single logical qubit. Black lines outline regions used to perform lattice surgery between the patches.  }
    \label{fig:qec}
\end{figure}

For the code distances required for utility scale quantum computing, inevitably we will run into instances of physical qubit failure. 
While early devices can require all qubits to be functional and within specification, scalable systems will require implementing vacancy mitigation strategies in the syndrome extraction circuit. 
Vacancy mitigation for the Hastings-Haah Floquet codes and pair-wise measurement-based surface code was described in Refs.~\onlinecite{Grans-Samuelsson23, aasen23}.  
Demonstrating these circuits in an actual device require logical qubit patches with fault distance of at least 5 so that error correction is still possible in the presence of a sparse number of failed components. 

A demonstration of scalable quantum error correction with tetrons should support both idle and logical operations below the code threshold, and utilize the vacancy mitigation in the syndrome extraction circuits~\cite{aasen23,Grans-Samuelsson23}.  
As such, we propose a tetron array supporting $2\times 1$ logical qubits.  To achieve sub-threshold operations in the presence of a sparse number ($\leq 2$) of failed qubits in either logical patch, we use the results of Ref.~\onlinecite{Grans-Samuelsson23} to estimate that code patches of fault distance 7 are sufficient.  Such a tetron array supporting the pair-wise measurement-based surface code is depicted in the right panel of \Cref{fig:roadmap}. 
At distance 7, each logical qubit patch is a $13\times 13$ array of tetrons, with a single row of tetrons between the patches to support lattice surgery; a $2\times 1$ tetron array within the larger grid has conceptually identical design to the two-qubit device for the roadmap.  The equivalent device supporting the 4.8.8 Hastings-Haah Floquet code has $196$ tetrons, roughly equivalent to a $14\times 14$ array.  
No additional tetrons are needed between the two patches to support logical operations~\cite{Hastings21, Haah2022, Gidney2021b, Paetznick23}. 
The mapping of both codes onto a tetron array is shown in \Cref{fig:qec}, where the colored faces indicate how the syndrome extraction circuits are constructed. 

For a utility-scale quantum computer of hundreds, if not thousands of logical qubits that is able to solve commercially relevant problems~\cite{beverland2022assessing,Gidney21}, the qubit approach laid out here has several key advantages. 
First, with a single qubit having an area of roughly $\SI{5}{\micro\meter} \times \SI{3}{\micro\meter}$, it is possible to fit millions of qubits onto a single wafer.
Secondly, physical operations can be performed on a $\SI{}{\micro\second}$ time scale, thus reducing the runtime of utility-scale calculations to the range of hours to days~\cite{beverland2022assessing}. 
Thirdly, topological protection allows a systematic, exponential reduction of many error mechanisms in dimensionless parameter ratios such as the topological gap over temperature $\Delta/k_B T$ and the wire length over topological coherence length $L/\xi$. 

To minimize the overhead associated with error correction, it is favorable to work far below threshold, such as an error rate of $10^{-4}$, satisfied by $\Delta/k_B T \sim 12$ and $L/\xi=20$, see \Cref{app:noise-model-details}. 
These values are within reach of existing material systems, {\it e.g.}, $\Delta/k_B T=12$ is achieved for the $\SI{50}{\milli\kelvin}$ temperature measured in Ref.~\onlinecite{Aghaee24} and a topological gap of $\Delta \sim \SI{50}{\ueV}$, which is in the range measured in Ref.~\onlinecite{Aghaee23}.
Exceeding these values combined with sufficient IR filtering and shielding, achieves a regime where physical error-rates are dictated by the SNR. 
Following \Cref{eqn:pa}, achieving $p_a=10^{-4}$ then amounts to achieving an SNR of $3.7$ in $\SI{1}{\micro\second}$.
This readout performance is achievable. 
For example, the SNR model in Ref.~\onlinecite{Aghaee24} predicts that a state-of-the art amplification chain with 5 quanta of added noise and critically-coupled readout resonators with $Q_c=500$ and parasitic capacitance of $\SI{200}{\femto\farad}$ would be sufficient (assuming the same signal size as measured in Ref.~\onlinecite{Aghaee24}).

Finally, measurement-based qubits offer significant advantages in terms of their control. 
While conventional qubits typically rely on precise shaping of control pulses, measurement-based topological qubits are more digital in nature: the pulses need to tune from an idle configuration to approximately the optimal measurement point, but the precise timing and shape of the pulse have negligible effect on the overall measurement performance.
This digital nature of the control pulses significantly simplifies tuning and control of the device. 
The energy scales lower- and upper-bounding the pulse rise/fall times are exponentially separated in $L/\xi$, indicating the robustness of the measurement-based topological qubits to control imperfections, see \Cref{app:noise-model-details}.
While the preparation of a physical $T$-state is sensitive to the details of the control pulse analogous to gate pulses for conventional qubits (see \Cref{app:T-gates}), the target error for this state preparation is significantly relaxed $\mathcal{O}(10^{-2})$ compared to that for topologically protected operations.

\begin{acknowledgments}
We are grateful to Michael Beverland and Nicolas Delfosse for early discussions of this roadmap, and to Sankar Das Sarma, Liang Fu, John Preskill, and Sagar Vijay for feedback on an initial draft of this manuscript.  
We also thank Connor Gilbert for help with preparation of the figures.
Correspondence and requests for materials should be addressed to Chetan Nayak (cnayak@microsoft.com).

All data associated with simulations in this paper is available from a Zenodo repository~\cite{Aasen25Zenodo}.

\end{acknowledgments}

\vspace{1cm}
$^\dagger${\small
David Aasen, 
Morteza Aghaee, 
Zulfi Alam, 
Mariusz Andrzejczuk, 
Andrey Antipov, 
Mikhail Astafev, 
Lukas Avilovas, 
Amin Barzegar, 
Bela Bauer, 
Jonathan Becker,
Juan M. Bello-Rivas,
Umesh Bhaskar, 
Alex Bocharov, 
Srini Boddapati, 
David Bohn, 
Jouri Bommer, 
Parsa Bonderson, 
Jan Borovsky, 
Leo Bourdet, 
Samuel Boutin, 
Tom Brown, 
Gary Campbell, 
Lucas Casparis, 
Srivatsa Chakravarthi, 
Rui Chao, 
Benjamin J. Chapman, 
Sohail Chatoor, 
Anna Wulff Christensen, 
Patrick Codd, 
William Cole, 
Paul Cooper, 
Fabiano Corsetti, 
Ajuan Cui, 
Wim van Dam, 
Tareq El Dandachi, 
Sahar Daraeizadeh, 
Adrian Dumitrascu, 
Andreas Ekefjärd, 
Saeed Fallahi, 
Luca Galletti, 
Geoff Gardner, 
Raghu Gatta, 
Haris Gavranovic, 
Michael Goulding, 
Deshan Govender, 
Flavio Griggio, 
Ruben Grigoryan, 
Sebastian Grijalva, 
Sergei Gronin, 
Jan Gukelberger, 
Jeongwan Haah, 
Marzie Hamdast, 
Esben Bork Hansen, 
Matthew Hastings, 
Sebastian Heedt, 
Samantha Ho, 
Justin Hogaboam, 
Laurens Holgaard, 
Kevin Van Hoogdalem, 
Jinnapat Indrapiromkul, 
Henrik Ingerslev, 
Lovro Ivancevic, 
Sarah Jablonski, 
Thomas Jensen, 
Jaspreet Jhoja, 
Jeffrey Jones, 
Kostya Kalashnikov, 
Ray Kallaher, 
Rachpon Kalra, 
Farhad Karimi, 
Torsten Karzig, 
Seth Kimes, 
Vadym Kliuchnikov, 
Maren Elisabeth Kloster, 
Christina Knapp, 
Derek Knee, 
Jonne Koski, 
Pasi Kostamo, 
Jamie Kuesel, 
Brad Lackey, 
Tom Laeven, 
Jeffrey Lai, 
Gijs de Lange, 
Thorvald Larsen, 
Jason Lee, 
Kyunghoon Lee, 
Grant Leum, 
Kongyi Li, 
Tyler Lindemann, 
Marijn Lucas, 
Roman Lutchyn, 
Morten Hannibal Madsen, 
Nash Madulid, 
Michael Manfra,
Signe Brynold Markussen, 
Esteban Martinez, 
Marco Mattila, 
Jake Mattinson, 
Robert McNeil, 
Antonio Rodolph Mei, 
Ryan V. Mishmash, 
Gopakumar Mohandas, 
Christian Mollgaard, 
Michiel de Moor, 
Trevor Morgan, 
George Moussa, 
Anirudh Narla, 
Chetan Nayak, 
Jens Hedegaard Nielsen, 
William Hvidtfelt Padkær Nielsen, 
Fr\'{e}d\'{e}ric Nolet, 
Mike Nystrom, 
Eoin O'Farrell, 
Keita Otani, 
Adam Paetznick, 
Camille Papon, 
Andres Paz, 
Karl Petersson, 
Luca Petit, 
Dima Pikulin, 
Diego Olivier Fernandez Pons, 
Sam Quinn, 
Mohana Rajpalke, 
Alejandro Alcaraz Ramirez, 
Katrine Rasmussen, 
David Razmadze, 
Ben Reichardt, 
Yuan Ren, 
Ken Reneris, 
Roy Riccomini, 
Ivan Sadovskyy, 
Lauri Sainiemi, 
Juan Carlos Estrada Salda\~{n}a,
Irene Sanlorenzo, 
Simon Schaal, 
Emma Schmidgall, 
Cristina Sfiligoj, 
Marcus P. da Silva,
Shilpi Singh,
Sarat Sinha, 
Mathias Soeken, 
Patrick Sohr, 
Tomas Stankevic, 
Lieuwe Stek, 
Patrick Strøm-Hansen, 
Eric Stuppard, 
Aarthi Sundaram,
Henri Suominen, 
Judith Suter, 
Satoshi Suzuki, 
Krysta Svore, 
Sam Teicher, 
Nivetha Thiyagarajah, 
Raj Tholapi, 
Mason Thomas, 
Dennis Tom, 
Emily Toomey, 
Josh Tracy, 
Matthias Troyer, 
Michelle Turley, 
Matthew D. Turner, 
Shivendra Upadhyay, 
Ivan Urban, 
Alexander Vaschillo, 
Dmitrii Viazmitinov, 
Dominik Vogel, 
Zhenghan Wang, 
John Watson, 
Alex Webster, 
Joseph Weston, 
Timothy Williamson, 
Georg W. Winkler, 
David J. van Woerkom, 
Brian Paquelet Wütz, 
Chung Kai Yang, 
Richard Yu, 
Emrah Yucelen, 
Jesús Herranz Zamorano, 
Roland Zeisel, 
Guoji Zheng, 
Justin Zilke, 
Andrew Zimmerman
}

\appendix
\crefalias{section}{appsec}

\section{Additional aspects of MBQB}
\label{app:MBQB-details}

We now discuss the measurement-based qubit benchmarking scheme more formally. Measurements can be described by a collection of non-deterministic linear transformations referred to as \emph{measurement instruments}~\cite{davies1970instruments,wiseman2010measuerment}: a measurement instrument $\instr{P}$ with outcomes $\pm1$ consists of the two quantum operations $\{ \mop{P}{+1}, \mop{P}{-1} \}$, each acting linearly on a given quantum state. The superscript $P$ indicates the intended Pauli basis of the measurement.  Each measurement instrument is {\em completely positive} (they map positive semidefinite matrices to positive semidefinite matrices even when acting on a subsystem), and {\em trace non-increasing} (they may reduce the trace).
Given some input state $\rho$, the probability of each measurement outcome is given by $\Pr(\mop{P}{s}; \rho) = \tr \mop{P}{s}(\rho)$, and the post-measurement state is given by $\mop{P}{s}(\rho)/\Pr(\mop{P}{s})$~\footnote{We write $\Pr(\mop{P}{s})$ instead of $\Pr(\mop{P}{s}; \rho)$ when the relevant quantum state is clear from context.}. Due to their linearity, we write the composition of quantum operations as a sequence of quantum operations, with sequence ordering going from right to left like matrix multiplication---thus $ \mop{P}{+1}\mop{Q}{-1}(\rho)$ corresponds to the (unnormalized) state obtained by starting with the state $\rho$, then measuring $\instr{Q}$ and obtaining outcome $-1$, then measuring $\instr{P}$ and obtaining outcome $+1$.

We define the conditional probabilities
\begin{equation} \label{eqn:cond-prob}
\Pr(\mop{P}{s}|\mop{Q}{r}; \rho) 
= \frac{\Pr(\mop{P}{s}\mop{Q}{r}; \rho)}{\Pr(\mop{Q}{r};\rho)}
= \frac{\tr \mop{P}{s}\mop{Q}{r}(\rho) }{\tr\mop{Q}{r}(\rho)},
\end{equation}
where $\instr{P},\instr{Q} \in \lbrace \instr{X},\instr{Z} \rbrace$ and $s,r = \pm 1$. This corresponds to the probability of outcome $s$ when measuring $\instr{P}$ immediately after a measurement of $\instr{Q}$ with outcome $r$.

In the ideal case, where $\instr{X}$ and $\instr{Z}$ correspond to projective measurements of the corresponding Pauli operators, we have that for all $\rho$ with $\Pr(\mop{P}{r};\rho)>0$ and $\Pr(\mop{Q}{r};\rho)>0$
\begin{align} \begin{split}\label{eqn:ideal-prob}
\Pr(\mop{P}{s}|\mop{P}{r}; \rho) & = \delta_{s,r}, \\
\Pr(\mop{P}{s}|\mop{Q}{r}; \rho) & = \frac{1}{2}
\end{split} \end{align}
where the second equality assumes $P\neq Q$.

The ability to prepare $\frac{1}{2}\tilde{\openone}$ (the experimental approximation to the maximally mixed state) allows for both outcomes of an instrument to be observed, at least in the regime where the instrument does not include any operations proportional to zero. This preparation does not need to be perfect, but crucially the probability of observing a particular outcome of an instrument must be non-zero, as otherwise the conditional probabilities are not well defined.
In practice, preparing $\frac{1}{2}\tilde{\openone}$ is done by applying a sequence of approximately non-commuting measurements without conditioning on the outcomes.  The action of this sequence, after averaging over outcomes and orderings, is 
$\sop{R}=\frac{1}{2}\left(\mop{X}{*} \mop{Z}{*}+\mop{Z}{*}\mop{X}{*}\right)$, 
where the $*$ refers to the fact that we do not care about that particular outcome. In the absence of errors $\sop{R}(\rho)=\frac{1}{2}\openone$, independent of $\rho$. For this reason, we call $\sop{R}$ the reset operation.

The sequences given in \Cref{eqn:rqm-seqs} can be written more explicitly as 
\begin{subequations}
\label{eqn:rqm-seqs2}
\begin{align}
\mop{X}{r} \mop{X}{s} & \mop{X}{*} \mop{Z}{*}, & \mop{X}{r} \mop{X}{s} & \mop{Z}{*} \mop{X}{*}, \label{eqn:a} \\ 
\mop{Z}{r} \mop{Z}{s} & \mop{X}{*} \mop{Z}{*}, & \mop{Z}{r} \mop{Z}{s} & \mop{Z}{*} \mop{X}{*}, \label{eqn:b}\\ 
\mop{X}{r} \mop{Z}{s} & \mop{X}{*} \mop{Z}{*}, & \mop{X}{r} \mop{Z}{s} & \mop{Z}{*} \mop{X}{*}, \label{eqn:c}\\ 
\mop{Z}{r} \mop{X}{s} & \mop{X}{*} \mop{Z}{*}, & \mop{Z}{r} \mop{X}{s} & \mop{Z}{*} \mop{X}{*}. \label{eqn:d}
\end{align}
\end{subequations}
One experiment is to run a long sequence of random instruments and count the separate occurrence frequencies of subsequences \Cref{eqn:rqm-seqs2}. 
For each error, we take the mean over state preparation ({\it i.e.}, over the two subsequences corresponding to the same row in the above expressions) and the maximum over the two bases ({\it i.e.}, the maximum over \Cref{eqn:a} and \Cref{eqn:b} for the operational assignment error, and the maximum over \Cref{eqn:c} and \Cref{eqn:d} for the operational bias error).
It may be more practical to consider pseudo-random sequences of instruments, in particular sequences that fairly sample all subsequences of a given length. 
This allows an experiment to store a relatively short random sequence and cycle through it in a continuous loop, all the while iterating the shorter subsequences evenly. 
Many sequences with this fair-sampling property exist---in particular, a construction by de Bruijn provides a plethora of such sequences, parameterized by the length of subsequences that must be sampled~\cite{debruijn1946}.

In addition to the operational metrics described above, one can reconstruct the action of different measurement operations for arbitrary input states using \emph{gate set tomography}~\cite{Nielsen21}.  As we assume the only operations available are measurements in $X$ and $Z$, and disallow more complex sequences or the use of ancillas, we can only reconstruct transformations that map the real part of the density matrix to itself ({\it i.e.,} operations on a {\em rebit}, not a qubit). Given some characterizations of the rebit part of an operation, the {\em no pancake theorem}~\cite[Chapter~3]{Preskill22} allows us to bound the remaining matrix elements due to complete positivity constraints, thereby bounding more general error metrics.
The full Clifford group is accessible if $Y$ measurements are introduced, as described in \Cref{sec:braiding-protocols}, so full state and process tomography can then be performed. Here we focus only on the rebit case.

The usual fixed preparation of a tomography experiment can be  approximated by the reset operation followed by a measurement and post-selecting on the outcome.  The estimation of the expectation of an observable is done by applying one of the measurements at hand and taking the expectation value of the outcome.  The behavior of any operation on the real part of a qubit can then be characterized by the operation's action on these 16 experiments in combination with 16 experiments for the trivial zero-duration ``no-op" operation (where preparation is immediately followed by the measurement of an observable). This illustrates that the experiments for $\assignerr$ and $\biaserr$ are sensitive to all features of the measurement operations, since they enable full reconstruction of these operations under relatively mild assumptions about Markovianity in the experimental setup.

\section{Noise model for tetrons}\label{app:noise-model-details}

A simple noise model for tetrons includes assignment errors (equivalently measurement bit-flip errors), single-qubit Pauli errors, and two-qubit Pauli errors, with respective error channels 
\begin{align}
    \sop{N}_{P,s}[p_a](\rho) &= (1-p_a) \mst{P}{s}(\rho) + p_a \mst{P}{-s}(\rho), \label{eq:noise-pa}
    \\ \sop{P}^{(1)}[p_1](\rho) &= (1-p_1 ) \rho +  \frac{p_1}{3}\sum_P  P \rho P, \label{eq:noise-p1}
    \\
    \begin{split}
        \sop{P}^{(2)}[p_2](\rho) &= (1-p_2 ) \rho \, + \\& \frac{p_2}{9}\sum_{P, Q \in X,Y,Z} (P\otimes Q) \, \rho \, (P \otimes Q) .
    \end{split} \label{eq:noise-p2}
\end{align}
Recall that $\mst{P}{s} = \frac{1}{2}(\mathds{1}+sP)$ projects onto the $P$ eigenstate with outcome $s$. 

During a time step where the qubit is idle, we apply 
\begin{align}
    \sop{E}^{(0)}[p_1](\rho) &= \sop{P}^{(1)}[p_1](\rho).
\end{align}
During a time step where the qubit is part of a single-qubit measurement in Pauli basis $P$, we apply 
\begin{align}
    \sop{E}^{(1)}_{P,s}[p_a, p_1](\rho) &= \sop{P}^{(1)}\left[\frac{p_1}{2}\right] 
    \sop{N}_{P,s}[p_a] \,
    \sop{P}^{(1)}\left[\frac{p_1}{2}\right] (\rho).
\end{align}
The above error channel heuristically captures the fact that the measurement outcome is sensitive to whether the qubit undergoes a state flip during the measurement.  
\begin{widetext}
Similarly, during a time step where two qubits are measured in Pauli basis $PQ$, we apply
\begin{align}
    \sop{E}^{(2)}_{PQ,s}[p_a, p_1, p_2](\rho) 
    &= \sop{P}^{(1)} \left[ \frac{p_1}{2} \right] \sop{P}^{(2)} \left[ \frac{p_2}{2} \right] \sop{N}_{P,s}[p_a]\,  \sop{P}^{(1)} \left[ \frac{p_1}{2} \right] \sop{P}^{(2)} \left[ \frac{p_2}{2} \right] (\rho).  
\end{align}
In the above expression, it is implicit that the error channel $\sop{P}^{(1)}$ is applied to both qubits.
\end{widetext}

The expression for $p_a$ is given in \Cref{eqn:pA}.  
For the error channel defined above, 
\begin{align}\label{eqn:p1-norm}
    p_1 &=\frac{3}{4} \left( 1-e^{-\tau_\text{meas}/T_\text{life}}\right)
\end{align}
so that the limit $T_\text{life}/\tau_\text{meas} \to 0$ returns the maximally mixed state.
Recall from \Cref{eqn:Z} and \Cref{eqn:X} that Pauli operators map to pairs of MZMs, thus errors that couple pairs of MZMs map to Pauli errors.  
This is the case for the two processes discussed in the main text contributing to the qubit lifetime $\Tlife$.  Exciting a fermion from the MZMs to an above-gap quasiparticle will likely be followed by the above-gap quasiparticle relaxing back to the MZMs; because above-gap quasiparticles are extended states there is a high probability of this process involving two distinct MZMs. 
We expect this error to be $Z$-biased due to the topological gap in the wire being smaller than the trivial superconducting gap in the backbone.
The corresponding contribution to $\Tlife$ can be estimated as
\begin{align}\label{eq:TlifeDelta}
    \Tlife^{(\Delta)} &=\tau_\text{el-ph} e^{-\Delta/k_B T},
\end{align}
for $\tau_\text{el-ph}\sim 50\,\text{ns}$~\cite{Knapp18a}.
Similarly, residual coupling between a pair of MZMs allows coupling to charge noise and thus can result in Pauli errors in the basis of the coupled MZMs; the basis of this error will depend on whether the residual coupling is through the topological wire or through the QDs (residual coupling through the qubit island of MZMs on different topological wires is further suppressed again by the trivial superconducting backbone). 
The contribution to $\Tlife$ is given by 
\begin{align}\label{eq:TlifeZeno}
    \Tlife^{(\varepsilon)} &= \left( \frac{\emst}{\eres}\right)^2 \left( S(\emst) + S(-\emst)\right)^{-1},
\end{align}
where $\emst$ is the energy splitting between the MZMs being measured, $\eres$ is the residual energy splitting between one of the MZMs being measured and one not being measured, and $S(\varepsilon)$ describes charge noise in the system.  
Note that the residual coupling through the topological wire is given by 
\begin{align}\label{eq:eres}
    \eres^{(\text{wire})} &\sim \Delta e^{-L/\xi}
\end{align}
while residual coupling through the quantum dots is suppressed exponentially in the width of the tunnel barrier.
Lastly, dynamic errors in the pulsing sequence can further increase the probability of single-qubit Pauli errors.  
For simplicity, we model all Pauli errors as occurring with equal probability, however this model can be refined by fitting simulations and experimental results to different Pauli-bias to extract which error process is limiting. 
Such simulations can be performed within the framework proposed in Ref.~\onlinecite{Boutin25}.

The correlated two-qubit Pauli errors can arise due to electron exchange during a two-qubit measurement, resulting in the charge states of the qubits changing after the measurement compared to before the measurement.  
While the lowest order contribution to this process only involves one MZM per qubit, this error can be mapped to a known Pauli error when combined with a heralding measurement of the total MZM parity on a qubit island ({\it e.g.,} by measuring the charge of the qubit island combined with the assumption that at low-enough temperature the island is in the ground state subspace).  
Higher-order two-qubit errors (corresponding to the remaining two-qubit Pauli errors) are additionally suppressed by the same mechanisms affecting single-qubit Pauli errors and thus occur with the probability of the lowest-order process multiplied by $p_1$.
Thus, as in the case of single-qubit Pauli errors, it is a simplification to set the probability of all two-qubit Pauli errors equal.
Accounting for the noise bias can improve the expected performance of the different milestone demonstrations simulated in the main text.

The probability of qubits exchanging electrons depends on the initial charge configuration of the islands; in the absence of heralding measurements only an initial electron poisoning event is suppressed (by $\mathcal{O}\left(e^{-2E_C/k_B T}\right)$ for qubit charging energy $E_C$ in thermal equilibrium), while subsequent two-qubit measurements can spread two-qubit errors with high probability. In the presence of heralding measurements ({\it e.g.,} of electron island charge), the Pauli errors can be tracked and corrected in software and $p_2$ is then approximated by the error of these heralding measurements, which is expected to be smaller than $p_1$.  This error source and mitigation strategy is discussed in detail in Ref.~\onlinecite{ICEP}.   

Measurement-based operation of tetrons is motivated both to avoid the challenge of physically moving MZMs, as well as to circumvent diabatic errors~\cite{bonderson2008measurement,Karzig17}, which are not in general exponentially suppressed in macroscopic parameter ratios of the system~\cite{Cheng09,Scheurer13,Knapp16}.  
Projective measurements reduce the effect of imperfections in the control pulse sequences to affecting the assignment error probability and Pauli error probabilities as discussed above.
However, measurement-based operations prevent working in a rotating frame and thus coherent rotations on the qubits contribute additional errors not captured by \Cref{eq:noise-pa}, \Cref{eq:noise-p1}, \Cref{eq:noise-p2}.  
If this residual coupling is limited by MZMs on the same wire, then the resulting coherent rotation corresponds to Krauss channel
\begin{align}\label{eqn:coh-rot}
    \Theta[\theta] &= e^{i\theta Z} \rho e^{-i\theta Z}
    \\ \theta &= \eres \tau_\text{mst}/\hbar, \label{eq:theta-def}
\end{align}
and the idle error channel becomes
\begin{align}\label{eqn:E0-coh-rot}
    \sop{E}^{(0)}[p_1, \theta](\rho) &= \sop{P}^{(1)}[p_1] \Theta [\theta](\rho).
\end{align}
(It is worth noting that $\sop{P}^{(1)}$ and $\Theta$ do not commute, and the ordering chosen above thus implies an approximation; one could write a more complicated operator that attempts to take this into account, but this would not affect conclusions in the low-error regime.)
These coherent rotations are suppressed by the Zeno effect when a qubit undergoes a measurement anticommuting with $Z$, however they do affect two-qubit measurements commuting with $Z$ by applying single-qubit rotations.  
\begin{widetext}
As such, the two-qubit measurement error channel becomes 
\begin{align}
    \sop{E}^{(2)}_{PQ,s}[p_a, p_1, p_2, \theta](\rho) 
    &= \sop{P}^{(1)} \left[ \frac{p_1}{2} \right] \sop{P}^{(2)} \left[ \frac{p_2}{2} \right] \Theta\left[\frac{\theta}{2}\right]\sop{N}_{P,s}[p_a]\,  \sop{P}^{(1)} \left[ \frac{p_1}{2} \right] \sop{P}^{(2)} \left[ \frac{p_2}{2} \right] \Theta\left[\frac{\theta}{2}\right](\rho),
\end{align}
where similarly to $\sop{P}^{(1)}$, it is implicit that $\Theta$ is applied to both qubits.
\end{widetext}
The Pauli twirl approximation~\cite{Kern05} suggests that in the quantum error correction syndrome extraction circuits, these errors can be approximated as Pauli errors and thus contribute towards $p_1$ with an amount $\mathcal{O}(\Theta^2)$ (for small $\Theta$). 

Targeting error rates of $10^{-4}$ as discussed in \Cref{sec:outlook} amounts to extracting parameter values for which the effective noise model parameters are set to $10^{-4}$.  
Setting $\theta=10^{-4}$ and assuming we are limited by $\eres^{(\text{wire})}$ bounds $L/\xi\gtrsim 20$.
(Assuming the Pauli twirl approximation would cut this bound in half.)
As $\Tlife^{(\varepsilon)}$ scales as $e^{-2L/\xi}$, we can assume $\Tlife$ is dominated by $\Tlife^{(\Delta)}$ (assuming that quasiparticle poisoning times are similar to those extracted in Ref.~\onlinecite{Aghaee24}) and use \Cref{eq:TlifeDelta} and \Cref{eqn:p1} to bound $\Delta/k_B T\gtrsim 12.$

Pulses in and out of the measurement configuration should be diabatic with respect to $\eres$ and adiabatic with respect to the avoided crossing in the measurement configuration, $t_C$ (corresponding to different charge states for the same MZM parity), and the topological gap $\Delta$.
Since qubit coherence requires $\hbar/\eres$ to be much longer than the duration of a single operation, the relevant bound in practice is that pulses must be slower than $\hbar/t_C \approx \SI{1}{\nano\second}$.  
The large separation in the bounds (growing exponentially with $L/\xi$) demonstrates the robustness of measurement-based qubits to control errors. 
More precise estimates can be obtained by studying the Landau-Zener dynamics across the avoided crossings, which indicates that pulse rise/fall times on the scale of $\SI{10}{\nano\second}$ easily satisfy both bounds.

\section{Tetron coherence times}\label{app:coh-times}

In conventional qubit platforms, the $Z$-basis qubit states---often referred to as ``computational basis states"---are encoded in two energy levels of the idle system and it is natural to characterize the qubit through the decay of diagonal and off-diagonal elements of the density matrix in the computational basis, commonly referred to as $T_1$ and $T_2^*$, respectively~\cite{Nielsen00}.
These are typically measured through Rabi and Ramsey protocols.

For ideal topological qubits, the qubit states are degenerate in the idle configuration, and the assignment of Pauli bases, as in \Cref{eqn:Z} and \Cref{eqn:X}, is arbitrary; thus the (basis-dependent) definitions of $T_1$ and $T_2$ are similarly arbitrary.  
However, for realistic implementations of the tetron devices proposed here, exponentially small residual coupling $\eres^{(ij)}$ between MZMs $\gamma_i$ and $\gamma_j$ will lead to some energy splitting $\Delta \varepsilon$ between the qubit states:
\begin{align}\label{eqn:H-idle}
    H_\text{idle} &= \sum_{i\neq j} \eres^{(ij)}\, i\gamma_i\gamma_j = \Delta \varepsilon\, \tilde{\sigma}_Z,
\end{align}
where $\tilde{\sigma}_Z$ is the diagonalized basis of $H_\text{idle}$. Under realistic assumptions, the dominant term is between MZMs on the same topological wire, such that $\tilde{\sigma}_Z$ approximately matches the operator defined in \Cref{eqn:Z}.
By repeatedly measuring the qubit in the $X$ or $Z$ basis and inserting idle times between the measurements, the coherence time in different bases can be extracted, as discussed e.g. in Refs.~\onlinecite{mishmash2020dephasing,Sau24}.

In many qubit platforms whose native operations are coherent rotations, $T_1$ and $T_2^*$ are important performance metrics not only for idle qubits, but also for gate operations as they often bound the achievable fidelity (in the limit where gates are ``coherence-limited''). This is not the case for tetrons, where the error channels that occur during single- or two-qubit measurements are not directly related to idle qubit lifetimes (see \Cref{app:noise-model-details}).

\section{Non-Clifford operations}\label{app:T-gates}

A common concern raised when discussing Majorana-based qubits is that their topologically protected Clifford operations (single- and two-qubit Pauli measurements in the context of tetrons) do not support universal quantum computation. As such, these operations need to be supplemented by a non-protected operation, usually chosen to be a physical rotation gate to prepare a magic state.
If these qubits were to be operated without quantum error correction, then one might worry that the performance of such a quantum computing platform is limited by the fidelity of this physical rotation gate, thus diminishing the benefit of using topologically protected qubits.
However, in a fault-tolerant quantum computer, the ability to perform universal gate operations to the \emph{encoded} qubits is that which matters. In this setting, the ability to apply physical $T$-gates with high fidelity is only advantageous if the error-correcting code supports transversal $T$-gates, such that applying a $T$-gate to all data qubits of a code patch fault-tolerantly applies a $T$-gate to the encoded qubit. This is not the case for the codes proposed in this roadmap~\cite{Eastin2009}, and instead one is forced to use magic state distillation~\cite{Knill04,Bravyi05}, which uses noisy $T$-gates and Clifford operations to perform higher-fidelity $T$-gates on the logical qubit. In such a scheme, a relatively small number of physical $T$ states is required, and their fidelity only weakly affects overall resource requirements as long as the error rate falls below a threshold that is on the order of several percent (with the details depending on the specific distillation scheme that is chosen)~\cite{Litinski2019,beverland2022assessing}. Therefore, the lack of universality of topologically protected operations does not impose a practical restriction on this fault-tolerant architecture.

We now discuss schemes to create physical $T$-states with sufficient fidelity to serve as input to such distillation schemes.
The simplest approach to preparing a $T$-state
\begin{align}
    \ket{\psi_T} &= \frac{1}{\sqrt{2}} \left( e^{-i\pi/8} \ket{0} + e^{i\pi/8}\ket{1}\right),
\end{align}
on a tetron is to couple a pair of MZMs for a precise amount of time.
This is the approach described for Ising anyons in the fractional quantum Hall effect in Refs.~\onlinecite{Bravyi2006,Freedman2006} and for MZMs in Ref.~\onlinecite{Sau10c}.
In essence, coupling the MZMs $\gamma_j$ and $\gamma_k$ by an amount $\varepsilon_T$ adds a term to the Hamiltonian $\varepsilon_T\,  i\gamma_j \gamma_k$, which introduces a dynamical phase between the two parity states of the MZM pair.  
The coupling can be done in the same way as for the Pauli measurements: opening cutter gates and tuning quantum dots to allow MZMs to couple along the same paths as for the $X$, $Y$, or $Z$ measurement loops.   
Thus, a pulse sequence beginning and ending in the idle configuration (all MZM energy splittings exponentially suppressed) implements a rotation gate on the axis of the coupled MZMs with phase
\begin{align}
    \varphi = \int dt \,\varepsilon_T(t).
\end{align}
When $\varphi = \pi/8$, this pulse sequence implements a $T$-gate.

The phase in the above approach is sensitive to the precise pulse sequence, thus the topological protection that makes physical error rate targets $\mathcal{O}\left(10^{-4}\right)$ feasible do not apply to this physical rotation gate.
However, given how infrequently the $T$-state is injected in the context of fault tolerant quantum computation, a target error rate of $0.01-0.05$ is sufficient to achieve the overheads discussed in Ref.~\onlinecite{beverland2022assessing}.

Other works discuss more complex $T$-state preparations compatible with tetrons that are predicted to achieve higher fidelity targets~\cite{Karzig15a,Karzig19} or are less sensitive to timing errors~\cite{Clarke16,Plugge16,Plugge17}.
When assessing the optimal protocol for fault-tolerant quantum computation with tetrons, the complexity of these protocols must be balanced against the possible reduction in spacetime overhead from using higher-fidelity $T$-gates.

\section{Additional aspects of measurement-based braiding transformations}
\label{app:braiding-details}

\begin{figure*}
    \centering
    \includegraphics[width=2\columnwidth]{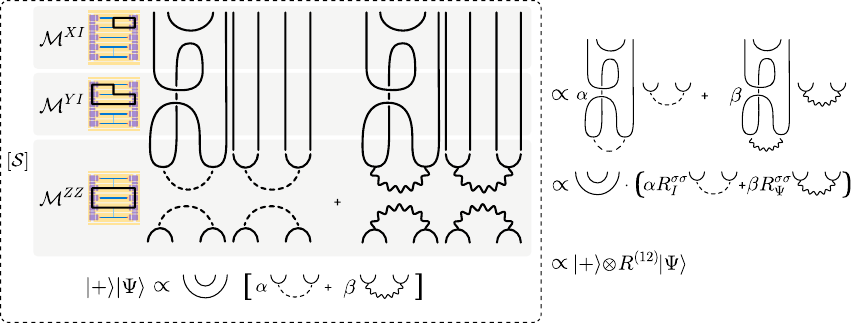}
    \caption{
    Mapping the measurement sequence for $[\sop{S}]$ to the braid of MZMs $1$ and $2$ on the computational qubit.  The solid lines correspond to MZMs labeled 1, 2, 4, 3 from left to right, dashed lines correspond to MZMs fusing to the trivial state $I$, squiggly lines correspond to MZMs fusing to the fermionic state $\psi$.
    {\it Left panel.} The state $\ket{+}\ket{\Psi}$ is the two-qubit state prepared after applying $\instr{XI}$ to a generic state.  The expression within the dashed-line box maps each measurement to its corresponding diagrammatic expression, with shaded sections separating different measurements.
    The sequence is read from bottom to top.
    {\it Right panel.} Applying the operator to the state on the left and collapsing the diagram results in the diagrammatic expressions shown, which can be rewritten as the braid on MZMs $1$ and $2$ applied to the state $\ket{+}\ket{\Psi}$.
    The diagrammatic calculus is explained in greater detail in  Refs.~\onlinecite{Bonderson_thesis} and \onlinecite{Kitaev06a} Appendix E. }
    \label{fig:braiding diagram}
\end{figure*}

Example measurement sequences to implement the braiding transformations associated with Pauli equivalence classes of single-qubit Clifford gates are given in \Cref{eqn:H}-\Cref{eqn:HS}.    
Tracking the measurement outcomes allows us to write the explicit Pauli corrections associated with the sequences.
Writing in terms of the ideal measurement projectors, we have 
\begin{align}
    \mst{XI}{s_3} \mst{YI}{s_2} \mst{ZY}{s_1} \mst{XI}{s_0} &= \mathcal{O}^{s_0}_{s_3} \otimes Y^{\frac{1+s_0 s_1 s_2}{2}} X H, 
    \\
    \mst{XI}{s_3} \mst{YI}{s_2} \mst{ZZ}{s_1} \mst{XI}{s_0} &= \mathcal{O}^{s_0}_{s_3} \otimes  Z^{\frac{1+ s_0 s_1 s_2} {2}} S 
    \\
    \mst{XI}{s_3} \mst{ZY}{s_2} \mst{ZZ}{s_1} \mst{XI}{s_0} &=\mathcal{O}^{s_0}_{s_3}  \otimes  Y^{\frac{1- s_0 s_3}{2}} X^{\frac{1+ s_1 s_2}{2}} HSH
    \\
    \mst{XI}{s_4} \mst{YI}{s_3} \mst{ZY}{s_2} \mst{ZZ}{s_1} \mst{XI}{s_0} &= \mathcal{O}^{s_0}_{s_4}  \otimes Y^{\frac{1- s_0 s_2 s_3}{2}} Z^{\frac{1- s_1 s_2}{2}} SH
    \\
    \mst{XI}{s_4} \mst{YI}{s_3} \mst{ZZ}{s_2} \mst{ZY}{s_1} \mst{XI}{s_0} &= \mathcal{O}^{s_0}_{s_4}  \otimes X^{\frac{1+ s_1 s_2}{2}} Z^{\frac{1+s_0 s_1 s_3}{2}} HS,
\end{align}
where we have written the operator on the auxiliary qubit $\mathcal{O}^{s_0}_{s_3}$ as a shorthand for $\ket{X_{s_3}}\bra{X_{s_0}}$.
The equality in these expressions are up to overall constants.

The connection between the measurement sequence and the braiding transformation applied to the computational qubit can be explicitly seen through diagrammatic calculus, as shown for $[\sop{S}]$ in \Cref{fig:braiding diagram}.
Recall that for a pair of MZMs in state $\ket{\Psi} = \alpha \ket{\sigma, \sigma; I} + \beta \ket{\sigma, \sigma; \psi}$, a braiding transformation $R^{\sigma \sigma}$ acts on the state as
\begin{align} \label{eqn:braid}
    R^{\sigma \sigma} \ket{\Psi} & = \alpha  R^{\sigma \sigma}_I \ket{\sigma, \sigma; I} + \beta R^{\sigma \sigma}_\psi \ket{\sigma, \sigma; \psi},
\end{align}
where $\{I, \psi, \sigma\}$ refer to the topological charges with non-Abelian fusion rules 
\begin{align}
    \psi \times \psi &= I
    \\ \psi \times \sigma &= \sigma 
    \\ \sigma \times \sigma  &= I+ \psi. 
\end{align}

The measurement sequence above is closely related to ``one-bit teleportation''~\cite{Zhou00}, a construction often used in the implementation of fault-tolerant gadgets (including the teleportation of non-Clifford gates using magic states as a resource) --- the main difference being that, instead of an entangling unitary gate, a two-body measurement is performed to entangle the two qubits. 
The implementation of unitary operations in this manner has been previously described as ``state transfer''~\cite{Perdrix2005}, and is also the basis for measurement-based implementations of CNOT operations via lattice surgery~\cite{Horsman12}.
The expression in \Cref{eqn:S} is also very similar to the measurement calculus formalism~\cite{Danos07} used to describe the one-way model of quantum computation~\cite{Raussendorf01}, although the qubit measurements in our Majorana platform are non-destructive (unlike typical measurements in optics-based platforms). These connections are elucidated by the concept of anyonic teleportation~\cite{Bonderson08a}.

\begin{figure}
    \centering
    \includegraphics[width=\columnwidth]{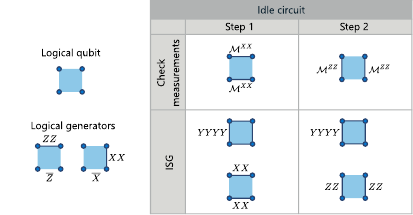}
    \caption{ {\it Top left.}  Four qubits (dots) form a logical qubit in the $2\times 2$ ladder code.  {\it Bottom left.} The logical generators for the ladder code are $ZZ$ between horizontally adjacent qubits and $XX$ between vertically adjacent qubits. {\it Right panel.} The table shows the check measurements and instantaneous stabilizer group (ISG) for each step of the idle ladder code.}
    \label{fig:lad-idle}
\end{figure}

\subsection{Simulation details} \label{sec:braiding_sim_app}

To estimate the fidelity of $[\sop{S}]$, we perform an exact simulation of the measurement sequence given in \Cref{eqn:S} subject to the effective noise model described in \Cref{app:noise-model-details} parameterized by $p_a, p_1, p_2$.
Beginning from a (two-qubit) initial state $\rho^{(0)}_{ab} = \ket{+}_a \bra{+}_a \otimes \ket{\psi}_b\bra{\psi}_b$, we run full density matrix simulations tracking the measurement outcomes through the corresponding noisy quantum instrument circuit~\cite{PRXQuantum.4.020303, kliuchnikov2023stabilizercircuitverification}.
We thus obtain two-qubit density matrix ``trajectories'' $\rho_{ab}[\{s_i\}]$, which correspond to the (unnormalized) density matrices for a given set of measurement outcomes $\{s_i=\pm1\}$.
Tracing over the auxiliary qubit's degrees of freedom we are left with the computational qubit's (unnormalized) density matrices $\rho_b[\{s_i\}]$ for measurement outcomes $\{s_i\}$.  
We can then apply the corresponding Pauli correction $P(\{s_i\}) = Z^{\frac{1+s_0s_1s_2}{2}}) $ (see \Cref{eqn:Sexplicit}), to find the computational qubit's Pauli-corrected density matrices
$\rho_b^{(c)}[\{s_i\}]$.
Summing these over all possible measurement outcomes $\rho_b^{(c)} = \sum_{\{s_i=\pm1\}} \rho_b^{(c)}[\{s_i\}]$ results in the density matrix associated with applying $\sop{S}$ to the initial state $\rho^{(0)}_{ab}$.

Following the discussion in Ref.~\onlinecite{Nielsen00}, we can use process tomography to extract the noise channel associated with the above process $\tilde{\sop{S}}$, and average its overlap with the ideal Clifford gate $S$ over all possible computational qubit initial states.
The associated fidelity $F_{[\sop{S}]}$ (\Cref{eqn:Fid-def}) is plotted in \Cref{fig:braiding_sim}.

\section{Additional aspects of quantum error detection in the eight-qubit device}
\label{app:protological-details}

\subsection{Ladder code summary}

The idle circuit for the ladder code on a $2\times2$ array of qubits is shown in \Cref{fig:lad-idle}.  The instantaneous stabilizer group after a given step of the circuit corresponds to two-qubit Paulis of the measurements performed in that step, in addition to $YYYY$ on all four qubits (whose eigenvalue is inferred from multiplying the measurement outcomes for the previous two steps of the circuit).  
Errors are detected when the stabilizer eigenvalues change.  The logical group generators are $\logical{Z}=ZZ$ between horizontal nearest neighbor qubits and $\logical{X}=XX$ between vertical nearest neighbor qubits.

\begin{figure}
    \centering
    \includegraphics[width=\columnwidth]{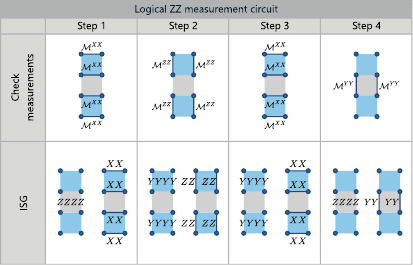}
    \caption{Check measurements and instantaneous stabilizer group for each step of the $\logical{ZZ}$ measurement circuit.  The logical qubits are indicated by the blue shading, grey regions indicate qubits used to perform lattice surgery (the $\logical{ZZ}$ measurement).  }
    \label{fig:lad-ZZ}
\end{figure}

A $\logical{ZZ}$ measurement in the ladder code corresponds to measuring the product of the $\logical{Z}$ on two adjacent logical qubits.  The circuit shown in Fig.~\ref{fig:lad-ZZ} achieves this: after steps 4 (and 1 when repeated) the product of $ZZZZ$ on the four middle qubits can be inferred from the measurement outcomes of the previous two steps.  (Each qubit's $YYYY$ stabilizer can be inferred after steps 2 and 3.)

\subsection{Repetition code decay experiment}

\begin{figure}
    \centering
    \includegraphics[width=\columnwidth]{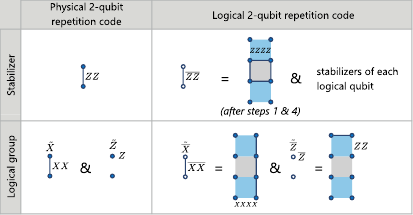}
    \caption{ The stabilizers and logical group generators for the repetition code on both the physical and logical qubits.}
    \label{fig:repetition-code}
\end{figure}

To compare the (post-selected) error rate of the logical $\logical{ZZ}$ measurement with the physical ${ZZ}$ measurement, we compare the repetition code run directly on a pair of physical qubits with the repetition code concatenated with the $2\times2$ ladder code (Fig.~\ref{fig:repetition-code}).  The repetition code consists of repeatedly measuring $ZZ$ and discarding any runs where the eigenvalue changes.  As such, the stabilizer is $ZZ$, and the logical group is generated by $\repcode{X}=XX$ and $\repcode{Z}=ZI$.  For the concatenated code, these correspond to $\repcode{\logical{X}}=\logical{XX} = XXXX$ along a column of physical qubits, and $\repcode{\logical{Z}}=\logical{ZI}=ZZ$ along a row of physical qubits.
While $\repcode{\logical{Z}}$ has weight on one more physical qubit than $\repcode{Z}$, $\repcode{\logical{X}}$ has weight on two more physical qubits than $\repcode{X}$; as such we would expect the latter to manifest logical improvement for a larger region of parameter space than the former.

A logical state of the repetition code can be written as $\ket{\psi_i} = \alpha\ket{00} + \beta \ket{11}$, where $\ket{0}$ and $\ket{1}$ correspond to $Z$-eigenstates of the repetition code qubits.  When the system is initialized in the state $\ket{\psi_i} = \frac{1}{\sqrt{2}} \left( \ket{00} + \ket{11}\right)$, a $\repcode{Z}$ error can be detected by the expectation value of $XX$:
\begin{align}
    \ket{\psi_f} &= ZI \ket{\psi_i}
    \\ \bra{\psi_f} ZZ \ket{\psi_f} &= 1
    \\ \bra{\psi_f} ZI \ket{\psi_f} &= 0
    \\ \bra{\psi_f} XX \ket{\psi_f} &= -1.
\end{align}
In experiment, the state can be prepared by measuring each qubit in $X$, post-selecting on the outcomes, and then running the (physical or logical) $ZZ$ measurement circuit.

When the system is initialized in the state $\ket{\psi_i} = \ket{00}$, a $\repcode{X}$ error can be detected by the expectation value of $ZI$:
\begin{align}
    \ket{\psi_f} &= XX \ket{\psi_i}
    \\ \bra{\psi_f} ZZ \ket{\psi_f} &= 1
    \\ \bra{\psi_f} ZI \ket{\psi_f} &= -1
    \\ \bra{\psi_f} XX \ket{\psi_f} &= 0.
\end{align}
The state can be prepared experimentally by measuring each qubit in $Z$, post-selecting on the outcomes, and then running the (physical or logical) $ZZ$ measurement.  

To find the average logical improvement, defined in \Cref{eqn:log-imp}, we compare the repetition code run directly on the physical qubits to the repetition code run on the logical qubits for both sets of initial conditions.

\subsection{Simulation details}

To estimate the logical improvement, we perform an \emph{exact} simulation (as in Sec.~\ref{sec:braiding_sim_app}) of the decay experiment described within the effective noise model described in \Cref{app:noise-model-details} (parameterized by $p_a, p_1, p_2$). Starting with a prescribed 8-qubit initial state $\rho_0 = \ket{\psi_i}\bra{\psi_i}$, we run full density matrix simulations again tracking the measurement outcomes through the noisy quantum instrument circuit, and thereby obtain density matrix ``trajectories'' $\rho[\{s_i\}]$ corresponding to the (unnormalized) density matrices for a given set of measurement outcomes $\{s_i=\pm1\}$.
In principle, the number of such density matrices in this trajectories map doubles with every measurement, leading to an exponential simulation cost in the length of the decay experiment protocol; however, we can discard those trajectories corresponding to outcome combinations for which an error is detected. Here, we use a decoder constructed from the ``checks'' or ``detectors'' of the corresponding outcome code~\cite{delfosse2023spacetimecodescliffordcircuits}. In addition, we can trace over those outcomes which are not needed in future checks. These optimizations keep the number of retained density matrices at a bounded, manageable number. Finally, the pertinent expectation values, as spelled out in the previous section, are computed using the system density matrix \emph{ignoring the outcomes}, {\it i.e.,} $\rho = \sum_{\{s_i=\pm1\}} \rho[\{s_i\}]$, and the post-selected acceptance rate for the protocol is given merely by $\tr(\rho)$.

\begin{figure}
    \centering
    \includegraphics[width=\columnwidth]{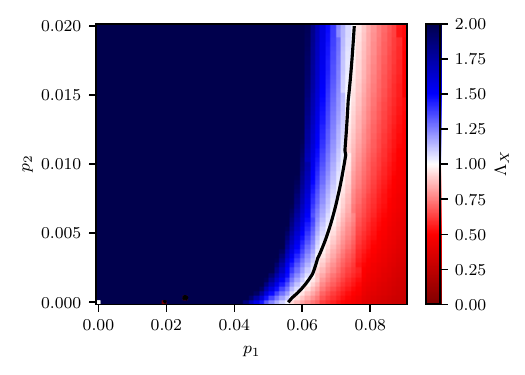}
    \includegraphics[width=\columnwidth]{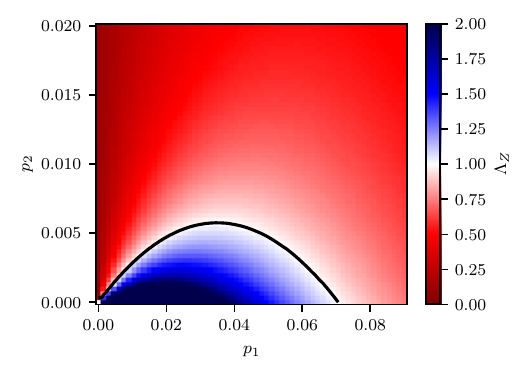}
    \caption{
    Ratio of physical to logical decay rates for $\repcode{X}$ (top panel) and $\repcode{Z}$ (bottom panel) for $p_a=0.01$, as defined in \Cref{eqn:LambdaXZ}. The region of logical improvement (blue) for $\repcode{X}$ is much larger than that of $\repcode{Z}$, as is expected from the weight of the corresponding physical operators.}
    \label{fig:decay_rates}
\end{figure}

In \Cref{fig:decay_rates} we plot the ratio of the physical to logical decay rates associated with $\repcode{X}$ and $\repcode{Z}$, defined by 
\begin{align}\label{eqn:LambdaXZ}
    \Lambda_X &= \frac{\decay{XX}}{\decay{\logical{XX}}}, & \Lambda_Z &= \frac{\decay{ZI}}{\decay{\logical{ZI}}}.
\end{align}  
As discussed above, the former has a larger region of logical improvement than the latter due to the respective weights of the operators: $\logical{XX}=XXXX$ acts on two additional physical qubits compared to $XX$, while $\logical{ZI}=ZZ$ acts on one additional physical qubit compared to $ZI$.

\bibliography{bibliography,mv11,qec}

\begin{thebibliography}{208}%
\makeatletter
\providecommand \@ifxundefined [1]{%
 \@ifx{#1\undefined}
}%
\providecommand \@ifnum [1]{%
 \ifnum #1\expandafter \@firstoftwo
 \else \expandafter \@secondoftwo
 \fi
}%
\providecommand \@ifx [1]{%
 \ifx #1\expandafter \@firstoftwo
 \else \expandafter \@secondoftwo
 \fi
}%
\providecommand \natexlab [1]{#1}%
\providecommand \enquote  [1]{``#1''}%
\providecommand \bibnamefont  [1]{#1}%
\providecommand \bibfnamefont [1]{#1}%
\providecommand \citenamefont [1]{#1}%
\providecommand \href@noop [0]{\@secondoftwo}%
\providecommand \href [0]{\begingroup \@sanitize@url \@href}%
\providecommand \@href[1]{\@@startlink{#1}\@@href}%
\providecommand \@@href[1]{\endgroup#1\@@endlink}%
\providecommand \@sanitize@url [0]{\catcode `\\12\catcode `\$12\catcode
  `\&12\catcode `\#12\catcode `\^12\catcode `\_12\catcode `\%12\relax}%
\providecommand \@@startlink[1]{}%
\providecommand \@@endlink[0]{}%
\providecommand \url  [0]{\begingroup\@sanitize@url \@url }%
\providecommand \@url [1]{\endgroup\@href {#1}{\urlprefix }}%
\providecommand \urlprefix  [0]{URL }%
\providecommand \Eprint [0]{\href }%
\providecommand \doibase [0]{https://doi.org/}%
\providecommand \selectlanguage [0]{\@gobble}%
\providecommand \bibinfo  [0]{\@secondoftwo}%
\providecommand \bibfield  [0]{\@secondoftwo}%
\providecommand \translation [1]{[#1]}%
\providecommand \BibitemOpen [0]{}%
\providecommand \bibitemStop [0]{}%
\providecommand \bibitemNoStop [0]{.\EOS\space}%
\providecommand \EOS [0]{\spacefactor3000\relax}%
\providecommand \BibitemShut  [1]{\csname bibitem#1\endcsname}%
\let\auto@bib@innerbib\@empty
\bibitem [{\citenamefont {Nielsen}\ and\ \citenamefont
  {Chuang}(2000)}]{Nielsen00}%
  \BibitemOpen
  \bibfield  {author} {\bibinfo {author} {\bibfnamefont {M.~A.}\ \bibnamefont
  {Nielsen}}\ and\ \bibinfo {author} {\bibfnamefont {I.~L.}\ \bibnamefont
  {Chuang}},\ }\href@noop {} {\emph {\bibinfo {title} {Quantum Computation and
  Quantum Information}}}\ (\bibinfo  {publisher} {Cambridge University Press},\
  \bibinfo {address} {Cambridge},\ \bibinfo {year} {2000})\BibitemShut
  {NoStop}%
\bibitem [{\citenamefont {Preskill}(2022)}]{Preskill22}%
  \BibitemOpen
  \bibfield  {author} {\bibinfo {author} {\bibfnamefont {J.}~\bibnamefont
  {Preskill}},\ }\href {http://theory.caltech.edu/~preskill/ph219/} {\bibinfo
  {title} {{L}ecture {N}otes for {P}h219/{CS}219: {Q}uantum {I}nformation}}
  (\bibinfo {year} {2022})\BibitemShut {NoStop}%
\bibitem [{\citenamefont {Karzig}\ \emph
  {et~al.}(2017{\natexlab{a}})\citenamefont {Karzig}, \citenamefont {Knapp},
  \citenamefont {Lutchyn}, \citenamefont {Bonderson}, \citenamefont {Hastings},
  \citenamefont {Nayak}, \citenamefont {Alicea}, \citenamefont {Flensberg},
  \citenamefont {Plugge}, \citenamefont {Oreg}, \citenamefont {Marcus},\ and\
  \citenamefont {Freedman}}]{Karzig17}%
  \BibitemOpen
  \bibfield  {author} {\bibinfo {author} {\bibfnamefont {T.}~\bibnamefont
  {Karzig}}, \bibinfo {author} {\bibfnamefont {C.}~\bibnamefont {Knapp}},
  \bibinfo {author} {\bibfnamefont {R.~M.}\ \bibnamefont {Lutchyn}}, \bibinfo
  {author} {\bibfnamefont {P.}~\bibnamefont {Bonderson}}, \bibinfo {author}
  {\bibfnamefont {M.~B.}\ \bibnamefont {Hastings}}, \bibinfo {author}
  {\bibfnamefont {C.}~\bibnamefont {Nayak}}, \bibinfo {author} {\bibfnamefont
  {J.}~\bibnamefont {Alicea}}, \bibinfo {author} {\bibfnamefont
  {K.}~\bibnamefont {Flensberg}}, \bibinfo {author} {\bibfnamefont
  {S.}~\bibnamefont {Plugge}}, \bibinfo {author} {\bibfnamefont
  {Y.}~\bibnamefont {Oreg}}, \bibinfo {author} {\bibfnamefont {C.~M.}\
  \bibnamefont {Marcus}},\ and\ \bibinfo {author} {\bibfnamefont {M.~H.}\
  \bibnamefont {Freedman}},\ }\bibfield  {title} {\bibinfo {title} {Scalable
  designs for quasiparticle-poisoning-protected topological quantum computation
  with {Majorana} zero modes},\ }\href
  {https://doi.org/10.1103/PhysRevB.95.235305} {\bibfield  {journal} {\bibinfo
  {journal} {Phys. Rev. B}\ }\textbf {\bibinfo {volume} {95}},\ \bibinfo
  {pages} {235305} (\bibinfo {year} {2017}{\natexlab{a}})}\BibitemShut
  {NoStop}%
\bibitem [{\citenamefont {Plugge}\ \emph {et~al.}(2017)\citenamefont {Plugge},
  \citenamefont {Rasmussen}, \citenamefont {Egger},\ and\ \citenamefont
  {Flensberg}}]{Plugge17}%
  \BibitemOpen
  \bibfield  {author} {\bibinfo {author} {\bibfnamefont {S.}~\bibnamefont
  {Plugge}}, \bibinfo {author} {\bibfnamefont {A.}~\bibnamefont {Rasmussen}},
  \bibinfo {author} {\bibfnamefont {R.}~\bibnamefont {Egger}},\ and\ \bibinfo
  {author} {\bibfnamefont {K.}~\bibnamefont {Flensberg}},\ }\bibfield  {title}
  {\bibinfo {title} {Majorana box qubits},\ }\href
  {https://doi.org/10.1088/1367-2630/aa54e1} {\bibfield  {journal} {\bibinfo
  {journal} {New J. Phys.}\ }\textbf {\bibinfo {volume} {19}},\ \bibinfo
  {pages} {012001} (\bibinfo {year} {2017})}\BibitemShut {NoStop}%
\bibitem [{\citenamefont {{Fidkowski}}\ \emph {et~al.}(2011)\citenamefont
  {{Fidkowski}}, \citenamefont {{Lutchyn}}, \citenamefont {{Nayak}},\ and\
  \citenamefont {{Fisher}}}]{Fidkowski11b}%
  \BibitemOpen
  \bibfield  {author} {\bibinfo {author} {\bibfnamefont {L.}~\bibnamefont
  {{Fidkowski}}}, \bibinfo {author} {\bibfnamefont {R.~M.}\ \bibnamefont
  {{Lutchyn}}}, \bibinfo {author} {\bibfnamefont {C.}~\bibnamefont {{Nayak}}},\
  and\ \bibinfo {author} {\bibfnamefont {M.~P.~A.}\ \bibnamefont {{Fisher}}},\
  }\bibfield  {title} {\bibinfo {title} {{Majorana zero modes in
  one-dimensional quantum wires without long-ranged superconducting order}},\
  }\href {https://doi.org/10.1103/PhysRevB.84.195436} {\bibfield  {journal}
  {\bibinfo  {journal} {\prb}\ }\textbf {\bibinfo {volume} {84}},\ \bibinfo
  {eid} {195436} (\bibinfo {year} {2011})},\ \Eprint
  {https://arxiv.org/abs/1106.2598} {arXiv:1106.2598} \BibitemShut {NoStop}%
\bibitem [{\citenamefont {Plugge}\ \emph {et~al.}(2016)\citenamefont {Plugge},
  \citenamefont {Landau}, \citenamefont {Sela}, \citenamefont {Altland},
  \citenamefont {Flensberg},\ and\ \citenamefont {Egger}}]{Plugge16}%
  \BibitemOpen
  \bibfield  {author} {\bibinfo {author} {\bibfnamefont {S.}~\bibnamefont
  {Plugge}}, \bibinfo {author} {\bibfnamefont {L.~A.}\ \bibnamefont {Landau}},
  \bibinfo {author} {\bibfnamefont {E.}~\bibnamefont {Sela}}, \bibinfo {author}
  {\bibfnamefont {A.}~\bibnamefont {Altland}}, \bibinfo {author} {\bibfnamefont
  {K.}~\bibnamefont {Flensberg}},\ and\ \bibinfo {author} {\bibfnamefont
  {R.}~\bibnamefont {Egger}},\ }\bibfield  {title} {\bibinfo {title} {Roadmap
  to {Majorana} surface codes},\ }\href
  {https://doi.org/10.1103/PhysRevB.94.174514} {\bibfield  {journal} {\bibinfo
  {journal} {Phys. Rev. B}\ }\textbf {\bibinfo {volume} {94}},\ \bibinfo
  {pages} {174514} (\bibinfo {year} {2016})}\BibitemShut {NoStop}%
\bibitem [{\citenamefont {{Vijay}}\ \emph {et~al.}(2015)\citenamefont
  {{Vijay}}, \citenamefont {{Hsieh}},\ and\ \citenamefont {{Fu}}}]{Vijay15}%
  \BibitemOpen
  \bibfield  {author} {\bibinfo {author} {\bibfnamefont {S.}~\bibnamefont
  {{Vijay}}}, \bibinfo {author} {\bibfnamefont {T.~H.}\ \bibnamefont
  {{Hsieh}}},\ and\ \bibinfo {author} {\bibfnamefont {L.}~\bibnamefont
  {{Fu}}},\ }\bibfield  {title} {\bibinfo {title} {Majorana fermion surface
  code for universal quantum computation},\ }\href
  {https://doi.org/10.1103/PhysRevX.5.041038} {\bibfield  {journal} {\bibinfo
  {journal} {Phys. Rev. X}\ }\textbf {\bibinfo {volume} {5}},\ \bibinfo {eid}
  {041038} (\bibinfo {year} {2015})},\ \Eprint
  {https://arxiv.org/abs/1504.01724} {arXiv:1504.01724} \BibitemShut {NoStop}%
\bibitem [{\citenamefont {Schrade}\ and\ \citenamefont {Fu}(2018)}]{Schrade18}%
  \BibitemOpen
  \bibfield  {author} {\bibinfo {author} {\bibfnamefont {C.}~\bibnamefont
  {Schrade}}\ and\ \bibinfo {author} {\bibfnamefont {L.}~\bibnamefont {Fu}},\
  }\bibfield  {title} {\bibinfo {title} {Majorana superconducting qubit},\
  }\bibfield  {journal} {\bibinfo  {journal} {Physical Review Letters}\
  }\textbf {\bibinfo {volume} {121}},\ \href
  {https://doi.org/10.1103/physrevlett.121.267002}
  {10.1103/physrevlett.121.267002} (\bibinfo {year} {2018})\BibitemShut
  {NoStop}%
\bibitem [{\citenamefont {Kitaev}(2001)}]{Kitaev01}%
  \BibitemOpen
  \bibfield  {author} {\bibinfo {author} {\bibfnamefont {A.~Y.}\ \bibnamefont
  {Kitaev}},\ }\bibfield  {title} {\bibinfo {title} {Unpaired {Majorana}
  fermions in quantum wires},\ }\href
  {https://doi.org/10.1070/1063-7869/44/10S/S29} {\bibfield  {journal}
  {\bibinfo  {journal} {Phys.-Usp.}\ }\textbf {\bibinfo {volume} {44}},\
  \bibinfo {pages} {31} (\bibinfo {year} {2001})},\ \Eprint
  {https://arxiv.org/abs/cond-mat/0010440} {arXiv:cond-mat/0010440}
  \BibitemShut {NoStop}%
\bibitem [{\citenamefont {{Lutchyn}}\ \emph {et~al.}(2010)\citenamefont
  {{Lutchyn}}, \citenamefont {{Sau}},\ and\ \citenamefont {{Das
  Sarma}}}]{Lutchyn10}%
  \BibitemOpen
  \bibfield  {author} {\bibinfo {author} {\bibfnamefont {R.~M.}\ \bibnamefont
  {{Lutchyn}}}, \bibinfo {author} {\bibfnamefont {J.~D.}\ \bibnamefont
  {{Sau}}},\ and\ \bibinfo {author} {\bibfnamefont {S.}~\bibnamefont {{Das
  Sarma}}},\ }\bibfield  {title} {\bibinfo {title} {Majorana {Fermions} and a
  topological phase transition in semiconductor-superconductor
  heterostructures},\ }\href {https://doi.org/10.1103/PhysRevLett.105.077001}
  {\bibfield  {journal} {\bibinfo  {journal} {Phys. Rev. Lett.}\ }\textbf
  {\bibinfo {volume} {105}},\ \bibinfo {pages} {077001} (\bibinfo {year}
  {2010})},\ \Eprint {https://arxiv.org/abs/1002.4033} {arXiv:1002.4033}
  \BibitemShut {NoStop}%
\bibitem [{\citenamefont {{Oreg}}\ \emph {et~al.}(2010)\citenamefont {{Oreg}},
  \citenamefont {{Refael}},\ and\ \citenamefont {{von Oppen}}}]{Oreg10}%
  \BibitemOpen
  \bibfield  {author} {\bibinfo {author} {\bibfnamefont {Y.}~\bibnamefont
  {{Oreg}}}, \bibinfo {author} {\bibfnamefont {G.}~\bibnamefont {{Refael}}},\
  and\ \bibinfo {author} {\bibfnamefont {F.}~\bibnamefont {{von Oppen}}},\
  }\bibfield  {title} {\bibinfo {title} {Helical liquids and {Majorana} bound
  states in quantum wires},\ }\href
  {https://doi.org/10.1103/PhysRevLett.105.177002} {\bibfield  {journal}
  {\bibinfo  {journal} {Phys. Rev. Lett.}\ }\textbf {\bibinfo {volume} {105}},\
  \bibinfo {pages} {177002} (\bibinfo {year} {2010})},\ \Eprint
  {https://arxiv.org/abs/1003.1145} {arXiv:1003.1145} \BibitemShut {NoStop}%
\bibitem [{\citenamefont {Aghaee}\ \emph
  {et~al.}(2025{\natexlab{a}})\citenamefont {Aghaee} \emph
  {et~al.}}]{Aghaee24}%
  \BibitemOpen
  \bibfield  {author} {\bibinfo {author} {\bibfnamefont {M.}~\bibnamefont
  {Aghaee}} \emph {et~al.},\ }\bibfield  {title} {\bibinfo {title}
  {{Interferometric Single-Shot Parity Measurement in an InAs-Al Hybrid
  Device}},\ }\href {https://doi.org/10.1038/s41586-024-08445-2} {\bibfield
  {journal} {\bibinfo  {journal} {Nature}\ }\textbf {\bibinfo {volume} {638}},\
  \bibinfo {pages} {651} (\bibinfo {year} {2025}{\natexlab{a}})}\BibitemShut
  {NoStop}%
\bibitem [{\citenamefont {Aghaee}\ \emph
  {et~al.}(2025{\natexlab{b}})\citenamefont {Aghaee}, \citenamefont {Alam},
  \citenamefont {Andersen}, \citenamefont {Andrzejczuk}, \citenamefont
  {Antipov}, \citenamefont {Astafev}, \citenamefont {Avilovas}, \citenamefont
  {Azizimanesh}, \citenamefont {Bauer}, \citenamefont {Becker}, \citenamefont
  {Bhaskar}, \citenamefont {Boa}, \citenamefont {Boddapati}, \citenamefont
  {Bohac}, \citenamefont {Bommer}, \citenamefont {Borovsky}, \citenamefont
  {Bourdet}, \citenamefont {Boutin}, \citenamefont {Casparis}, \citenamefont
  {Chakravarthi}, \citenamefont {Chalabi}, \citenamefont {Chapman},
  \citenamefont {Chatzaras}, \citenamefont {Chien}, \citenamefont {Cho},
  \citenamefont {Codd}, \citenamefont {Cole}, \citenamefont {Cooper},
  \citenamefont {Corsetti}, \citenamefont {Cui}, \citenamefont {Dandachi},
  \citenamefont {Dinesen}, \citenamefont {Ekefjärd}, \citenamefont {Fallahi},
  \citenamefont {Galletti}, \citenamefont {Gardner}, \citenamefont {Gonzalez},
  \citenamefont {Govender}, \citenamefont {Griggio}, \citenamefont {Grigoryan},
  \citenamefont {Grijalva}, \citenamefont {Gronin}, \citenamefont
  {Gukelberger}, \citenamefont {Hamdast}, \citenamefont {Hamida}, \citenamefont
  {Hansen}, \citenamefont {Hansen}, \citenamefont {Heedt}, \citenamefont {Ho},
  \citenamefont {Holgaard}, \citenamefont {van Hoogdalem}, \citenamefont
  {Hornibrook}, \citenamefont {Ivancevic}, \citenamefont {Jantos},
  \citenamefont {Jensen}, \citenamefont {Jhoja}, \citenamefont {Jones},
  \citenamefont {Joshi}, \citenamefont {Kalashnikov}, \citenamefont {Kallaher},
  \citenamefont {Kalra}, \citenamefont {Karimi}, \citenamefont {Karzig},
  \citenamefont {Kimes}, \citenamefont {King}, \citenamefont {Kloster},
  \citenamefont {Knapp}, \citenamefont {Koski}, \citenamefont {Kostamo},
  \citenamefont {Laeven}, \citenamefont {Lai}, \citenamefont {de~Lange},
  \citenamefont {Larsen}, \citenamefont {Lee}, \citenamefont {Li},
  \citenamefont {Li}, \citenamefont {Liang}, \citenamefont {Lindemann},
  \citenamefont {Looij}, \citenamefont {Lucas}, \citenamefont {Lutchyn},
  \citenamefont {Madsen}, \citenamefont {Madulid}, \citenamefont {Manfra},
  \citenamefont {Manjunath}, \citenamefont {Markussen}, \citenamefont
  {Martinez}, \citenamefont {Mattila}, \citenamefont {Mattinson}, \citenamefont
  {McNeil}, \citenamefont {Millan}, \citenamefont {Mishmash}, \citenamefont
  {Mittal}, \citenamefont {Møllgaard}, \citenamefont {de~Moor}, \citenamefont
  {Morejon}, \citenamefont {Morgan}, \citenamefont {Moussa}, \citenamefont
  {Nabar}, \citenamefont {Narla}, \citenamefont {Nayak}, \citenamefont
  {Nielsen}, \citenamefont {Nielsen}, \citenamefont {Nolet}, \citenamefont
  {Nystrom}, \citenamefont {O'Farrell}, \citenamefont {Ohki}, \citenamefont
  {Papon}, \citenamefont {Petersson}, \citenamefont {Petit}, \citenamefont
  {Pikulin}, \citenamefont {Rajpalke}, \citenamefont {Ramirez}, \citenamefont
  {Razmadze}, \citenamefont {Sadovskyy}, \citenamefont {Sainiemi},
  \citenamefont {Saldaña}, \citenamefont {Sanlorenzo}, \citenamefont {dos
  Santos}, \citenamefont {Schaal}, \citenamefont {Schack}, \citenamefont
  {Schmidgall}, \citenamefont {Sfetsou}, \citenamefont {Sfiligoj},
  \citenamefont {Sinha}, \citenamefont {Sohr}, \citenamefont {Sørensen},
  \citenamefont {Spiegelhauer}, \citenamefont {Stankević}, \citenamefont
  {Stek}, \citenamefont {Strøm-Hansen}, \citenamefont {Suominen},
  \citenamefont {Suter}, \citenamefont {Teicher}, \citenamefont {Tholapi},
  \citenamefont {Thomas}, \citenamefont {Tom}, \citenamefont {Toomey},
  \citenamefont {Tracy}, \citenamefont {Turley}, \citenamefont {Turner},
  \citenamefont {Upadhyay}, \citenamefont {Urban}, \citenamefont {Viazmitinov},
  \citenamefont {Viazmitinova}, \citenamefont {Viegas}, \citenamefont {Vogel},
  \citenamefont {Watson}, \citenamefont {Webster}, \citenamefont {Weston},
  \citenamefont {Williamson}, \citenamefont {Winkler}, \citenamefont {van
  Woerkom}, \citenamefont {Wuetz}, \citenamefont {Yang}, \citenamefont
  {Shang-Jyun}, \citenamefont {Yu}, \citenamefont {Yucelen}, \citenamefont
  {Zamorano}, \citenamefont {Zeisel}, \citenamefont {Zheng},\ and\
  \citenamefont {Zimmerman}}]{Aghaee25}%
  \BibitemOpen
  \bibfield  {author} {\bibinfo {author} {\bibfnamefont {M.}~\bibnamefont
  {Aghaee}}, \bibinfo {author} {\bibfnamefont {Z.}~\bibnamefont {Alam}},
  \bibinfo {author} {\bibfnamefont {R.}~\bibnamefont {Andersen}}, \bibinfo
  {author} {\bibfnamefont {M.}~\bibnamefont {Andrzejczuk}}, \bibinfo {author}
  {\bibfnamefont {A.}~\bibnamefont {Antipov}}, \bibinfo {author} {\bibfnamefont
  {M.}~\bibnamefont {Astafev}}, \bibinfo {author} {\bibfnamefont
  {L.}~\bibnamefont {Avilovas}}, \bibinfo {author} {\bibfnamefont
  {A.}~\bibnamefont {Azizimanesh}}, \bibinfo {author} {\bibfnamefont
  {B.}~\bibnamefont {Bauer}}, \bibinfo {author} {\bibfnamefont
  {J.}~\bibnamefont {Becker}}, \bibinfo {author} {\bibfnamefont {U.~K.}\
  \bibnamefont {Bhaskar}}, \bibinfo {author} {\bibfnamefont {A.~G.}\
  \bibnamefont {Boa}}, \bibinfo {author} {\bibfnamefont {S.}~\bibnamefont
  {Boddapati}}, \bibinfo {author} {\bibfnamefont {N.}~\bibnamefont {Bohac}},
  \bibinfo {author} {\bibfnamefont {J.~D.~S.}\ \bibnamefont {Bommer}}, \bibinfo
  {author} {\bibfnamefont {J.}~\bibnamefont {Borovsky}}, \bibinfo {author}
  {\bibfnamefont {L.}~\bibnamefont {Bourdet}}, \bibinfo {author} {\bibfnamefont
  {S.}~\bibnamefont {Boutin}}, \bibinfo {author} {\bibfnamefont
  {L.}~\bibnamefont {Casparis}}, \bibinfo {author} {\bibfnamefont
  {S.}~\bibnamefont {Chakravarthi}}, \bibinfo {author} {\bibfnamefont
  {H.}~\bibnamefont {Chalabi}}, \bibinfo {author} {\bibfnamefont {B.~J.}\
  \bibnamefont {Chapman}}, \bibinfo {author} {\bibfnamefont {N.}~\bibnamefont
  {Chatzaras}}, \bibinfo {author} {\bibfnamefont {T.-C.}\ \bibnamefont
  {Chien}}, \bibinfo {author} {\bibfnamefont {J.}~\bibnamefont {Cho}}, \bibinfo
  {author} {\bibfnamefont {P.}~\bibnamefont {Codd}}, \bibinfo {author}
  {\bibfnamefont {W.}~\bibnamefont {Cole}}, \bibinfo {author} {\bibfnamefont
  {P.~W.}\ \bibnamefont {Cooper}}, \bibinfo {author} {\bibfnamefont
  {F.}~\bibnamefont {Corsetti}}, \bibinfo {author} {\bibfnamefont
  {A.}~\bibnamefont {Cui}}, \bibinfo {author} {\bibfnamefont {T.~E.}\
  \bibnamefont {Dandachi}}, \bibinfo {author} {\bibfnamefont {C.}~\bibnamefont
  {Dinesen}}, \bibinfo {author} {\bibfnamefont {A.}~\bibnamefont {Ekefjärd}},
  \bibinfo {author} {\bibfnamefont {S.}~\bibnamefont {Fallahi}}, \bibinfo
  {author} {\bibfnamefont {L.}~\bibnamefont {Galletti}}, \bibinfo {author}
  {\bibfnamefont {G.~C.}\ \bibnamefont {Gardner}}, \bibinfo {author}
  {\bibfnamefont {G.~L.}\ \bibnamefont {Gonzalez}}, \bibinfo {author}
  {\bibfnamefont {D.}~\bibnamefont {Govender}}, \bibinfo {author}
  {\bibfnamefont {F.}~\bibnamefont {Griggio}}, \bibinfo {author} {\bibfnamefont
  {R.}~\bibnamefont {Grigoryan}}, \bibinfo {author} {\bibfnamefont
  {S.}~\bibnamefont {Grijalva}}, \bibinfo {author} {\bibfnamefont
  {S.}~\bibnamefont {Gronin}}, \bibinfo {author} {\bibfnamefont
  {J.}~\bibnamefont {Gukelberger}}, \bibinfo {author} {\bibfnamefont
  {M.}~\bibnamefont {Hamdast}}, \bibinfo {author} {\bibfnamefont {A.~B.}\
  \bibnamefont {Hamida}}, \bibinfo {author} {\bibfnamefont {E.~B.}\
  \bibnamefont {Hansen}}, \bibinfo {author} {\bibfnamefont {C.~T.}\
  \bibnamefont {Hansen}}, \bibinfo {author} {\bibfnamefont {S.}~\bibnamefont
  {Heedt}}, \bibinfo {author} {\bibfnamefont {S.}~\bibnamefont {Ho}}, \bibinfo
  {author} {\bibfnamefont {L.}~\bibnamefont {Holgaard}}, \bibinfo {author}
  {\bibfnamefont {K.}~\bibnamefont {van Hoogdalem}}, \bibinfo {author}
  {\bibfnamefont {J.}~\bibnamefont {Hornibrook}}, \bibinfo {author}
  {\bibfnamefont {L.}~\bibnamefont {Ivancevic}}, \bibinfo {author}
  {\bibfnamefont {M.}~\bibnamefont {Jantos}}, \bibinfo {author} {\bibfnamefont
  {T.}~\bibnamefont {Jensen}}, \bibinfo {author} {\bibfnamefont {J.~S.}\
  \bibnamefont {Jhoja}}, \bibinfo {author} {\bibfnamefont {J.~C.}\ \bibnamefont
  {Jones}}, \bibinfo {author} {\bibfnamefont {V.}~\bibnamefont {Joshi}},
  \bibinfo {author} {\bibfnamefont {K.~V.}\ \bibnamefont {Kalashnikov}},
  \bibinfo {author} {\bibfnamefont {R.}~\bibnamefont {Kallaher}}, \bibinfo
  {author} {\bibfnamefont {R.}~\bibnamefont {Kalra}}, \bibinfo {author}
  {\bibfnamefont {F.}~\bibnamefont {Karimi}}, \bibinfo {author} {\bibfnamefont
  {T.}~\bibnamefont {Karzig}}, \bibinfo {author} {\bibfnamefont
  {S.}~\bibnamefont {Kimes}}, \bibinfo {author} {\bibfnamefont
  {E.}~\bibnamefont {King}}, \bibinfo {author} {\bibfnamefont {M.~E.}\
  \bibnamefont {Kloster}}, \bibinfo {author} {\bibfnamefont {C.}~\bibnamefont
  {Knapp}}, \bibinfo {author} {\bibfnamefont {J.~V.}\ \bibnamefont {Koski}},
  \bibinfo {author} {\bibfnamefont {P.}~\bibnamefont {Kostamo}}, \bibinfo
  {author} {\bibfnamefont {T.}~\bibnamefont {Laeven}}, \bibinfo {author}
  {\bibfnamefont {J.}~\bibnamefont {Lai}}, \bibinfo {author} {\bibfnamefont
  {G.}~\bibnamefont {de~Lange}}, \bibinfo {author} {\bibfnamefont {T.~W.}\
  \bibnamefont {Larsen}}, \bibinfo {author} {\bibfnamefont {K.}~\bibnamefont
  {Lee}}, \bibinfo {author} {\bibfnamefont {K.}~\bibnamefont {Li}}, \bibinfo
  {author} {\bibfnamefont {G.}~\bibnamefont {Li}}, \bibinfo {author}
  {\bibfnamefont {S.}~\bibnamefont {Liang}}, \bibinfo {author} {\bibfnamefont
  {T.}~\bibnamefont {Lindemann}}, \bibinfo {author} {\bibfnamefont
  {M.}~\bibnamefont {Looij}}, \bibinfo {author} {\bibfnamefont
  {M.}~\bibnamefont {Lucas}}, \bibinfo {author} {\bibfnamefont
  {R.}~\bibnamefont {Lutchyn}}, \bibinfo {author} {\bibfnamefont {M.~H.}\
  \bibnamefont {Madsen}}, \bibinfo {author} {\bibfnamefont {N.}~\bibnamefont
  {Madulid}}, \bibinfo {author} {\bibfnamefont {M.~J.}\ \bibnamefont {Manfra}},
  \bibinfo {author} {\bibfnamefont {L.}~\bibnamefont {Manjunath}}, \bibinfo
  {author} {\bibfnamefont {S.}~\bibnamefont {Markussen}}, \bibinfo {author}
  {\bibfnamefont {E.}~\bibnamefont {Martinez}}, \bibinfo {author}
  {\bibfnamefont {M.}~\bibnamefont {Mattila}}, \bibinfo {author} {\bibfnamefont
  {J.~R.}\ \bibnamefont {Mattinson}}, \bibinfo {author} {\bibfnamefont
  {R.~P.~G.}\ \bibnamefont {McNeil}}, \bibinfo {author} {\bibfnamefont {A.~P.}\
  \bibnamefont {Millan}}, \bibinfo {author} {\bibfnamefont {R.~V.}\
  \bibnamefont {Mishmash}}, \bibinfo {author} {\bibfnamefont {S.}~\bibnamefont
  {Mittal}}, \bibinfo {author} {\bibfnamefont {C.}~\bibnamefont {Møllgaard}},
  \bibinfo {author} {\bibfnamefont {M.~W.~A.}\ \bibnamefont {de~Moor}},
  \bibinfo {author} {\bibfnamefont {E.~P.}\ \bibnamefont {Morejon}}, \bibinfo
  {author} {\bibfnamefont {T.}~\bibnamefont {Morgan}}, \bibinfo {author}
  {\bibfnamefont {G.}~\bibnamefont {Moussa}}, \bibinfo {author} {\bibfnamefont
  {B.~P.}\ \bibnamefont {Nabar}}, \bibinfo {author} {\bibfnamefont
  {A.}~\bibnamefont {Narla}}, \bibinfo {author} {\bibfnamefont
  {C.}~\bibnamefont {Nayak}}, \bibinfo {author} {\bibfnamefont {J.~H.}\
  \bibnamefont {Nielsen}}, \bibinfo {author} {\bibfnamefont {W.~H.~P.}\
  \bibnamefont {Nielsen}}, \bibinfo {author} {\bibfnamefont {F.}~\bibnamefont
  {Nolet}}, \bibinfo {author} {\bibfnamefont {M.~J.}\ \bibnamefont {Nystrom}},
  \bibinfo {author} {\bibfnamefont {E.}~\bibnamefont {O'Farrell}}, \bibinfo
  {author} {\bibfnamefont {T.~A.}\ \bibnamefont {Ohki}}, \bibinfo {author}
  {\bibfnamefont {C.}~\bibnamefont {Papon}}, \bibinfo {author} {\bibfnamefont
  {K.~D.}\ \bibnamefont {Petersson}}, \bibinfo {author} {\bibfnamefont
  {L.}~\bibnamefont {Petit}}, \bibinfo {author} {\bibfnamefont
  {D.}~\bibnamefont {Pikulin}}, \bibinfo {author} {\bibfnamefont
  {M.}~\bibnamefont {Rajpalke}}, \bibinfo {author} {\bibfnamefont {A.~A.}\
  \bibnamefont {Ramirez}}, \bibinfo {author} {\bibfnamefont {D.}~\bibnamefont
  {Razmadze}}, \bibinfo {author} {\bibfnamefont {I.}~\bibnamefont {Sadovskyy}},
  \bibinfo {author} {\bibfnamefont {L.}~\bibnamefont {Sainiemi}}, \bibinfo
  {author} {\bibfnamefont {J.~C.~E.}\ \bibnamefont {Saldaña}}, \bibinfo
  {author} {\bibfnamefont {I.}~\bibnamefont {Sanlorenzo}}, \bibinfo {author}
  {\bibfnamefont {T.~P.}\ \bibnamefont {dos Santos}}, \bibinfo {author}
  {\bibfnamefont {S.}~\bibnamefont {Schaal}}, \bibinfo {author} {\bibfnamefont
  {J.}~\bibnamefont {Schack}}, \bibinfo {author} {\bibfnamefont {E.~R.}\
  \bibnamefont {Schmidgall}}, \bibinfo {author} {\bibfnamefont
  {C.}~\bibnamefont {Sfetsou}}, \bibinfo {author} {\bibfnamefont
  {C.}~\bibnamefont {Sfiligoj}}, \bibinfo {author} {\bibfnamefont
  {S.}~\bibnamefont {Sinha}}, \bibinfo {author} {\bibfnamefont
  {P.}~\bibnamefont {Sohr}}, \bibinfo {author} {\bibfnamefont {T.~L.}\
  \bibnamefont {Sørensen}}, \bibinfo {author} {\bibfnamefont {K.}~\bibnamefont
  {Spiegelhauer}}, \bibinfo {author} {\bibfnamefont {T.}~\bibnamefont
  {Stankević}}, \bibinfo {author} {\bibfnamefont {L.~J.}\ \bibnamefont
  {Stek}}, \bibinfo {author} {\bibfnamefont {P.}~\bibnamefont {Strøm-Hansen}},
  \bibinfo {author} {\bibfnamefont {H.~J.}\ \bibnamefont {Suominen}}, \bibinfo
  {author} {\bibfnamefont {J.}~\bibnamefont {Suter}}, \bibinfo {author}
  {\bibfnamefont {S.~M.~L.}\ \bibnamefont {Teicher}}, \bibinfo {author}
  {\bibfnamefont {R.}~\bibnamefont {Tholapi}}, \bibinfo {author} {\bibfnamefont
  {M.}~\bibnamefont {Thomas}}, \bibinfo {author} {\bibfnamefont {D.~W.}\
  \bibnamefont {Tom}}, \bibinfo {author} {\bibfnamefont {E.}~\bibnamefont
  {Toomey}}, \bibinfo {author} {\bibfnamefont {J.}~\bibnamefont {Tracy}},
  \bibinfo {author} {\bibfnamefont {M.}~\bibnamefont {Turley}}, \bibinfo
  {author} {\bibfnamefont {M.~D.}\ \bibnamefont {Turner}}, \bibinfo {author}
  {\bibfnamefont {S.}~\bibnamefont {Upadhyay}}, \bibinfo {author}
  {\bibfnamefont {I.}~\bibnamefont {Urban}}, \bibinfo {author} {\bibfnamefont
  {D.~V.}\ \bibnamefont {Viazmitinov}}, \bibinfo {author} {\bibfnamefont
  {A.~W.}\ \bibnamefont {Viazmitinova}}, \bibinfo {author} {\bibfnamefont
  {B.}~\bibnamefont {Viegas}}, \bibinfo {author} {\bibfnamefont {D.~J.}\
  \bibnamefont {Vogel}}, \bibinfo {author} {\bibfnamefont {J.}~\bibnamefont
  {Watson}}, \bibinfo {author} {\bibfnamefont {A.}~\bibnamefont {Webster}},
  \bibinfo {author} {\bibfnamefont {J.}~\bibnamefont {Weston}}, \bibinfo
  {author} {\bibfnamefont {T.}~\bibnamefont {Williamson}}, \bibinfo {author}
  {\bibfnamefont {G.~W.}\ \bibnamefont {Winkler}}, \bibinfo {author}
  {\bibfnamefont {D.~J.}\ \bibnamefont {van Woerkom}}, \bibinfo {author}
  {\bibfnamefont {B.~P.}\ \bibnamefont {Wuetz}}, \bibinfo {author}
  {\bibfnamefont {C.-K.}\ \bibnamefont {Yang}}, \bibinfo {author} {\bibnamefont
  {Shang-Jyun}}, \bibinfo {author} {\bibnamefont {Yu}}, \bibinfo {author}
  {\bibfnamefont {E.}~\bibnamefont {Yucelen}}, \bibinfo {author} {\bibfnamefont
  {J.~H.}\ \bibnamefont {Zamorano}}, \bibinfo {author} {\bibfnamefont
  {R.}~\bibnamefont {Zeisel}}, \bibinfo {author} {\bibfnamefont
  {G.}~\bibnamefont {Zheng}},\ and\ \bibinfo {author} {\bibfnamefont {A.~M.}\
  \bibnamefont {Zimmerman}},\ }\href {https://arxiv.org/abs/2507.08795}
  {\bibinfo {title} {Distinct lifetimes for $x$ and $z$ loop measurements in a
  majorana tetron device}} (\bibinfo {year} {2025}{\natexlab{b}}),\ \Eprint
  {https://arxiv.org/abs/2507.08795} {arXiv:2507.08795 [cond-mat.mes-hall]}
  \BibitemShut {NoStop}%
\bibitem [{\citenamefont {{Kitaev}}(2003)}]{Kitaev97}%
  \BibitemOpen
  \bibfield  {author} {\bibinfo {author} {\bibfnamefont {A.~Y.}\ \bibnamefont
  {{Kitaev}}},\ }\bibfield  {title} {\bibinfo {title} {{Fault-tolerant quantum
  computation by anyons}},\ }\href
  {https://doi.org/10.1016/S0003-4916(02)00018-0} {\bibfield  {journal}
  {\bibinfo  {journal} {Ann. Phys.}\ }\textbf {\bibinfo {volume} {303}},\
  \bibinfo {pages} {2} (\bibinfo {year} {2003})},\ \Eprint
  {https://arxiv.org/abs/quant-ph/9707021} {quant-ph/9707021} \BibitemShut
  {NoStop}%
\bibitem [{\citenamefont {Nayak}\ \emph {et~al.}(2008)\citenamefont {Nayak},
  \citenamefont {Simon}, \citenamefont {Stern}, \citenamefont {Freedman},\ and\
  \citenamefont {Das~Sarma}}]{Nayak08}%
  \BibitemOpen
  \bibfield  {author} {\bibinfo {author} {\bibfnamefont {C.}~\bibnamefont
  {Nayak}}, \bibinfo {author} {\bibfnamefont {S.~H.}\ \bibnamefont {Simon}},
  \bibinfo {author} {\bibfnamefont {A.}~\bibnamefont {Stern}}, \bibinfo
  {author} {\bibfnamefont {M.}~\bibnamefont {Freedman}},\ and\ \bibinfo
  {author} {\bibfnamefont {S.}~\bibnamefont {Das~Sarma}},\ }\bibfield  {title}
  {\bibinfo {title} {Non-{Abelian} anyons and topological quantum
  computation},\ }\href {https://doi.org/10.1103/RevModPhys.80.1083} {\bibfield
   {journal} {\bibinfo  {journal} {Rev. Mod. Phys.}\ }\textbf {\bibinfo
  {volume} {80}},\ \bibinfo {pages} {1083} (\bibinfo {year} {2008})},\ \Eprint
  {https://arxiv.org/abs/0707.1889} {arXiv:0707.1889} \BibitemShut {NoStop}%
\bibitem [{\citenamefont {Cheng}\ \emph {et~al.}(2009)\citenamefont {Cheng},
  \citenamefont {Lutchyn}, \citenamefont {Galitski},\ and\ \citenamefont
  {Das~Sarma}}]{Cheng09}%
  \BibitemOpen
  \bibfield  {author} {\bibinfo {author} {\bibfnamefont {M.}~\bibnamefont
  {Cheng}}, \bibinfo {author} {\bibfnamefont {R.~M.}\ \bibnamefont {Lutchyn}},
  \bibinfo {author} {\bibfnamefont {V.}~\bibnamefont {Galitski}},\ and\
  \bibinfo {author} {\bibfnamefont {S.}~\bibnamefont {Das~Sarma}},\ }\bibfield
  {title} {\bibinfo {title} {Splitting of majorana-fermion modes due to
  intervortex tunneling in a ${p}_{x}+i{p}_{y}$ superconductor},\ }\href
  {https://doi.org/10.1103/PhysRevLett.103.107001} {\bibfield  {journal}
  {\bibinfo  {journal} {Phys. Rev. Lett.}\ }\textbf {\bibinfo {volume} {103}},\
  \bibinfo {pages} {107001} (\bibinfo {year} {2009})}\BibitemShut {NoStop}%
\bibitem [{\citenamefont {Knapp}\ \emph {et~al.}(2018)\citenamefont {Knapp},
  \citenamefont {Karzig}, \citenamefont {Lutchyn},\ and\ \citenamefont
  {Nayak}}]{Knapp18a}%
  \BibitemOpen
  \bibfield  {author} {\bibinfo {author} {\bibfnamefont {C.}~\bibnamefont
  {Knapp}}, \bibinfo {author} {\bibfnamefont {T.}~\bibnamefont {Karzig}},
  \bibinfo {author} {\bibfnamefont {R.~M.}\ \bibnamefont {Lutchyn}},\ and\
  \bibinfo {author} {\bibfnamefont {C.}~\bibnamefont {Nayak}},\ }\bibfield
  {title} {\bibinfo {title} {Dephasing of {{Majorana-based}} qubits},\ }\href
  {https://doi.org/10.1103/PhysRevB.97.125404} {\bibfield  {journal} {\bibinfo
  {journal} {Phys. Rev. B}\ }\textbf {\bibinfo {volume} {97}},\ \bibinfo
  {pages} {125404} (\bibinfo {year} {2018})}\BibitemShut {NoStop}%
\bibitem [{\citenamefont {Mishmash}\ \emph {et~al.}(2020)\citenamefont
  {Mishmash}, \citenamefont {Bauer}, \citenamefont {von Oppen},\ and\
  \citenamefont {Alicea}}]{mishmash2020dephasing}%
  \BibitemOpen
  \bibfield  {author} {\bibinfo {author} {\bibfnamefont {R.~V.}\ \bibnamefont
  {Mishmash}}, \bibinfo {author} {\bibfnamefont {B.}~\bibnamefont {Bauer}},
  \bibinfo {author} {\bibfnamefont {F.}~\bibnamefont {von Oppen}},\ and\
  \bibinfo {author} {\bibfnamefont {J.}~\bibnamefont {Alicea}},\ }\bibfield
  {title} {\bibinfo {title} {Dephasing and leakage dynamics of noisy
  majorana-based qubits: Topological versus andreev},\ }\href@noop {}
  {\bibfield  {journal} {\bibinfo  {journal} {Physical Review B}\ }\textbf
  {\bibinfo {volume} {101}},\ \bibinfo {pages} {075404} (\bibinfo {year}
  {2020})}\BibitemShut {NoStop}%
\bibitem [{\citenamefont {Litinski}\ and\ \citenamefont {von
  Oppen}(2018)}]{Litinski18}%
  \BibitemOpen
  \bibfield  {author} {\bibinfo {author} {\bibfnamefont {D.}~\bibnamefont
  {Litinski}}\ and\ \bibinfo {author} {\bibfnamefont {F.}~\bibnamefont {von
  Oppen}},\ }\bibfield  {title} {\bibinfo {title} {Quantum computing with
  majorana fermion codes},\ }\href {https://doi.org/10.1103/PhysRevB.97.205404}
  {\bibfield  {journal} {\bibinfo  {journal} {Phys. Rev. B}\ }\textbf {\bibinfo
  {volume} {97}},\ \bibinfo {pages} {205404} (\bibinfo {year}
  {2018})}\BibitemShut {NoStop}%
\bibitem [{\citenamefont {Hastings}\ and\ \citenamefont
  {Haah}(2021)}]{Hastings21}%
  \BibitemOpen
  \bibfield  {author} {\bibinfo {author} {\bibfnamefont {M.~B.}\ \bibnamefont
  {Hastings}}\ and\ \bibinfo {author} {\bibfnamefont {J.}~\bibnamefont
  {Haah}},\ }\bibfield  {title} {\bibinfo {title} {Dynamically generated
  logical qubits},\ }\href {https://doi.org/10.22331/q-2021-10-19-564}
  {\bibfield  {journal} {\bibinfo  {journal} {Quantum}\ }\textbf {\bibinfo
  {volume} {5}},\ \bibinfo {pages} {564} (\bibinfo {year} {2021})}\BibitemShut
  {NoStop}%
\bibitem [{\citenamefont {Gidney}(2023)}]{Gidney23}%
  \BibitemOpen
  \bibfield  {author} {\bibinfo {author} {\bibfnamefont {C.}~\bibnamefont
  {Gidney}},\ }\bibfield  {title} {\bibinfo {title} {A {P}air {M}easurement
  {S}urface {C}ode on {P}entagons},\ }\href
  {https://doi.org/10.22331/q-2023-10-25-1156} {\bibfield  {journal} {\bibinfo
  {journal} {{Quantum}}\ }\textbf {\bibinfo {volume} {7}},\ \bibinfo {pages}
  {1156} (\bibinfo {year} {2023})}\BibitemShut {NoStop}%
\bibitem [{\citenamefont {Grans-Samuelsson}\ \emph {et~al.}(2023)\citenamefont
  {Grans-Samuelsson}, \citenamefont {Mishmash}, \citenamefont {Aasen},
  \citenamefont {Knapp}, \citenamefont {Bauer}, \citenamefont {Lackey},
  \citenamefont {da~Silva},\ and\ \citenamefont
  {Bonderson}}]{Grans-Samuelsson23}%
  \BibitemOpen
  \bibfield  {author} {\bibinfo {author} {\bibfnamefont {L.}~\bibnamefont
  {Grans-Samuelsson}}, \bibinfo {author} {\bibfnamefont {R.~V.}\ \bibnamefont
  {Mishmash}}, \bibinfo {author} {\bibfnamefont {D.}~\bibnamefont {Aasen}},
  \bibinfo {author} {\bibfnamefont {C.}~\bibnamefont {Knapp}}, \bibinfo
  {author} {\bibfnamefont {B.}~\bibnamefont {Bauer}}, \bibinfo {author}
  {\bibfnamefont {B.}~\bibnamefont {Lackey}}, \bibinfo {author} {\bibfnamefont
  {M.~P.}\ \bibnamefont {da~Silva}},\ and\ \bibinfo {author} {\bibfnamefont
  {P.}~\bibnamefont {Bonderson}},\ }\href@noop {} {\bibinfo {title} {Improved
  pairwise measurement-based surface code}} (\bibinfo {year} {2023}),\ \Eprint
  {https://arxiv.org/abs/2310.12981} {arXiv:2310.12981} \BibitemShut {NoStop}%
\bibitem [{\citenamefont {{Hassler}}\ \emph {et~al.}(2011)\citenamefont
  {{Hassler}}, \citenamefont {{Akhmerov}},\ and\ \citenamefont
  {{Beenakker}}}]{Hassler11}%
  \BibitemOpen
  \bibfield  {author} {\bibinfo {author} {\bibfnamefont {F.}~\bibnamefont
  {{Hassler}}}, \bibinfo {author} {\bibfnamefont {A.~R.}\ \bibnamefont
  {{Akhmerov}}},\ and\ \bibinfo {author} {\bibfnamefont {C.~W.~J.}\
  \bibnamefont {{Beenakker}}},\ }\bibfield  {title} {\bibinfo {title} {{The
  top-transmon: a hybrid superconducting qubit for parity-protected quantum
  computation}},\ }\href {https://doi.org/10.1088/1367-2630/13/9/095004}
  {\bibfield  {journal} {\bibinfo  {journal} {New J. Phys.}\ }\textbf {\bibinfo
  {volume} {13}},\ \bibinfo {eid} {095004} (\bibinfo {year} {2011})},\ \Eprint
  {https://arxiv.org/abs/1105.0315} {arXiv:1105.0315} \BibitemShut {NoStop}%
\bibitem [{\citenamefont {{van Heck}}\ \emph {et~al.}(2012)\citenamefont {{van
  Heck}}, \citenamefont {{Akhmerov}}, \citenamefont {{Hassler}}, \citenamefont
  {{Burrello}},\ and\ \citenamefont {{Beenakker}}}]{Heck12}%
  \BibitemOpen
  \bibfield  {author} {\bibinfo {author} {\bibfnamefont {B.}~\bibnamefont {{van
  Heck}}}, \bibinfo {author} {\bibfnamefont {A.~R.}\ \bibnamefont
  {{Akhmerov}}}, \bibinfo {author} {\bibfnamefont {F.}~\bibnamefont
  {{Hassler}}}, \bibinfo {author} {\bibfnamefont {M.}~\bibnamefont
  {{Burrello}}},\ and\ \bibinfo {author} {\bibfnamefont {C.~W.~J.}\
  \bibnamefont {{Beenakker}}},\ }\bibfield  {title} {\bibinfo {title}
  {{Coulomb-assisted braiding of {Majorana} fermions in a {Josephson} junction
  array}},\ }\href {https://doi.org/10.1088/1367-2630/14/3/035019} {\bibfield
  {journal} {\bibinfo  {journal} {New J. Phys.}\ }\textbf {\bibinfo {volume}
  {14}},\ \bibinfo {eid} {035019} (\bibinfo {year} {2012})},\ \Eprint
  {https://arxiv.org/abs/1111.6001} {arXiv:1111.6001} \BibitemShut {NoStop}%
\bibitem [{\citenamefont {{Hyart}}\ \emph {et~al.}(2013)\citenamefont
  {{Hyart}}, \citenamefont {{van Heck}}, \citenamefont {{Fulga}}, \citenamefont
  {{Burrello}}, \citenamefont {{Akhmerov}},\ and\ \citenamefont
  {{Beenakker}}}]{Hyart13}%
  \BibitemOpen
  \bibfield  {author} {\bibinfo {author} {\bibfnamefont {T.}~\bibnamefont
  {{Hyart}}}, \bibinfo {author} {\bibfnamefont {B.}~\bibnamefont {{van Heck}}},
  \bibinfo {author} {\bibfnamefont {I.~C.}\ \bibnamefont {{Fulga}}}, \bibinfo
  {author} {\bibfnamefont {M.}~\bibnamefont {{Burrello}}}, \bibinfo {author}
  {\bibfnamefont {A.~R.}\ \bibnamefont {{Akhmerov}}},\ and\ \bibinfo {author}
  {\bibfnamefont {C.~W.~J.}\ \bibnamefont {{Beenakker}}},\ }\bibfield  {title}
  {\bibinfo {title} {{Flux-controlled quantum computation with {Majorana}
  fermions}},\ }\href {https://doi.org/10.1103/PhysRevB.88.035121} {\bibfield
  {journal} {\bibinfo  {journal} {\prb}\ }\textbf {\bibinfo {volume} {88}},\
  \bibinfo {eid} {035121} (\bibinfo {year} {2013})},\ \Eprint
  {https://arxiv.org/abs/1303.4379} {arXiv:1303.4379} \BibitemShut {NoStop}%
\bibitem [{\citenamefont {Aasen}\ \emph {et~al.}(2016)\citenamefont {Aasen},
  \citenamefont {Hell}, \citenamefont {Mishmash}, \citenamefont {Higginbotham},
  \citenamefont {Danon}, \citenamefont {Leijnse}, \citenamefont {Jespersen},
  \citenamefont {Folk}, \citenamefont {Marcus}, \citenamefont {Flensberg},\
  and\ \citenamefont {Alicea}}]{Aasen16}%
  \BibitemOpen
  \bibfield  {author} {\bibinfo {author} {\bibfnamefont {D.}~\bibnamefont
  {Aasen}}, \bibinfo {author} {\bibfnamefont {M.}~\bibnamefont {Hell}},
  \bibinfo {author} {\bibfnamefont {R.~V.}\ \bibnamefont {Mishmash}}, \bibinfo
  {author} {\bibfnamefont {A.}~\bibnamefont {Higginbotham}}, \bibinfo {author}
  {\bibfnamefont {J.}~\bibnamefont {Danon}}, \bibinfo {author} {\bibfnamefont
  {M.}~\bibnamefont {Leijnse}}, \bibinfo {author} {\bibfnamefont {T.~S.}\
  \bibnamefont {Jespersen}}, \bibinfo {author} {\bibfnamefont {J.~A.}\
  \bibnamefont {Folk}}, \bibinfo {author} {\bibfnamefont {C.~M.}\ \bibnamefont
  {Marcus}}, \bibinfo {author} {\bibfnamefont {K.}~\bibnamefont {Flensberg}},\
  and\ \bibinfo {author} {\bibfnamefont {J.}~\bibnamefont {Alicea}},\
  }\bibfield  {title} {\bibinfo {title} {Milestones toward {Majorana}-based
  quantum computing},\ }\href {https://doi.org/10.1103/PhysRevX.6.031016}
  {\bibfield  {journal} {\bibinfo  {journal} {Phys. Rev. X}\ }\textbf {\bibinfo
  {volume} {6}},\ \bibinfo {pages} {031016} (\bibinfo {year} {2016})},\ \Eprint
  {https://arxiv.org/abs/arXiv:1511.05153} {arXiv:1511.05153} \BibitemShut
  {NoStop}%
\bibitem [{\citenamefont {{Alicea}}\ \emph {et~al.}(2011)\citenamefont
  {{Alicea}}, \citenamefont {{Oreg}}, \citenamefont {{Refael}}, \citenamefont
  {{von Oppen}},\ and\ \citenamefont {{Fisher}}}]{Alicea11}%
  \BibitemOpen
  \bibfield  {author} {\bibinfo {author} {\bibfnamefont {J.}~\bibnamefont
  {{Alicea}}}, \bibinfo {author} {\bibfnamefont {Y.}~\bibnamefont {{Oreg}}},
  \bibinfo {author} {\bibfnamefont {G.}~\bibnamefont {{Refael}}}, \bibinfo
  {author} {\bibfnamefont {F.}~\bibnamefont {{von Oppen}}},\ and\ \bibinfo
  {author} {\bibfnamefont {M.~P.~A.}\ \bibnamefont {{Fisher}}},\ }\bibfield
  {title} {\bibinfo {title} {Non-{Abelian} statistics and topological quantum
  information processing in 1d wire networks},\ }\href
  {https://doi.org/10.1038/nphys1915} {\bibfield  {journal} {\bibinfo
  {journal} {Nat. Phys.}\ }\textbf {\bibinfo {volume} {7}},\ \bibinfo {pages}
  {412} (\bibinfo {year} {2011})},\ \Eprint {https://arxiv.org/abs/1006.4395}
  {arXiv:1006.4395} \BibitemShut {NoStop}%
\bibitem [{\citenamefont {Vijay}\ and\ \citenamefont {Fu}(2016)}]{Vijay16b}%
  \BibitemOpen
  \bibfield  {author} {\bibinfo {author} {\bibfnamefont {S.}~\bibnamefont
  {Vijay}}\ and\ \bibinfo {author} {\bibfnamefont {L.}~\bibnamefont {Fu}},\
  }\bibfield  {title} {\bibinfo {title} {Teleportation-based quantum
  information processing with {Majorana} zero modes},\ }\href
  {https://doi.org/10.1103/PhysRevB.94.235446} {\bibfield  {journal} {\bibinfo
  {journal} {Phys. Rev. B}\ }\textbf {\bibinfo {volume} {94}},\ \bibinfo
  {pages} {235446} (\bibinfo {year} {2016})},\ \Eprint
  {https://arxiv.org/abs/1609.00950} {arXiv:1609.00950} \BibitemShut {NoStop}%
\bibitem [{\citenamefont {{Vijay}}\ and\ \citenamefont
  {{Fu}}(2016)}]{Vijay16a}%
  \BibitemOpen
  \bibfield  {author} {\bibinfo {author} {\bibfnamefont {S.}~\bibnamefont
  {{Vijay}}}\ and\ \bibinfo {author} {\bibfnamefont {L.}~\bibnamefont {{Fu}}},\
  }\bibfield  {title} {\bibinfo {title} {Physical implementation of a
  {Majorana} fermion surface code for fault-tolerant quantum computation},\
  }\href {https://doi.org/10.1088/0031-8949/T168/1/014002} {\bibfield
  {journal} {\bibinfo  {journal} {Phys. Scr.}\ }\textbf {\bibinfo {volume}
  {T168}},\ \bibinfo {eid} {014002} (\bibinfo {year} {2016})},\ \Eprint
  {https://arxiv.org/abs/1509.08134} {arXiv:1509.08134} \BibitemShut {NoStop}%
\bibitem [{\citenamefont {Paetznick}\ \emph {et~al.}(2023)\citenamefont
  {Paetznick}, \citenamefont {Knapp}, \citenamefont {Delfosse}, \citenamefont
  {Bauer}, \citenamefont {Haah}, \citenamefont {Hastings},\ and\ \citenamefont
  {da~Silva}}]{Paetznick23}%
  \BibitemOpen
  \bibfield  {author} {\bibinfo {author} {\bibfnamefont {A.}~\bibnamefont
  {Paetznick}}, \bibinfo {author} {\bibfnamefont {C.}~\bibnamefont {Knapp}},
  \bibinfo {author} {\bibfnamefont {N.}~\bibnamefont {Delfosse}}, \bibinfo
  {author} {\bibfnamefont {B.}~\bibnamefont {Bauer}}, \bibinfo {author}
  {\bibfnamefont {J.}~\bibnamefont {Haah}}, \bibinfo {author} {\bibfnamefont
  {M.~B.}\ \bibnamefont {Hastings}},\ and\ \bibinfo {author} {\bibfnamefont
  {M.~P.}\ \bibnamefont {da~Silva}},\ }\bibfield  {title} {\bibinfo {title}
  {Performance of planar {Floquet} codes with {Majorana}-based qubits},\ }\href
  {https://doi.org/10.1103/PRXQuantum.4.010310} {\bibfield  {journal} {\bibinfo
   {journal} {PRX Quantum}\ }\textbf {\bibinfo {volume} {4}},\ \bibinfo {pages}
  {010310} (\bibinfo {year} {2023})}\BibitemShut {NoStop}%
\bibitem [{Note1()}]{Note1}%
  \BibitemOpen
  \bibinfo {note} {The measurement bases are labeled by the corresponding Pauli
  operators. Even a two qubit measurement will still yield only two possible
  outcomes, corresponding to the different eigenvalues of the corresponding
  (two-qubit) Pauli operator. Equivalently, we sometimes refer to these
  measurements as measuring only the parity of the operators, since outcomes on
  individual qubits are not revealed, only the product of the corresponding
  single qubit $\pm 1$ eigenvalues of each tensor factor.}\BibitemShut {Stop}%
\bibitem [{\citenamefont {{Sau}}\ and\ \citenamefont {{Das
  Sarma}}(2024)}]{Sau24}%
  \BibitemOpen
  \bibfield  {author} {\bibinfo {author} {\bibfnamefont {J.~D.}\ \bibnamefont
  {{Sau}}}\ and\ \bibinfo {author} {\bibfnamefont {S.}~\bibnamefont {{Das
  Sarma}}},\ }\bibfield  {title} {\bibinfo {title} {{Capacitance-based Fermion
  parity read-out and predicted Rabi oscillations in a Majorana nanowire}},\
  }\href {https://doi.org/10.48550/arXiv.2406.18080} {\bibfield  {journal}
  {\bibinfo  {journal} {arXiv e-prints}\ ,\ \bibinfo {eid} {arXiv:2406.18080}}
  (\bibinfo {year} {2024})},\ \Eprint {https://arxiv.org/abs/2406.18080}
  {arXiv:2406.18080 [cond-mat.mes-hall]} \BibitemShut {NoStop}%
\bibitem [{\citenamefont {Gerlach}\ and\ \citenamefont
  {Stern}(1922)}]{Gerlach1922}%
  \BibitemOpen
  \bibfield  {author} {\bibinfo {author} {\bibfnamefont {W.}~\bibnamefont
  {Gerlach}}\ and\ \bibinfo {author} {\bibfnamefont {O.}~\bibnamefont
  {Stern}},\ }\bibfield  {title} {\bibinfo {title} {{Der experimentelle
  Nachweis der Richtungsquantelung im Magnetfeld}},\ }\href@noop {} {\bibfield
  {journal} {\bibinfo  {journal} {Z. Physik}\ }\textbf {\bibinfo {volume}
  {9}},\ \bibinfo {pages} {349} (\bibinfo {year} {1922})}\BibitemShut {NoStop}%
\bibitem [{Note2()}]{Note2}%
  \BibitemOpen
  \bibinfo {note} {The idea of verifying the existence of a qubit through two
  anticommuting measurements has been introduced before---Chao {\protect \it et
  al}.~\cite {chao2017overlapping} \protect \emph {define} a qubit in this way,
  and discuss the relation to more conventional notions of a qubit. Similarly,
  stronger guarantees about the action of the measurements can be obtained
  through {\protect \em self-testing} protocols~\cite {Irfan20}, although this
  is beyond the scope of this roadmap.}\BibitemShut {Stop}%
\bibitem [{\citenamefont {Huie}\ \emph {et~al.}(2023)\citenamefont {Huie},
  \citenamefont {Li}, \citenamefont {Chen}, \citenamefont {Hu}, \citenamefont
  {Jia}, \citenamefont {Sun},\ and\ \citenamefont {Covey}}]{Huie23}%
  \BibitemOpen
  \bibfield  {author} {\bibinfo {author} {\bibfnamefont {W.}~\bibnamefont
  {Huie}}, \bibinfo {author} {\bibfnamefont {L.}~\bibnamefont {Li}}, \bibinfo
  {author} {\bibfnamefont {N.}~\bibnamefont {Chen}}, \bibinfo {author}
  {\bibfnamefont {X.}~\bibnamefont {Hu}}, \bibinfo {author} {\bibfnamefont
  {Z.}~\bibnamefont {Jia}}, \bibinfo {author} {\bibfnamefont {W.~K.~C.}\
  \bibnamefont {Sun}},\ and\ \bibinfo {author} {\bibfnamefont {J.~P.}\
  \bibnamefont {Covey}},\ }\bibfield  {title} {\bibinfo {title} {Repetitive
  readout and real-time control of nuclear spin qubits in ${}^{171}\mathrm{Yb}$
  atoms},\ }\href {https://doi.org/10.1103/PRXQuantum.4.030337} {\bibfield
  {journal} {\bibinfo  {journal} {PRX Quantum}\ }\textbf {\bibinfo {volume}
  {4}},\ \bibinfo {pages} {030337} (\bibinfo {year} {2023})}\BibitemShut
  {NoStop}%
\bibitem [{Note3()}]{Note3}%
  \BibitemOpen
  \bibinfo {note} {Note that the same protocols could be applied to logical
  qubits.}\BibitemShut {Stop}%
\bibitem [{Note4()}]{Note4}%
  \BibitemOpen
  \bibinfo {note} {We emphasize the distinction between the usual definition of
  assignment error, which is the probability that a measurement classifier will
  assign an outcome to the observed measurement response that is different from
  the (inaccessible) true measurement outcome. Unlike the usual definition, the
  operational assignment error is directly measurable, but also accounts for
  state transitions during the measurement, not simply the classification
  errors.}\BibitemShut {Stop}%
\bibitem [{Note5()}]{Note5}%
  \BibitemOpen
  \bibinfo {note} {\protect \Cref {eqn:pA,eqn:pM} can be obtained by noting $
  \Pr \left ({\protect \mathcal {M}^{P}_{-1}}\leavevmode@ifvmode {\hbox {$\left
  |\vcenter to1.5\big@size {}\right .\nulldelimiterspace \z@ \mathsurround \z@
  $}}{\protect \mathcal {M}^{Q}_{s}};\protect \frac {1}{2}\protect \tilde
  {\protect \openone }\right )=1- \Pr \left ({\protect \mathcal
  {M}^{P}_{+1}}\leavevmode@ifvmode {\hbox {$\left |\vcenter to1.5\big@size
  {}\right .\nulldelimiterspace \z@ \mathsurround \z@ $}}{\protect \mathcal
  {M}^{Q}_{s}};\protect \frac {1}{2}\protect \tilde {\protect \openone }\right
  )$.}\BibitemShut {Stop}%
\bibitem [{\citenamefont {Schwinger}(1960)}]{Schwinger60}%
  \BibitemOpen
  \bibfield  {author} {\bibinfo {author} {\bibfnamefont {J.}~\bibnamefont
  {Schwinger}},\ }\bibfield  {title} {\bibinfo {title} {Unitary operator
  bases},\ }\href {https://doi.org/10.1073/pnas.46.4.570} {\bibfield  {journal}
  {\bibinfo  {journal} {Proceedings of the National Academy of Sciences}\
  }\textbf {\bibinfo {volume} {46}},\ \bibinfo {pages} {570} (\bibinfo {year}
  {1960})},\ \Eprint
  {https://arxiv.org/abs/https://www.pnas.org/doi/pdf/10.1073/pnas.46.4.570}
  {https://www.pnas.org/doi/pdf/10.1073/pnas.46.4.570} \BibitemShut {NoStop}%
\bibitem [{\citenamefont {Wootters}\ and\ \citenamefont
  {Fields}(1989)}]{Wooters89}%
  \BibitemOpen
  \bibfield  {author} {\bibinfo {author} {\bibfnamefont {W.~K.}\ \bibnamefont
  {Wootters}}\ and\ \bibinfo {author} {\bibfnamefont {B.~D.}\ \bibnamefont
  {Fields}},\ }\bibfield  {title} {\bibinfo {title} {Optimal
  state-determination by mutually unbiased measurements},\ }\href
  {https://doi.org/https://doi.org/10.1016/0003-4916(89)90322-9} {\bibfield
  {journal} {\bibinfo  {journal} {Annals of Physics}\ }\textbf {\bibinfo
  {volume} {191}},\ \bibinfo {pages} {363} (\bibinfo {year}
  {1989})}\BibitemShut {NoStop}%
\bibitem [{\citenamefont {Barenco}\ \emph {et~al.}(1995)\citenamefont
  {Barenco}, \citenamefont {Bennett}, \citenamefont {Cleve}, \citenamefont
  {DiVincenzo}, \citenamefont {Margolus}, \citenamefont {Shor}, \citenamefont
  {Sleator}, \citenamefont {Smolin},\ and\ \citenamefont
  {Weinfurter}}]{Barenco95}%
  \BibitemOpen
  \bibfield  {author} {\bibinfo {author} {\bibfnamefont {A.}~\bibnamefont
  {Barenco}}, \bibinfo {author} {\bibfnamefont {C.~H.}\ \bibnamefont
  {Bennett}}, \bibinfo {author} {\bibfnamefont {R.}~\bibnamefont {Cleve}},
  \bibinfo {author} {\bibfnamefont {D.~P.}\ \bibnamefont {DiVincenzo}},
  \bibinfo {author} {\bibfnamefont {N.}~\bibnamefont {Margolus}}, \bibinfo
  {author} {\bibfnamefont {P.}~\bibnamefont {Shor}}, \bibinfo {author}
  {\bibfnamefont {T.}~\bibnamefont {Sleator}}, \bibinfo {author} {\bibfnamefont
  {J.~A.}\ \bibnamefont {Smolin}},\ and\ \bibinfo {author} {\bibfnamefont
  {H.}~\bibnamefont {Weinfurter}},\ }\bibfield  {title} {\bibinfo {title}
  {Elementary gates for quantum computation},\ }\href
  {https://doi.org/10.1103/PhysRevA.52.3457} {\bibfield  {journal} {\bibinfo
  {journal} {Phys. Rev. A}\ }\textbf {\bibinfo {volume} {52}},\ \bibinfo
  {pages} {3457} (\bibinfo {year} {1995})}\BibitemShut {NoStop}%
\bibitem [{\citenamefont {Knill}(2004)}]{Knill04}%
  \BibitemOpen
  \bibfield  {author} {\bibinfo {author} {\bibfnamefont {E.}~\bibnamefont
  {Knill}},\ }\href@noop {} {\bibinfo {title} {Fault-tolerant postselected
  quantum computation: Schemes}} (\bibinfo {year} {2004}),\ \Eprint
  {https://arxiv.org/abs/quant-ph/0402171} {arXiv:quant-ph/0402171}
  \BibitemShut {NoStop}%
\bibitem [{\citenamefont {{Bravyi}}\ and\ \citenamefont
  {{Kitaev}}(2005)}]{Bravyi05}%
  \BibitemOpen
  \bibfield  {author} {\bibinfo {author} {\bibfnamefont {S.}~\bibnamefont
  {{Bravyi}}}\ and\ \bibinfo {author} {\bibfnamefont {A.}~\bibnamefont
  {{Kitaev}}},\ }\bibfield  {title} {\bibinfo {title} {{Universal quantum
  computation with ideal Clifford gates and noisy ancillas}},\ }\href
  {https://doi.org/10.1103/PhysRevA.71.022316} {\bibfield  {journal} {\bibinfo
  {journal} {\pra}\ }\textbf {\bibinfo {volume} {71}},\ \bibinfo {eid} {022316}
  (\bibinfo {year} {2005})},\ \Eprint {https://arxiv.org/abs/quant-ph/0403025}
  {arXiv:quant-ph/0403025} \BibitemShut {NoStop}%
\bibitem [{\citenamefont {Nichele}\ \emph {et~al.}(2017)\citenamefont
  {Nichele}, \citenamefont {Drachmann}, \citenamefont {Whiticar}, \citenamefont
  {O'Farrell}, \citenamefont {Suominen}, \citenamefont {Fornieri},
  \citenamefont {Wang}, \citenamefont {Gardner}, \citenamefont {Thomas},
  \citenamefont {Hatke}, \citenamefont {Krogstrup}, \citenamefont {Manfra},
  \citenamefont {Flensberg},\ and\ \citenamefont {Marcus}}]{Nichele17}%
  \BibitemOpen
  \bibfield  {author} {\bibinfo {author} {\bibfnamefont {F.}~\bibnamefont
  {Nichele}}, \bibinfo {author} {\bibfnamefont {A.~C.~C.}\ \bibnamefont
  {Drachmann}}, \bibinfo {author} {\bibfnamefont {A.~M.}\ \bibnamefont
  {Whiticar}}, \bibinfo {author} {\bibfnamefont {E.~C.~T.}\ \bibnamefont
  {O'Farrell}}, \bibinfo {author} {\bibfnamefont {H.~J.}\ \bibnamefont
  {Suominen}}, \bibinfo {author} {\bibfnamefont {A.}~\bibnamefont {Fornieri}},
  \bibinfo {author} {\bibfnamefont {T.}~\bibnamefont {Wang}}, \bibinfo {author}
  {\bibfnamefont {G.~C.}\ \bibnamefont {Gardner}}, \bibinfo {author}
  {\bibfnamefont {C.}~\bibnamefont {Thomas}}, \bibinfo {author} {\bibfnamefont
  {A.~T.}\ \bibnamefont {Hatke}}, \bibinfo {author} {\bibfnamefont
  {P.}~\bibnamefont {Krogstrup}}, \bibinfo {author} {\bibfnamefont {M.~J.}\
  \bibnamefont {Manfra}}, \bibinfo {author} {\bibfnamefont {K.}~\bibnamefont
  {Flensberg}},\ and\ \bibinfo {author} {\bibfnamefont {C.~M.}\ \bibnamefont
  {Marcus}},\ }\bibfield  {title} {\bibinfo {title} {Scaling of {Majorana}
  zero-bias conductance peaks},\ }\href
  {https://doi.org/10.1103/PhysRevLett.119.136803} {\bibfield  {journal}
  {\bibinfo  {journal} {Phys. Rev. Lett.}\ }\textbf {\bibinfo {volume} {119}},\
  \bibinfo {pages} {136803} (\bibinfo {year} {2017})}\BibitemShut {NoStop}%
\bibitem [{\citenamefont {Suominen}\ \emph {et~al.}(2017)\citenamefont
  {Suominen}, \citenamefont {Kjaergaard}, \citenamefont {Hamilton},
  \citenamefont {Shabani}, \citenamefont {Palmstr\o{}m}, \citenamefont
  {Marcus},\ and\ \citenamefont {Nichele}}]{Suominen17}%
  \BibitemOpen
  \bibfield  {author} {\bibinfo {author} {\bibfnamefont {H.~J.}\ \bibnamefont
  {Suominen}}, \bibinfo {author} {\bibfnamefont {M.}~\bibnamefont
  {Kjaergaard}}, \bibinfo {author} {\bibfnamefont {A.~R.}\ \bibnamefont
  {Hamilton}}, \bibinfo {author} {\bibfnamefont {J.}~\bibnamefont {Shabani}},
  \bibinfo {author} {\bibfnamefont {C.~J.}\ \bibnamefont {Palmstr\o{}m}},
  \bibinfo {author} {\bibfnamefont {C.~M.}\ \bibnamefont {Marcus}},\ and\
  \bibinfo {author} {\bibfnamefont {F.}~\bibnamefont {Nichele}},\ }\bibfield
  {title} {\bibinfo {title} {Zero-energy modes from coalescing {Andreev} states
  in a two-dimensional semiconductor-superconductor hybrid platform},\ }\href
  {https://doi.org/10.1103/PhysRevLett.119.176805} {\bibfield  {journal}
  {\bibinfo  {journal} {Phys. Rev. Lett.}\ }\textbf {\bibinfo {volume} {119}},\
  \bibinfo {pages} {176805} (\bibinfo {year} {2017})}\BibitemShut {NoStop}%
\bibitem [{\citenamefont {Aghaee}\ \emph {et~al.}(2023)\citenamefont {Aghaee}
  \emph {et~al.}}]{Aghaee23}%
  \BibitemOpen
  \bibfield  {author} {\bibinfo {author} {\bibfnamefont {M.}~\bibnamefont
  {Aghaee}} \emph {et~al.} (\bibinfo {collaboration} {Microsoft Quantum}),\
  }\bibfield  {title} {\bibinfo {title} {{InAs-Al} hybrid devices passing the
  topological gap protocol},\ }\href
  {https://doi.org/10.1103/PhysRevB.107.245423} {\bibfield  {journal} {\bibinfo
   {journal} {Phys. Rev. B}\ }\textbf {\bibinfo {volume} {107}},\ \bibinfo
  {pages} {245423} (\bibinfo {year} {2023})}\BibitemShut {NoStop}%
\bibitem [{\citenamefont {{Sau}}\ \emph {et~al.}(2011)\citenamefont {{Sau}},
  \citenamefont {{Halperin}}, \citenamefont {{Flensberg}},\ and\ \citenamefont
  {{Das Sarma}}}]{Sau11b}%
  \BibitemOpen
  \bibfield  {author} {\bibinfo {author} {\bibfnamefont {J.~D.}\ \bibnamefont
  {{Sau}}}, \bibinfo {author} {\bibfnamefont {B.~I.}\ \bibnamefont
  {{Halperin}}}, \bibinfo {author} {\bibfnamefont {K.}~\bibnamefont
  {{Flensberg}}},\ and\ \bibinfo {author} {\bibfnamefont {S.}~\bibnamefont
  {{Das Sarma}}},\ }\href@noop {} {\bibinfo {title} {{A number conserving
  theory for topologically protected degeneracy in one-dimensional fermions}}}
  (\bibinfo {year} {2011}),\ \Eprint {https://arxiv.org/abs/1106.4014}
  {arXiv:1106.4014} \BibitemShut {NoStop}%
\bibitem [{\citenamefont {Knapp}\ \emph {et~al.}(2020)\citenamefont {Knapp},
  \citenamefont {Väyrynen},\ and\ \citenamefont {Lutchyn}}]{Knapp20}%
  \BibitemOpen
  \bibfield  {author} {\bibinfo {author} {\bibfnamefont {C.}~\bibnamefont
  {Knapp}}, \bibinfo {author} {\bibfnamefont {J.~I.}\ \bibnamefont
  {Väyrynen}},\ and\ \bibinfo {author} {\bibfnamefont {R.~M.}\ \bibnamefont
  {Lutchyn}},\ }\bibfield  {title} {\bibinfo {title} {Number-conserving
  analysis of measurement-based braiding with majorana zero modes},\ }\bibfield
   {journal} {\bibinfo  {journal} {Physical Review B}\ }\textbf {\bibinfo
  {volume} {101}},\ \href {https://doi.org/10.1103/physrevb.101.125108}
  {10.1103/physrevb.101.125108} (\bibinfo {year} {2020})\BibitemShut {NoStop}%
\bibitem [{\citenamefont {Antipov}\ \emph {et~al.}(2018)\citenamefont
  {Antipov}, \citenamefont {Bargerbos}, \citenamefont {Winkler}, \citenamefont
  {Bauer}, \citenamefont {Rossi},\ and\ \citenamefont {Lutchyn}}]{Antipov18}%
  \BibitemOpen
  \bibfield  {author} {\bibinfo {author} {\bibfnamefont {A.~E.}\ \bibnamefont
  {Antipov}}, \bibinfo {author} {\bibfnamefont {A.}~\bibnamefont {Bargerbos}},
  \bibinfo {author} {\bibfnamefont {G.~W.}\ \bibnamefont {Winkler}}, \bibinfo
  {author} {\bibfnamefont {B.}~\bibnamefont {Bauer}}, \bibinfo {author}
  {\bibfnamefont {E.}~\bibnamefont {Rossi}},\ and\ \bibinfo {author}
  {\bibfnamefont {R.~M.}\ \bibnamefont {Lutchyn}},\ }\bibfield  {title}
  {\bibinfo {title} {Effects of gate-induced electric fields on semiconductor
  {Majorana} nanowires},\ }\href {https://doi.org/10.1103/PhysRevX.8.031041}
  {\bibfield  {journal} {\bibinfo  {journal} {Phys. Rev. X}\ }\textbf {\bibinfo
  {volume} {8}},\ \bibinfo {pages} {031041} (\bibinfo {year}
  {2018})}\BibitemShut {NoStop}%
\bibitem [{\citenamefont {Winkler}\ \emph {et~al.}(2019)\citenamefont
  {Winkler}, \citenamefont {Antipov}, \citenamefont {van Heck}, \citenamefont
  {Soluyanov}, \citenamefont {Glazman}, \citenamefont {Wimmer},\ and\
  \citenamefont {Lutchyn}}]{Winkler19}%
  \BibitemOpen
  \bibfield  {author} {\bibinfo {author} {\bibfnamefont {G.~W.}\ \bibnamefont
  {Winkler}}, \bibinfo {author} {\bibfnamefont {A.~E.}\ \bibnamefont
  {Antipov}}, \bibinfo {author} {\bibfnamefont {B.}~\bibnamefont {van Heck}},
  \bibinfo {author} {\bibfnamefont {A.~A.}\ \bibnamefont {Soluyanov}}, \bibinfo
  {author} {\bibfnamefont {L.~I.}\ \bibnamefont {Glazman}}, \bibinfo {author}
  {\bibfnamefont {M.}~\bibnamefont {Wimmer}},\ and\ \bibinfo {author}
  {\bibfnamefont {R.~M.}\ \bibnamefont {Lutchyn}},\ }\bibfield  {title}
  {\bibinfo {title} {Unified numerical approach to topological
  semiconductor-superconductor heterostructures},\ }\href
  {https://doi.org/10.1103/PhysRevB.99.245408} {\bibfield  {journal} {\bibinfo
  {journal} {Phys. Rev. B}\ }\textbf {\bibinfo {volume} {99}},\ \bibinfo
  {pages} {245408} (\bibinfo {year} {2019})}\BibitemShut {NoStop}%
\bibitem [{\citenamefont {Motrunich}\ \emph {et~al.}(2001)\citenamefont
  {Motrunich}, \citenamefont {Damle},\ and\ \citenamefont
  {Huse}}]{Motrunich01}%
  \BibitemOpen
  \bibfield  {author} {\bibinfo {author} {\bibfnamefont {O.}~\bibnamefont
  {Motrunich}}, \bibinfo {author} {\bibfnamefont {K.}~\bibnamefont {Damle}},\
  and\ \bibinfo {author} {\bibfnamefont {D.~A.}\ \bibnamefont {Huse}},\
  }\bibfield  {title} {\bibinfo {title} {Griffiths effects and quantum critical
  points in dirty superconductors without spin-rotation invariance:
  One-dimensional examples},\ }\href
  {https://doi.org/10.1103/PhysRevB.63.224204} {\bibfield  {journal} {\bibinfo
  {journal} {Phys. Rev. B}\ }\textbf {\bibinfo {volume} {63}},\ \bibinfo
  {pages} {224204} (\bibinfo {year} {2001})}\BibitemShut {NoStop}%
\bibitem [{\citenamefont {Brouwer}\ \emph
  {et~al.}(2011{\natexlab{a}})\citenamefont {Brouwer}, \citenamefont
  {Duckheim}, \citenamefont {Romito},\ and\ \citenamefont {von
  Oppen}}]{Brouwer11a}%
  \BibitemOpen
  \bibfield  {author} {\bibinfo {author} {\bibfnamefont {P.~W.}\ \bibnamefont
  {Brouwer}}, \bibinfo {author} {\bibfnamefont {M.}~\bibnamefont {Duckheim}},
  \bibinfo {author} {\bibfnamefont {A.}~\bibnamefont {Romito}},\ and\ \bibinfo
  {author} {\bibfnamefont {F.}~\bibnamefont {von Oppen}},\ }\bibfield  {title}
  {\bibinfo {title} {Topological superconducting phases in disordered quantum
  wires with strong spin-orbit coupling},\ }\href
  {https://doi.org/10.1103/PhysRevB.84.144526} {\bibfield  {journal} {\bibinfo
  {journal} {Phys. Rev. B}\ }\textbf {\bibinfo {volume} {84}},\ \bibinfo
  {pages} {144526} (\bibinfo {year} {2011}{\natexlab{a}})}\BibitemShut
  {NoStop}%
\bibitem [{\citenamefont {Brouwer}\ \emph
  {et~al.}(2011{\natexlab{b}})\citenamefont {Brouwer}, \citenamefont
  {Duckheim}, \citenamefont {Romito},\ and\ \citenamefont {von
  Oppen}}]{Brouwer11b}%
  \BibitemOpen
  \bibfield  {author} {\bibinfo {author} {\bibfnamefont {P.~W.}\ \bibnamefont
  {Brouwer}}, \bibinfo {author} {\bibfnamefont {M.}~\bibnamefont {Duckheim}},
  \bibinfo {author} {\bibfnamefont {A.}~\bibnamefont {Romito}},\ and\ \bibinfo
  {author} {\bibfnamefont {F.}~\bibnamefont {von Oppen}},\ }\bibfield  {title}
  {\bibinfo {title} {Probability distribution of {Majorana} end-state energies
  in disordered wires},\ }\href
  {https://doi.org/10.1103/PhysRevLett.107.196804} {\bibfield  {journal}
  {\bibinfo  {journal} {Phys. Rev. Lett.}\ }\textbf {\bibinfo {volume} {107}},\
  \bibinfo {pages} {196804} (\bibinfo {year} {2011}{\natexlab{b}})}\BibitemShut
  {NoStop}%
\bibitem [{\citenamefont {Akhmerov}\ \emph {et~al.}(2011)\citenamefont
  {Akhmerov}, \citenamefont {Dahlhaus}, \citenamefont {Hassler}, \citenamefont
  {Wimmer},\ and\ \citenamefont {Beenakker}}]{Akhmerov11}%
  \BibitemOpen
  \bibfield  {author} {\bibinfo {author} {\bibfnamefont {A.~R.}\ \bibnamefont
  {Akhmerov}}, \bibinfo {author} {\bibfnamefont {J.~P.}\ \bibnamefont
  {Dahlhaus}}, \bibinfo {author} {\bibfnamefont {F.}~\bibnamefont {Hassler}},
  \bibinfo {author} {\bibfnamefont {M.}~\bibnamefont {Wimmer}},\ and\ \bibinfo
  {author} {\bibfnamefont {C.~W.~J.}\ \bibnamefont {Beenakker}},\ }\bibfield
  {title} {\bibinfo {title} {Quantized conductance at the {Majorana} phase
  transition in a disordered superconducting wire},\ }\href
  {https://doi.org/10.1103/PhysRevLett.106.057001} {\bibfield  {journal}
  {\bibinfo  {journal} {Phys. Rev. Lett.}\ }\textbf {\bibinfo {volume} {106}},\
  \bibinfo {pages} {057001} (\bibinfo {year} {2011})}\BibitemShut {NoStop}%
\bibitem [{\citenamefont {{Stanescu}}\ \emph {et~al.}(2011)\citenamefont
  {{Stanescu}}, \citenamefont {{Lutchyn}},\ and\ \citenamefont {{Das
  Sarma}}}]{Stanescu11}%
  \BibitemOpen
  \bibfield  {author} {\bibinfo {author} {\bibfnamefont {T.~D.}\ \bibnamefont
  {{Stanescu}}}, \bibinfo {author} {\bibfnamefont {R.~M.}\ \bibnamefont
  {{Lutchyn}}},\ and\ \bibinfo {author} {\bibfnamefont {S.}~\bibnamefont {{Das
  Sarma}}},\ }\bibfield  {title} {\bibinfo {title} {{Majorana fermions in
  semiconductor nanowires}},\ }\href
  {https://doi.org/10.1103/PhysRevB.84.144522} {\bibfield  {journal} {\bibinfo
  {journal} {\prb}\ }\textbf {\bibinfo {volume} {84}},\ \bibinfo {eid} {144522}
  (\bibinfo {year} {2011})},\ \Eprint {https://arxiv.org/abs/1106.3078}
  {arXiv:1106.3078} \BibitemShut {NoStop}%
\bibitem [{\citenamefont {Kells}\ \emph {et~al.}(2012)\citenamefont {Kells},
  \citenamefont {Meidan},\ and\ \citenamefont {Brouwer}}]{Kells12}%
  \BibitemOpen
  \bibfield  {author} {\bibinfo {author} {\bibfnamefont {G.}~\bibnamefont
  {Kells}}, \bibinfo {author} {\bibfnamefont {D.}~\bibnamefont {Meidan}},\ and\
  \bibinfo {author} {\bibfnamefont {P.~W.}\ \bibnamefont {Brouwer}},\
  }\bibfield  {title} {\bibinfo {title} {Near-zero-energy end states in
  topologically trivial spin-orbit coupled superconducting nanowires with a
  smooth confinement},\ }\href {https://doi.org/10.1103/PhysRevB.86.100503}
  {\bibfield  {journal} {\bibinfo  {journal} {Phys. Rev. B}\ }\textbf {\bibinfo
  {volume} {86}},\ \bibinfo {pages} {100503} (\bibinfo {year}
  {2012})}\BibitemShut {NoStop}%
\bibitem [{\citenamefont {Prada}\ \emph {et~al.}(2012)\citenamefont {Prada},
  \citenamefont {San-Jose},\ and\ \citenamefont {Aguado}}]{Prada12}%
  \BibitemOpen
  \bibfield  {author} {\bibinfo {author} {\bibfnamefont {E.}~\bibnamefont
  {Prada}}, \bibinfo {author} {\bibfnamefont {P.}~\bibnamefont {San-Jose}},\
  and\ \bibinfo {author} {\bibfnamefont {R.}~\bibnamefont {Aguado}},\
  }\bibfield  {title} {\bibinfo {title} {Transport spectroscopy of ${NS}$
  nanowire junctions with {Majorana} fermions},\ }\href
  {https://doi.org/10.1103/PhysRevB.86.180503} {\bibfield  {journal} {\bibinfo
  {journal} {Phys. Rev. B}\ }\textbf {\bibinfo {volume} {86}},\ \bibinfo
  {pages} {180503} (\bibinfo {year} {2012})},\ \Eprint
  {https://arxiv.org/abs/1203.4488} {arXiv:1203.4488} \BibitemShut {NoStop}%
\bibitem [{\citenamefont {DeGottardi}\ \emph {et~al.}(2013)\citenamefont
  {DeGottardi}, \citenamefont {Sen},\ and\ \citenamefont
  {Vishveshwara}}]{DeGottardi13}%
  \BibitemOpen
  \bibfield  {author} {\bibinfo {author} {\bibfnamefont {W.}~\bibnamefont
  {DeGottardi}}, \bibinfo {author} {\bibfnamefont {D.}~\bibnamefont {Sen}},\
  and\ \bibinfo {author} {\bibfnamefont {S.}~\bibnamefont {Vishveshwara}},\
  }\bibfield  {title} {\bibinfo {title} {Majorana fermions in superconducting
  {1D} systems having periodic, quasiperiodic, and disordered potentials},\
  }\href {https://doi.org/10.1103/PhysRevLett.110.146404} {\bibfield  {journal}
  {\bibinfo  {journal} {Phys. Rev. Lett.}\ }\textbf {\bibinfo {volume} {110}},\
  \bibinfo {pages} {146404} (\bibinfo {year} {2013})}\BibitemShut {NoStop}%
\bibitem [{\citenamefont {Adagideli}\ \emph {et~al.}(2014)\citenamefont
  {Adagideli}, \citenamefont {Wimmer},\ and\ \citenamefont
  {Teker}}]{Adagideli14}%
  \BibitemOpen
  \bibfield  {author} {\bibinfo {author} {\bibfnamefont {I.}~\bibnamefont
  {Adagideli}}, \bibinfo {author} {\bibfnamefont {M.}~\bibnamefont {Wimmer}},\
  and\ \bibinfo {author} {\bibfnamefont {A.}~\bibnamefont {Teker}},\ }\bibfield
   {title} {\bibinfo {title} {Effects of electron scattering on the topological
  properties of nanowires: {Majorana} fermions from disorder and
  superlattices},\ }\href {https://doi.org/10.1103/PhysRevB.89.144506}
  {\bibfield  {journal} {\bibinfo  {journal} {Phys. Rev. B}\ }\textbf {\bibinfo
  {volume} {89}},\ \bibinfo {pages} {144506} (\bibinfo {year}
  {2014})}\BibitemShut {NoStop}%
\bibitem [{\citenamefont {Reeg}\ \emph {et~al.}(2018)\citenamefont {Reeg},
  \citenamefont {Dmytruk}, \citenamefont {Chevallier}, \citenamefont {Loss},\
  and\ \citenamefont {Klinovaja}}]{Reeg18b}%
  \BibitemOpen
  \bibfield  {author} {\bibinfo {author} {\bibfnamefont {C.}~\bibnamefont
  {Reeg}}, \bibinfo {author} {\bibfnamefont {O.}~\bibnamefont {Dmytruk}},
  \bibinfo {author} {\bibfnamefont {D.}~\bibnamefont {Chevallier}}, \bibinfo
  {author} {\bibfnamefont {D.}~\bibnamefont {Loss}},\ and\ \bibinfo {author}
  {\bibfnamefont {J.}~\bibnamefont {Klinovaja}},\ }\bibfield  {title} {\bibinfo
  {title} {Zero-energy {Andreev} bound states from quantum dots in proximitized
  {Rashba} nanowires},\ }\href {https://doi.org/10.1103/PhysRevB.98.245407}
  {\bibfield  {journal} {\bibinfo  {journal} {Phys. Rev. B}\ }\textbf {\bibinfo
  {volume} {98}},\ \bibinfo {pages} {245407} (\bibinfo {year}
  {2018})}\BibitemShut {NoStop}%
\bibitem [{\citenamefont {Vuik}\ \emph {et~al.}(2019)\citenamefont {Vuik},
  \citenamefont {Nijholt}, \citenamefont {Akhmerov},\ and\ \citenamefont
  {Wimmer}}]{Vuik19}%
  \BibitemOpen
  \bibfield  {author} {\bibinfo {author} {\bibfnamefont {A.}~\bibnamefont
  {Vuik}}, \bibinfo {author} {\bibfnamefont {B.}~\bibnamefont {Nijholt}},
  \bibinfo {author} {\bibfnamefont {A.~R.}\ \bibnamefont {Akhmerov}},\ and\
  \bibinfo {author} {\bibfnamefont {M.}~\bibnamefont {Wimmer}},\ }\bibfield
  {title} {\bibinfo {title} {Reproducing topological properties with
  quasi-{Majorana} states},\ }\href
  {https://doi.org/10.21468/SciPostPhys.7.5.061} {\bibfield  {journal}
  {\bibinfo  {journal} {SciPost Phys.}\ }\textbf {\bibinfo {volume} {7}},\
  \bibinfo {pages} {061} (\bibinfo {year} {2019})}\BibitemShut {NoStop}%
\bibitem [{\citenamefont {Woods}\ \emph {et~al.}(2021)\citenamefont {Woods},
  \citenamefont {Das~Sarma},\ and\ \citenamefont {Stanescu}}]{Woods21}%
  \BibitemOpen
  \bibfield  {author} {\bibinfo {author} {\bibfnamefont {B.~D.}\ \bibnamefont
  {Woods}}, \bibinfo {author} {\bibfnamefont {S.}~\bibnamefont {Das~Sarma}},\
  and\ \bibinfo {author} {\bibfnamefont {T.~D.}\ \bibnamefont {Stanescu}},\
  }\bibfield  {title} {\bibinfo {title} {Charge-impurity effects in hybrid
  {Majorana} nanowires},\ }\href
  {https://doi.org/10.1103/PhysRevApplied.16.054053} {\bibfield  {journal}
  {\bibinfo  {journal} {Phys. Rev. Appl.}\ }\textbf {\bibinfo {volume} {16}},\
  \bibinfo {pages} {054053} (\bibinfo {year} {2021})}\BibitemShut {NoStop}%
\bibitem [{\citenamefont {Pan}\ and\ \citenamefont {Das~Sarma}(2020)}]{Pan20b}%
  \BibitemOpen
  \bibfield  {author} {\bibinfo {author} {\bibfnamefont {H.}~\bibnamefont
  {Pan}}\ and\ \bibinfo {author} {\bibfnamefont {S.}~\bibnamefont
  {Das~Sarma}},\ }\bibfield  {title} {\bibinfo {title} {Physical mechanisms for
  zero-bias conductance peaks in majorana nanowires},\ }\href
  {https://doi.org/10.1103/PhysRevResearch.2.013377} {\bibfield  {journal}
  {\bibinfo  {journal} {Phys. Rev. Res.}\ }\textbf {\bibinfo {volume} {2}},\
  \bibinfo {pages} {013377} (\bibinfo {year} {2020})}\BibitemShut {NoStop}%
\bibitem [{\citenamefont {Das~Sarma}\ and\ \citenamefont
  {Pan}(2021)}]{DasSarma21}%
  \BibitemOpen
  \bibfield  {author} {\bibinfo {author} {\bibfnamefont {S.}~\bibnamefont
  {Das~Sarma}}\ and\ \bibinfo {author} {\bibfnamefont {H.}~\bibnamefont
  {Pan}},\ }\bibfield  {title} {\bibinfo {title} {Disorder-induced zero-bias
  peaks in majorana nanowires},\ }\href
  {https://doi.org/10.1103/PhysRevB.103.195158} {\bibfield  {journal} {\bibinfo
   {journal} {Phys. Rev. B}\ }\textbf {\bibinfo {volume} {103}},\ \bibinfo
  {pages} {195158} (\bibinfo {year} {2021})}\BibitemShut {NoStop}%
\bibitem [{\citenamefont {Das~Sarma}\ and\ \citenamefont
  {Pan}(2023)}]{DasSarma23}%
  \BibitemOpen
  \bibfield  {author} {\bibinfo {author} {\bibfnamefont {S.}~\bibnamefont
  {Das~Sarma}}\ and\ \bibinfo {author} {\bibfnamefont {H.}~\bibnamefont
  {Pan}},\ }\bibfield  {title} {\bibinfo {title} {Density of states, transport,
  and topology in disordered majorana nanowires},\ }\href
  {https://doi.org/10.1103/PhysRevB.108.085415} {\bibfield  {journal} {\bibinfo
   {journal} {Phys. Rev. B}\ }\textbf {\bibinfo {volume} {108}},\ \bibinfo
  {pages} {085415} (\bibinfo {year} {2023})}\BibitemShut {NoStop}%
\bibitem [{\citenamefont {Das~Sarma}\ \emph {et~al.}(2023)\citenamefont
  {Das~Sarma}, \citenamefont {Sau},\ and\ \citenamefont
  {Stanescu}}]{DasSarma23b}%
  \BibitemOpen
  \bibfield  {author} {\bibinfo {author} {\bibfnamefont {S.}~\bibnamefont
  {Das~Sarma}}, \bibinfo {author} {\bibfnamefont {J.~D.}\ \bibnamefont {Sau}},\
  and\ \bibinfo {author} {\bibfnamefont {T.~D.}\ \bibnamefont {Stanescu}},\
  }\bibfield  {title} {\bibinfo {title} {Spectral properties, topological
  patches, and effective phase diagrams of finite disordered majorana
  nanowires},\ }\href {https://doi.org/10.1103/PhysRevB.108.085416} {\bibfield
  {journal} {\bibinfo  {journal} {Phys. Rev. B}\ }\textbf {\bibinfo {volume}
  {108}},\ \bibinfo {pages} {085416} (\bibinfo {year} {2023})}\BibitemShut
  {NoStop}%
\bibitem [{\citenamefont {Pan}\ and\ \citenamefont {Das~Sarma}(2024)}]{Pan24}%
  \BibitemOpen
  \bibfield  {author} {\bibinfo {author} {\bibfnamefont {H.}~\bibnamefont
  {Pan}}\ and\ \bibinfo {author} {\bibfnamefont {S.}~\bibnamefont
  {Das~Sarma}},\ }\bibfield  {title} {\bibinfo {title} {Disordered majorana
  nanowires: Studying disorder without any disorder},\ }\bibfield  {journal}
  {\bibinfo  {journal} {Physical Review B}\ }\textbf {\bibinfo {volume}
  {110}},\ \href {https://doi.org/10.1103/physrevb.110.075401}
  {10.1103/physrevb.110.075401} (\bibinfo {year} {2024})\BibitemShut {NoStop}%
\bibitem [{\citenamefont {Antipov}\ \emph {et~al.}(2025)\citenamefont
  {Antipov}, \citenamefont {Bauer}, \citenamefont {Boutin}, \citenamefont
  {Cole}, \citenamefont {Gukelberger}, \citenamefont {Karimi}, \citenamefont
  {Lutchyn},\ and\ \citenamefont {Winkler}}]{Antipov25}%
  \BibitemOpen
  \bibfield  {author} {\bibinfo {author} {\bibfnamefont {A.}~\bibnamefont
  {Antipov}}, \bibinfo {author} {\bibfnamefont {B.}~\bibnamefont {Bauer}},
  \bibinfo {author} {\bibfnamefont {S.}~\bibnamefont {Boutin}}, \bibinfo
  {author} {\bibfnamefont {W.}~\bibnamefont {Cole}}, \bibinfo {author}
  {\bibfnamefont {J.}~\bibnamefont {Gukelberger}}, \bibinfo {author}
  {\bibfnamefont {F.}~\bibnamefont {Karimi}}, \bibinfo {author} {\bibfnamefont
  {R.~M.}\ \bibnamefont {Lutchyn}},\ and\ \bibinfo {author} {\bibfnamefont
  {G.}~\bibnamefont {Winkler}},\ }\href@noop {} {\bibinfo {title} {Correlated
  disorder in one-dimensional topological superconductors}} (\bibinfo {year}
  {2025}),\ \bibinfo {note} {in preparation}\BibitemShut {NoStop}%
\bibitem [{\citenamefont {{Akhmerov}}\ \emph {et~al.}(2009)\citenamefont
  {{Akhmerov}}, \citenamefont {{Nilsson}},\ and\ \citenamefont
  {{Beenakker}}}]{Akhmerov09}%
  \BibitemOpen
  \bibfield  {author} {\bibinfo {author} {\bibfnamefont {A.~R.}\ \bibnamefont
  {{Akhmerov}}}, \bibinfo {author} {\bibfnamefont {J.}~\bibnamefont
  {{Nilsson}}},\ and\ \bibinfo {author} {\bibfnamefont {C.~W.~J.}\ \bibnamefont
  {{Beenakker}}},\ }\bibfield  {title} {\bibinfo {title} {Electrically detected
  interferometry of {Majorana} fermions in a topological insulator},\ }\href
  {https://doi.org/10.1103/PhysRevLett.102.216404} {\bibfield  {journal}
  {\bibinfo  {journal} {Phys. Rev. Lett.}\ }\textbf {\bibinfo {volume} {102}},\
  \bibinfo {eid} {216404} (\bibinfo {year} {2009})},\ \Eprint
  {https://arxiv.org/abs/0903.2196} {arXiv:0903.2196} \BibitemShut {NoStop}%
\bibitem [{\citenamefont {Fu}\ and\ \citenamefont
  {Kane}(2009{\natexlab{a}})}]{Fu09a}%
  \BibitemOpen
  \bibfield  {author} {\bibinfo {author} {\bibfnamefont {L.}~\bibnamefont
  {Fu}}\ and\ \bibinfo {author} {\bibfnamefont {C.~L.}\ \bibnamefont {Kane}},\
  }\bibfield  {title} {\bibinfo {title} {Probing neutral {Majorana} fermion
  edge modes with charge transport},\ }\href
  {https://doi.org/10.1103/PhysRevLett.102.216403} {\bibfield  {journal}
  {\bibinfo  {journal} {Phys. Rev. Lett.}\ }\textbf {\bibinfo {volume} {102}},\
  \bibinfo {pages} {216403} (\bibinfo {year} {2009}{\natexlab{a}})}\BibitemShut
  {NoStop}%
\bibitem [{\citenamefont {{Fu}}(2010)}]{Fu10}%
  \BibitemOpen
  \bibfield  {author} {\bibinfo {author} {\bibfnamefont {L.}~\bibnamefont
  {{Fu}}},\ }\bibfield  {title} {\bibinfo {title} {Electron teleportation via
  {Majorana} bound states in a mesoscopic superconductor},\ }\href
  {https://doi.org/10.1103/PhysRevLett.104.056402} {\bibfield  {journal}
  {\bibinfo  {journal} {Phys. Rev. Lett.}\ }\textbf {\bibinfo {volume} {104}},\
  \bibinfo {eid} {056402} (\bibinfo {year} {2010})},\ \Eprint
  {https://arxiv.org/abs/0909.5172} {arXiv:0909.5172} \BibitemShut {NoStop}%
\bibitem [{\citenamefont {Burnell}\ \emph {et~al.}(2013)\citenamefont
  {Burnell}, \citenamefont {Shnirman},\ and\ \citenamefont {Oreg}}]{Burnell13}%
  \BibitemOpen
  \bibfield  {author} {\bibinfo {author} {\bibfnamefont {F.~J.}\ \bibnamefont
  {Burnell}}, \bibinfo {author} {\bibfnamefont {A.}~\bibnamefont {Shnirman}},\
  and\ \bibinfo {author} {\bibfnamefont {Y.}~\bibnamefont {Oreg}},\ }\bibfield
  {title} {\bibinfo {title} {Measuring fermion parity correlations and
  relaxation rates in one-dimensional topological superconducting wires},\
  }\bibfield  {journal} {\bibinfo  {journal} {Physical Review B}\ }\textbf
  {\bibinfo {volume} {88}},\ \href {https://doi.org/10.1103/physrevb.88.224507}
  {10.1103/physrevb.88.224507} (\bibinfo {year} {2013})\BibitemShut {NoStop}%
\bibitem [{\citenamefont {Houzet}\ \emph {et~al.}(2013)\citenamefont {Houzet},
  \citenamefont {Meyer}, \citenamefont {Badiane},\ and\ \citenamefont
  {Glazman}}]{Houzet13}%
  \BibitemOpen
  \bibfield  {author} {\bibinfo {author} {\bibfnamefont {M.}~\bibnamefont
  {Houzet}}, \bibinfo {author} {\bibfnamefont {J.~S.}\ \bibnamefont {Meyer}},
  \bibinfo {author} {\bibfnamefont {D.~M.}\ \bibnamefont {Badiane}},\ and\
  \bibinfo {author} {\bibfnamefont {L.~I.}\ \bibnamefont {Glazman}},\
  }\bibfield  {title} {\bibinfo {title} {Dynamics of {Majorana} states in a
  topological {Josephson} junction},\ }\href
  {https://doi.org/10.1103/PhysRevLett.111.046401} {\bibfield  {journal}
  {\bibinfo  {journal} {Phys. Rev. Lett.}\ }\textbf {\bibinfo {volume} {111}},\
  \bibinfo {pages} {046401} (\bibinfo {year} {2013})}\BibitemShut {NoStop}%
\bibitem [{\citenamefont {Cheng}\ and\ \citenamefont
  {Lutchyn}(2015)}]{Cheng15}%
  \BibitemOpen
  \bibfield  {author} {\bibinfo {author} {\bibfnamefont {M.}~\bibnamefont
  {Cheng}}\ and\ \bibinfo {author} {\bibfnamefont {R.}~\bibnamefont
  {Lutchyn}},\ }\bibfield  {title} {\bibinfo {title} {Fractional {Josephson}
  effect in number-conserving systems},\ }\href
  {https://doi.org/10.1103/PhysRevB.92.134516} {\bibfield  {journal} {\bibinfo
  {journal} {Phys. Rev. B}\ }\textbf {\bibinfo {volume} {92}},\ \bibinfo
  {pages} {134516} (\bibinfo {year} {2015})}\BibitemShut {NoStop}%
\bibitem [{\citenamefont {Thamm}\ and\ \citenamefont
  {Rosenow}(2021)}]{Thamm21}%
  \BibitemOpen
  \bibfield  {author} {\bibinfo {author} {\bibfnamefont {M.}~\bibnamefont
  {Thamm}}\ and\ \bibinfo {author} {\bibfnamefont {B.}~\bibnamefont
  {Rosenow}},\ }\bibfield  {title} {\bibinfo {title} {Transmission amplitude
  through a coulomb blockaded majorana wire},\ }\href
  {https://doi.org/10.1103/PhysRevResearch.3.023221} {\bibfield  {journal}
  {\bibinfo  {journal} {Phys. Rev. Res.}\ }\textbf {\bibinfo {volume} {3}},\
  \bibinfo {pages} {023221} (\bibinfo {year} {2021})}\BibitemShut {NoStop}%
\bibitem [{\citenamefont {Drechsler}\ \emph {et~al.}(2024)\citenamefont
  {Drechsler}, \citenamefont {Lesser},\ and\ \citenamefont
  {Oreg}}]{Drechsler24}%
  \BibitemOpen
  \bibfield  {author} {\bibinfo {author} {\bibfnamefont {N.}~\bibnamefont
  {Drechsler}}, \bibinfo {author} {\bibfnamefont {O.}~\bibnamefont {Lesser}},\
  and\ \bibinfo {author} {\bibfnamefont {Y.}~\bibnamefont {Oreg}},\ }\bibfield
  {title} {\bibinfo {title} {Electron interference as a probe of majorana zero
  modes},\ }\href {https://doi.org/10.1103/PhysRevB.110.134503} {\bibfield
  {journal} {\bibinfo  {journal} {Phys. Rev. B}\ }\textbf {\bibinfo {volume}
  {110}},\ \bibinfo {pages} {134503} (\bibinfo {year} {2024})}\BibitemShut
  {NoStop}%
\bibitem [{\citenamefont {Nadj-Perge}\ \emph {et~al.}(2013)\citenamefont
  {Nadj-Perge}, \citenamefont {Drozdov}, \citenamefont {Bernevig},\ and\
  \citenamefont {Yazdani}}]{Nadj-Perge13}%
  \BibitemOpen
  \bibfield  {author} {\bibinfo {author} {\bibfnamefont {S.}~\bibnamefont
  {Nadj-Perge}}, \bibinfo {author} {\bibfnamefont {I.~K.}\ \bibnamefont
  {Drozdov}}, \bibinfo {author} {\bibfnamefont {B.~A.}\ \bibnamefont
  {Bernevig}},\ and\ \bibinfo {author} {\bibfnamefont {A.}~\bibnamefont
  {Yazdani}},\ }\bibfield  {title} {\bibinfo {title} {Proposal for realizing
  {Majorana} fermions in chains of magnetic atoms on a superconductor},\ }\href
  {https://doi.org/10.1103/PhysRevB.88.020407} {\bibfield  {journal} {\bibinfo
  {journal} {Phys. Rev. B}\ }\textbf {\bibinfo {volume} {88}},\ \bibinfo
  {pages} {020407} (\bibinfo {year} {2013})}\BibitemShut {NoStop}%
\bibitem [{\citenamefont {Klinovaja}\ \emph {et~al.}(2013)\citenamefont
  {Klinovaja}, \citenamefont {Stano}, \citenamefont {Yazdani},\ and\
  \citenamefont {Loss}}]{Klinovaja13}%
  \BibitemOpen
  \bibfield  {author} {\bibinfo {author} {\bibfnamefont {J.}~\bibnamefont
  {Klinovaja}}, \bibinfo {author} {\bibfnamefont {P.}~\bibnamefont {Stano}},
  \bibinfo {author} {\bibfnamefont {A.}~\bibnamefont {Yazdani}},\ and\ \bibinfo
  {author} {\bibfnamefont {D.}~\bibnamefont {Loss}},\ }\bibfield  {title}
  {\bibinfo {title} {Topological superconductivity and {Majorana} fermions in
  {RKKY} systems},\ }\href {https://doi.org/10.1103/PhysRevLett.111.186805}
  {\bibfield  {journal} {\bibinfo  {journal} {Phys. Rev. Lett.}\ }\textbf
  {\bibinfo {volume} {111}},\ \bibinfo {pages} {186805} (\bibinfo {year}
  {2013})}\BibitemShut {NoStop}%
\bibitem [{\citenamefont {Braunecker}\ and\ \citenamefont
  {Simon}(2013)}]{Braunecker13}%
  \BibitemOpen
  \bibfield  {author} {\bibinfo {author} {\bibfnamefont {B.}~\bibnamefont
  {Braunecker}}\ and\ \bibinfo {author} {\bibfnamefont {P.}~\bibnamefont
  {Simon}},\ }\bibfield  {title} {\bibinfo {title} {Interplay between classical
  magnetic moments and superconductivity in quantum one-dimensional conductors:
  Toward a self-sustained topological {Majorana} phase},\ }\href
  {https://doi.org/10.1103/PhysRevLett.111.147202} {\bibfield  {journal}
  {\bibinfo  {journal} {Phys. Rev. Lett.}\ }\textbf {\bibinfo {volume} {111}},\
  \bibinfo {pages} {147202} (\bibinfo {year} {2013})}\BibitemShut {NoStop}%
\bibitem [{\citenamefont {Pientka}\ \emph {et~al.}(2013)\citenamefont
  {Pientka}, \citenamefont {Glazman},\ and\ \citenamefont {von
  Oppen}}]{Pientka13}%
  \BibitemOpen
  \bibfield  {author} {\bibinfo {author} {\bibfnamefont {F.}~\bibnamefont
  {Pientka}}, \bibinfo {author} {\bibfnamefont {L.~I.}\ \bibnamefont
  {Glazman}},\ and\ \bibinfo {author} {\bibfnamefont {F.}~\bibnamefont {von
  Oppen}},\ }\bibfield  {title} {\bibinfo {title} {Topological superconducting
  phase in helical shiba chains},\ }\href
  {https://doi.org/10.1103/PhysRevB.88.155420} {\bibfield  {journal} {\bibinfo
  {journal} {Phys. Rev. B}\ }\textbf {\bibinfo {volume} {88}},\ \bibinfo
  {pages} {155420} (\bibinfo {year} {2013})}\BibitemShut {NoStop}%
\bibitem [{\citenamefont {{Nadj-Perge}}\ \emph {et~al.}(2014)\citenamefont
  {{Nadj-Perge}}, \citenamefont {{Drozdov}}, \citenamefont {{Li}},
  \citenamefont {{Chen}}, \citenamefont {{Jeon}}, \citenamefont {{Seo}},
  \citenamefont {{MacDonald}}, \citenamefont {{Bernevig}},\ and\ \citenamefont
  {{Yazdani}}}]{Nadj-Perge14}%
  \BibitemOpen
  \bibfield  {author} {\bibinfo {author} {\bibfnamefont {S.}~\bibnamefont
  {{Nadj-Perge}}}, \bibinfo {author} {\bibfnamefont {I.~K.}\ \bibnamefont
  {{Drozdov}}}, \bibinfo {author} {\bibfnamefont {J.}~\bibnamefont {{Li}}},
  \bibinfo {author} {\bibfnamefont {H.}~\bibnamefont {{Chen}}}, \bibinfo
  {author} {\bibfnamefont {S.}~\bibnamefont {{Jeon}}}, \bibinfo {author}
  {\bibfnamefont {J.}~\bibnamefont {{Seo}}}, \bibinfo {author} {\bibfnamefont
  {A.~H.}\ \bibnamefont {{MacDonald}}}, \bibinfo {author} {\bibfnamefont
  {B.~A.}\ \bibnamefont {{Bernevig}}},\ and\ \bibinfo {author} {\bibfnamefont
  {A.}~\bibnamefont {{Yazdani}}},\ }\bibfield  {title} {\bibinfo {title}
  {{Observation of {Majorana} fermions in ferromagnetic atomic chains on a
  superconductor}},\ }\href {https://doi.org/10.1126/science.1259327}
  {\bibfield  {journal} {\bibinfo  {journal} {Science}\ }\textbf {\bibinfo
  {volume} {346}},\ \bibinfo {pages} {602} (\bibinfo {year} {2014})},\ \Eprint
  {https://arxiv.org/abs/1410.0682} {arXiv:1410.0682} \BibitemShut {NoStop}%
\bibitem [{\citenamefont {Zhang}\ \emph {et~al.}(2016)\citenamefont {Zhang},
  \citenamefont {Kim}, \citenamefont {Rossi},\ and\ \citenamefont
  {Lutchyn}}]{Zhang16b}%
  \BibitemOpen
  \bibfield  {author} {\bibinfo {author} {\bibfnamefont {J.}~\bibnamefont
  {Zhang}}, \bibinfo {author} {\bibfnamefont {Y.}~\bibnamefont {Kim}}, \bibinfo
  {author} {\bibfnamefont {E.}~\bibnamefont {Rossi}},\ and\ \bibinfo {author}
  {\bibfnamefont {R.~M.}\ \bibnamefont {Lutchyn}},\ }\bibfield  {title}
  {\bibinfo {title} {Topological superconductivity in a multichannel
  yu-shiba-rusinov chain},\ }\href {https://doi.org/10.1103/PhysRevB.93.024507}
  {\bibfield  {journal} {\bibinfo  {journal} {Phys. Rev. B}\ }\textbf {\bibinfo
  {volume} {93}},\ \bibinfo {pages} {024507} (\bibinfo {year}
  {2016})}\BibitemShut {NoStop}%
\bibitem [{\citenamefont {Cook}\ and\ \citenamefont {Franz}(2011)}]{Cook11}%
  \BibitemOpen
  \bibfield  {author} {\bibinfo {author} {\bibfnamefont {A.}~\bibnamefont
  {Cook}}\ and\ \bibinfo {author} {\bibfnamefont {M.}~\bibnamefont {Franz}},\
  }\bibfield  {title} {\bibinfo {title} {Majorana fermions in a
  topological-insulator nanowire proximity-coupled to an $s$-wave
  superconductor},\ }\href {https://doi.org/10.1103/PhysRevB.84.201105}
  {\bibfield  {journal} {\bibinfo  {journal} {Phys. Rev. B}\ }\textbf {\bibinfo
  {volume} {84}},\ \bibinfo {pages} {201105} (\bibinfo {year}
  {2011})}\BibitemShut {NoStop}%
\bibitem [{\citenamefont {Hosur}\ \emph {et~al.}(2011)\citenamefont {Hosur},
  \citenamefont {Ghaemi}, \citenamefont {Mong},\ and\ \citenamefont
  {Vishwanath}}]{Hosur11}%
  \BibitemOpen
  \bibfield  {author} {\bibinfo {author} {\bibfnamefont {P.}~\bibnamefont
  {Hosur}}, \bibinfo {author} {\bibfnamefont {P.}~\bibnamefont {Ghaemi}},
  \bibinfo {author} {\bibfnamefont {R.~S.~K.}\ \bibnamefont {Mong}},\ and\
  \bibinfo {author} {\bibfnamefont {A.}~\bibnamefont {Vishwanath}},\ }\bibfield
   {title} {\bibinfo {title} {Majorana modes at the ends of superconductor
  vortices in doped topological insulators},\ }\href
  {https://doi.org/10.1103/PhysRevLett.107.097001} {\bibfield  {journal}
  {\bibinfo  {journal} {Phys. Rev. Lett.}\ }\textbf {\bibinfo {volume} {107}},\
  \bibinfo {pages} {097001} (\bibinfo {year} {2011})}\BibitemShut {NoStop}%
\bibitem [{\citenamefont {Vaitiekenas}\ \emph {et~al.}(2020)\citenamefont
  {Vaitiekenas}, \citenamefont {Winkler}, \citenamefont {van Heck},
  \citenamefont {Karzig}, \citenamefont {Deng}, \citenamefont {Flensberg},
  \citenamefont {Glazman}, \citenamefont {Nayak}, \citenamefont {Krogstrup},
  \citenamefont {Lutchyn},\ and\ \citenamefont {Marcus}}]{Vaitieknas20}%
  \BibitemOpen
  \bibfield  {author} {\bibinfo {author} {\bibfnamefont {S.}~\bibnamefont
  {Vaitiekenas}}, \bibinfo {author} {\bibfnamefont {G.~W.}\ \bibnamefont
  {Winkler}}, \bibinfo {author} {\bibfnamefont {B.}~\bibnamefont {van Heck}},
  \bibinfo {author} {\bibfnamefont {T.}~\bibnamefont {Karzig}}, \bibinfo
  {author} {\bibfnamefont {M.-T.}\ \bibnamefont {Deng}}, \bibinfo {author}
  {\bibfnamefont {K.}~\bibnamefont {Flensberg}}, \bibinfo {author}
  {\bibfnamefont {L.~I.}\ \bibnamefont {Glazman}}, \bibinfo {author}
  {\bibfnamefont {C.}~\bibnamefont {Nayak}}, \bibinfo {author} {\bibfnamefont
  {P.}~\bibnamefont {Krogstrup}}, \bibinfo {author} {\bibfnamefont {R.~M.}\
  \bibnamefont {Lutchyn}},\ and\ \bibinfo {author} {\bibfnamefont {C.~M.}\
  \bibnamefont {Marcus}},\ }\bibfield  {title} {\bibinfo {title} {Flux-induced
  topological superconductivity in full-shell nanowires},\ }\href
  {https://doi.org/10.1126/science.aav3392} {\bibfield  {journal} {\bibinfo
  {journal} {Science}\ }\textbf {\bibinfo {volume} {367}},\ \bibinfo {pages}
  {eaav3392} (\bibinfo {year} {2020})}\BibitemShut {NoStop}%
\bibitem [{\citenamefont {Wang}\ \emph {et~al.}(2018)\citenamefont {Wang},
  \citenamefont {Kong}, \citenamefont {Fan}, \citenamefont {Chen},
  \citenamefont {Zhu}, \citenamefont {Liu}, \citenamefont {Cao}, \citenamefont
  {Sun}, \citenamefont {Du}, \citenamefont {Schneeloch}, \citenamefont {Zhong},
  \citenamefont {Gu}, \citenamefont {Fu}, \citenamefont {Ding},\ and\
  \citenamefont {Gao}}]{Wang18}%
  \BibitemOpen
  \bibfield  {author} {\bibinfo {author} {\bibfnamefont {D.}~\bibnamefont
  {Wang}}, \bibinfo {author} {\bibfnamefont {L.}~\bibnamefont {Kong}}, \bibinfo
  {author} {\bibfnamefont {P.}~\bibnamefont {Fan}}, \bibinfo {author}
  {\bibfnamefont {H.}~\bibnamefont {Chen}}, \bibinfo {author} {\bibfnamefont
  {S.}~\bibnamefont {Zhu}}, \bibinfo {author} {\bibfnamefont {W.}~\bibnamefont
  {Liu}}, \bibinfo {author} {\bibfnamefont {L.}~\bibnamefont {Cao}}, \bibinfo
  {author} {\bibfnamefont {Y.}~\bibnamefont {Sun}}, \bibinfo {author}
  {\bibfnamefont {S.}~\bibnamefont {Du}}, \bibinfo {author} {\bibfnamefont
  {J.}~\bibnamefont {Schneeloch}}, \bibinfo {author} {\bibfnamefont
  {R.}~\bibnamefont {Zhong}}, \bibinfo {author} {\bibfnamefont
  {G.}~\bibnamefont {Gu}}, \bibinfo {author} {\bibfnamefont {L.}~\bibnamefont
  {Fu}}, \bibinfo {author} {\bibfnamefont {H.}~\bibnamefont {Ding}},\ and\
  \bibinfo {author} {\bibfnamefont {H.-J.}\ \bibnamefont {Gao}},\ }\bibfield
  {title} {\bibinfo {title} {Evidence for {Majorana} bound states in an
  iron-based superconductor},\ }\href {https://doi.org/10.1126/science.aao1797}
  {\bibfield  {journal} {\bibinfo  {journal} {Science}\ }\textbf {\bibinfo
  {volume} {362}},\ \bibinfo {pages} {333} (\bibinfo {year}
  {2018})}\BibitemShut {NoStop}%
\bibitem [{\citenamefont {Kong}\ \emph {et~al.}(2019)\citenamefont {Kong},
  \citenamefont {Zhu}, \citenamefont {Papaj}, \citenamefont {Chen},
  \citenamefont {Cao}, \citenamefont {Isobe}, \citenamefont {Xing},
  \citenamefont {Liu}, \citenamefont {Wang}, \citenamefont {Fan}, \citenamefont
  {Sun}, \citenamefont {Du}, \citenamefont {Schneeloch}, \citenamefont {Zhong},
  \citenamefont {Gu}, \citenamefont {Fu}, \citenamefont {Gao},\ and\
  \citenamefont {Ding}}]{Kong19}%
  \BibitemOpen
  \bibfield  {author} {\bibinfo {author} {\bibfnamefont {L.}~\bibnamefont
  {Kong}}, \bibinfo {author} {\bibfnamefont {S.}~\bibnamefont {Zhu}}, \bibinfo
  {author} {\bibfnamefont {M.}~\bibnamefont {Papaj}}, \bibinfo {author}
  {\bibfnamefont {H.}~\bibnamefont {Chen}}, \bibinfo {author} {\bibfnamefont
  {L.}~\bibnamefont {Cao}}, \bibinfo {author} {\bibfnamefont {H.}~\bibnamefont
  {Isobe}}, \bibinfo {author} {\bibfnamefont {Y.}~\bibnamefont {Xing}},
  \bibinfo {author} {\bibfnamefont {W.}~\bibnamefont {Liu}}, \bibinfo {author}
  {\bibfnamefont {D.}~\bibnamefont {Wang}}, \bibinfo {author} {\bibfnamefont
  {P.}~\bibnamefont {Fan}}, \bibinfo {author} {\bibfnamefont {Y.}~\bibnamefont
  {Sun}}, \bibinfo {author} {\bibfnamefont {S.}~\bibnamefont {Du}}, \bibinfo
  {author} {\bibfnamefont {J.}~\bibnamefont {Schneeloch}}, \bibinfo {author}
  {\bibfnamefont {R.}~\bibnamefont {Zhong}}, \bibinfo {author} {\bibfnamefont
  {G.}~\bibnamefont {Gu}}, \bibinfo {author} {\bibfnamefont {L.}~\bibnamefont
  {Fu}}, \bibinfo {author} {\bibfnamefont {H.-J.}\ \bibnamefont {Gao}},\ and\
  \bibinfo {author} {\bibfnamefont {H.}~\bibnamefont {Ding}},\ }\bibfield
  {title} {\bibinfo {title} {Half-integer level shift of vortex bound states in
  an iron-based superconductor},\ }\href
  {https://doi.org/10.1038/s41567-019-0630-5} {\bibfield  {journal} {\bibinfo
  {journal} {Nat. Phys.}\ }\textbf {\bibinfo {volume} {15}},\ \bibinfo {pages}
  {1181} (\bibinfo {year} {2019})}\BibitemShut {NoStop}%
\bibitem [{\citenamefont {Read}\ and\ \citenamefont {Green}(2000)}]{Read00}%
  \BibitemOpen
  \bibfield  {author} {\bibinfo {author} {\bibfnamefont {N.}~\bibnamefont
  {Read}}\ and\ \bibinfo {author} {\bibfnamefont {D.}~\bibnamefont {Green}},\
  }\bibfield  {title} {\bibinfo {title} {Paired states of fermions in two
  dimensions with breaking of parity and time-reversal symmetries and the
  fractional quantum {Hall} effect},\ }\href
  {https://doi.org/10.1103/PhysRevB.61.10267} {\bibfield  {journal} {\bibinfo
  {journal} {Phys. Rev. B}\ }\textbf {\bibinfo {volume} {61}},\ \bibinfo
  {pages} {10267} (\bibinfo {year} {2000})}\BibitemShut {NoStop}%
\bibitem [{\citenamefont {{Fu}}\ and\ \citenamefont {{Kane}}(2008)}]{Fu08}%
  \BibitemOpen
  \bibfield  {author} {\bibinfo {author} {\bibfnamefont {L.}~\bibnamefont
  {{Fu}}}\ and\ \bibinfo {author} {\bibfnamefont {C.~L.}\ \bibnamefont
  {{Kane}}},\ }\bibfield  {title} {\bibinfo {title} {Superconducting proximity
  effect and {Majorana} fermions at the surface of a topological insulator},\
  }\href {https://doi.org/10.1103/PhysRevLett.100.096407} {\bibfield  {journal}
  {\bibinfo  {journal} {Phys. Rev. Lett.}\ }\textbf {\bibinfo {volume} {100}},\
  \bibinfo {pages} {096407} (\bibinfo {year} {2008})},\ \Eprint
  {https://arxiv.org/abs/0707.1692} {arXiv:0707.1692} \BibitemShut {NoStop}%
\bibitem [{\citenamefont {Fu}\ and\ \citenamefont
  {Kane}(2009{\natexlab{b}})}]{Fu09}%
  \BibitemOpen
  \bibfield  {author} {\bibinfo {author} {\bibfnamefont {L.}~\bibnamefont
  {Fu}}\ and\ \bibinfo {author} {\bibfnamefont {C.~L.}\ \bibnamefont {Kane}},\
  }\bibfield  {title} {\bibinfo {title} {{Josephson} current and noise at a
  superconductor/quantum-spin-{Hall}-insulator/superconductor junction},\
  }\href {https://doi.org/10.1103/PhysRevB.79.161408} {\bibfield  {journal}
  {\bibinfo  {journal} {Phys. Rev. B}\ }\textbf {\bibinfo {volume} {79}},\
  \bibinfo {pages} {161408(R)} (\bibinfo {year}
  {2009}{\natexlab{b}})}\BibitemShut {NoStop}%
\bibitem [{\citenamefont {Hasan}\ and\ \citenamefont {Kane}(2010)}]{Hasan10}%
  \BibitemOpen
  \bibfield  {author} {\bibinfo {author} {\bibfnamefont {M.~Z.}\ \bibnamefont
  {Hasan}}\ and\ \bibinfo {author} {\bibfnamefont {C.~L.}\ \bibnamefont
  {Kane}},\ }\bibfield  {title} {\bibinfo {title} {Colloquium: Topological
  insulators},\ }\href {https://doi.org/10.1103/RevModPhys.82.3045} {\bibfield
  {journal} {\bibinfo  {journal} {Rev. Mod. Phys.}\ }\textbf {\bibinfo {volume}
  {82}},\ \bibinfo {pages} {3045} (\bibinfo {year} {2010})}\BibitemShut
  {NoStop}%
\bibitem [{\citenamefont {Sau}\ \emph {et~al.}(2010)\citenamefont {Sau},
  \citenamefont {Lutchyn}, \citenamefont {Tewari},\ and\ \citenamefont
  {Das~Sarma}}]{Sau10a}%
  \BibitemOpen
  \bibfield  {author} {\bibinfo {author} {\bibfnamefont {J.~D.}\ \bibnamefont
  {Sau}}, \bibinfo {author} {\bibfnamefont {R.~M.}\ \bibnamefont {Lutchyn}},
  \bibinfo {author} {\bibfnamefont {S.}~\bibnamefont {Tewari}},\ and\ \bibinfo
  {author} {\bibfnamefont {S.}~\bibnamefont {Das~Sarma}},\ }\bibfield  {title}
  {\bibinfo {title} {Generic new platform for topological quantum computation
  using semiconductor heterostructures},\ }\href
  {https://doi.org/10.1103/PhysRevLett.104.040502} {\bibfield  {journal}
  {\bibinfo  {journal} {Phys. Rev. Lett.}\ }\textbf {\bibinfo {volume} {104}},\
  \bibinfo {pages} {040502} (\bibinfo {year} {2010})},\ \Eprint
  {https://arxiv.org/abs/0907.2239} {arXiv:0907.2239} \BibitemShut {NoStop}%
\bibitem [{\citenamefont {Alicea}(2010)}]{Alicea10}%
  \BibitemOpen
  \bibfield  {author} {\bibinfo {author} {\bibfnamefont {J.}~\bibnamefont
  {Alicea}},\ }\bibfield  {title} {\bibinfo {title} {Majorana fermions in a
  tunable semiconductor device},\ }\href
  {https://doi.org/10.1103/PhysRevB.81.125318} {\bibfield  {journal} {\bibinfo
  {journal} {Phys. Rev. B}\ }\textbf {\bibinfo {volume} {81}},\ \bibinfo
  {pages} {125318} (\bibinfo {year} {2010})}\BibitemShut {NoStop}%
\bibitem [{\citenamefont {{Sau}}\ \emph {et~al.}(2010)\citenamefont {{Sau}},
  \citenamefont {{Tewari}}, \citenamefont {{Lutchyn}}, \citenamefont
  {{Stanescu}},\ and\ \citenamefont {{Das Sarma}}}]{Sau10b}%
  \BibitemOpen
  \bibfield  {author} {\bibinfo {author} {\bibfnamefont {J.~D.}\ \bibnamefont
  {{Sau}}}, \bibinfo {author} {\bibfnamefont {S.}~\bibnamefont {{Tewari}}},
  \bibinfo {author} {\bibfnamefont {R.~M.}\ \bibnamefont {{Lutchyn}}}, \bibinfo
  {author} {\bibfnamefont {T.~D.}\ \bibnamefont {{Stanescu}}},\ and\ \bibinfo
  {author} {\bibfnamefont {S.}~\bibnamefont {{Das Sarma}}},\ }\bibfield
  {title} {\bibinfo {title} {Non-{Abelian} quantum order in spin-orbit-coupled
  semiconductors: Search for topological {Majorana} particles in solid-state
  systems},\ }\href {https://doi.org/10.1103/PhysRevB.82.214509} {\bibfield
  {journal} {\bibinfo  {journal} {\prb}\ }\textbf {\bibinfo {volume} {82}},\
  \bibinfo {eid} {214509} (\bibinfo {year} {2010})},\ \Eprint
  {https://arxiv.org/abs/1006.2829} {arXiv:1006.2829} \BibitemShut {NoStop}%
\bibitem [{\citenamefont {Chung}\ \emph {et~al.}(2011)\citenamefont {Chung},
  \citenamefont {Zhang}, \citenamefont {Qi},\ and\ \citenamefont
  {Zhang}}]{Chung11}%
  \BibitemOpen
  \bibfield  {author} {\bibinfo {author} {\bibfnamefont {S.~B.}\ \bibnamefont
  {Chung}}, \bibinfo {author} {\bibfnamefont {H.-J.}\ \bibnamefont {Zhang}},
  \bibinfo {author} {\bibfnamefont {X.-L.}\ \bibnamefont {Qi}},\ and\ \bibinfo
  {author} {\bibfnamefont {S.-C.}\ \bibnamefont {Zhang}},\ }\bibfield  {title}
  {\bibinfo {title} {Topological superconducting phase and {Majorana} fermions
  in half-metal/superconductor heterostructures},\ }\href
  {https://doi.org/10.1103/PhysRevB.84.060510} {\bibfield  {journal} {\bibinfo
  {journal} {Phys. Rev. B}\ }\textbf {\bibinfo {volume} {84}},\ \bibinfo
  {pages} {060510} (\bibinfo {year} {2011})}\BibitemShut {NoStop}%
\bibitem [{\citenamefont {Duckheim}\ and\ \citenamefont
  {Brouwer}(2011)}]{Duckheim11}%
  \BibitemOpen
  \bibfield  {author} {\bibinfo {author} {\bibfnamefont {M.}~\bibnamefont
  {Duckheim}}\ and\ \bibinfo {author} {\bibfnamefont {P.~W.}\ \bibnamefont
  {Brouwer}},\ }\bibfield  {title} {\bibinfo {title} {Andreev reflection from
  noncentrosymmetric superconductors and {Majorana} bound-state generation in
  half-metallic ferromagnets},\ }\href
  {https://doi.org/10.1103/PhysRevB.83.054513} {\bibfield  {journal} {\bibinfo
  {journal} {Phys. Rev. B}\ }\textbf {\bibinfo {volume} {83}},\ \bibinfo
  {pages} {054513} (\bibinfo {year} {2011})}\BibitemShut {NoStop}%
\bibitem [{\citenamefont {Potter}\ and\ \citenamefont {Lee}(2012)}]{Potter12}%
  \BibitemOpen
  \bibfield  {author} {\bibinfo {author} {\bibfnamefont {A.~C.}\ \bibnamefont
  {Potter}}\ and\ \bibinfo {author} {\bibfnamefont {P.~A.}\ \bibnamefont
  {Lee}},\ }\bibfield  {title} {\bibinfo {title} {Topological superconductivity
  and {Majorana} fermions in metallic surface states},\ }\href
  {https://doi.org/10.1103/PhysRevB.85.094516} {\bibfield  {journal} {\bibinfo
  {journal} {Phys. Rev. B}\ }\textbf {\bibinfo {volume} {85}},\ \bibinfo
  {pages} {094516} (\bibinfo {year} {2012})}\BibitemShut {NoStop}%
\bibitem [{\citenamefont {Lutchyn}\ \emph {et~al.}(2018)\citenamefont
  {Lutchyn}, \citenamefont {Bakkers}, \citenamefont {Kouwenhoven},
  \citenamefont {Krogstrup}, \citenamefont {Marcus},\ and\ \citenamefont
  {Oreg}}]{Lutchyn18}%
  \BibitemOpen
  \bibfield  {author} {\bibinfo {author} {\bibfnamefont {R.~M.}\ \bibnamefont
  {Lutchyn}}, \bibinfo {author} {\bibfnamefont {E.~P. A.~M.}\ \bibnamefont
  {Bakkers}}, \bibinfo {author} {\bibfnamefont {L.~P.}\ \bibnamefont
  {Kouwenhoven}}, \bibinfo {author} {\bibfnamefont {P.}~\bibnamefont
  {Krogstrup}}, \bibinfo {author} {\bibfnamefont {C.~M.}\ \bibnamefont
  {Marcus}},\ and\ \bibinfo {author} {\bibfnamefont {Y.}~\bibnamefont {Oreg}},\
  }\bibfield  {title} {\bibinfo {title} {{Majorana} zero modes in
  superconductor-semiconductor heterostructures},\ }\href
  {https://doi.org/10.1038/s41578-018-0003-1} {\bibfield  {journal} {\bibinfo
  {journal} {Nat. Rev. Mater.}\ }\textbf {\bibinfo {volume} {3}},\ \bibinfo
  {pages} {52} (\bibinfo {year} {2018})}\BibitemShut {NoStop}%
\bibitem [{\citenamefont {{Liu}}\ \emph {et~al.}(2020)\citenamefont {{Liu}},
  \citenamefont {{Vaitiek{\.{e}}nas}}, \citenamefont
  {{Mart{\'\i}-S{\'a}nchez}}, \citenamefont {{Koch}}, \citenamefont {{Hart}},
  \citenamefont {{Cui}}, \citenamefont {{Kanne}}, \citenamefont {{Khan}},
  \citenamefont {{Tanta}}, \citenamefont {{Upadhyay}}, \citenamefont
  {{Cachaza}}, \citenamefont {{Marcus}}, \citenamefont {{Arbiol}},
  \citenamefont {{Moler}},\ and\ \citenamefont {{Krogstrup}}}]{Liu19a}%
  \BibitemOpen
  \bibfield  {author} {\bibinfo {author} {\bibfnamefont {Y.}~\bibnamefont
  {{Liu}}}, \bibinfo {author} {\bibfnamefont {S.}~\bibnamefont
  {{Vaitiek{\.{e}}nas}}}, \bibinfo {author} {\bibfnamefont {S.}~\bibnamefont
  {{Mart{\'\i}-S{\'a}nchez}}}, \bibinfo {author} {\bibfnamefont
  {C.}~\bibnamefont {{Koch}}}, \bibinfo {author} {\bibfnamefont
  {S.}~\bibnamefont {{Hart}}}, \bibinfo {author} {\bibfnamefont
  {Z.}~\bibnamefont {{Cui}}}, \bibinfo {author} {\bibfnamefont
  {T.}~\bibnamefont {{Kanne}}}, \bibinfo {author} {\bibfnamefont {S.~A.}\
  \bibnamefont {{Khan}}}, \bibinfo {author} {\bibfnamefont {R.}~\bibnamefont
  {{Tanta}}}, \bibinfo {author} {\bibfnamefont {S.}~\bibnamefont {{Upadhyay}}},
  \bibinfo {author} {\bibfnamefont {M.~E.}\ \bibnamefont {{Cachaza}}}, \bibinfo
  {author} {\bibfnamefont {C.~M.}\ \bibnamefont {{Marcus}}}, \bibinfo {author}
  {\bibfnamefont {J.}~\bibnamefont {{Arbiol}}}, \bibinfo {author}
  {\bibfnamefont {K.~A.}\ \bibnamefont {{Moler}}},\ and\ \bibinfo {author}
  {\bibfnamefont {P.}~\bibnamefont {{Krogstrup}}},\ }\bibfield  {title}
  {\bibinfo {title} {{Semiconductor-Ferromagnetic Insulator-Superconductor
  Nanowires: Stray Field and Exchange Field}},\ }\href
  {https://doi.org/10.1021/acs.nanolett.9b04187} {\bibfield  {journal}
  {\bibinfo  {journal} {Nano Letters}\ }\textbf {\bibinfo {volume} {20}},\
  \bibinfo {pages} {456} (\bibinfo {year} {2020})},\ \Eprint
  {https://arxiv.org/abs/1910.03364} {arXiv:1910.03364} \BibitemShut {NoStop}%
\bibitem [{\citenamefont {Laubscher}\ \emph {et~al.}(2024)\citenamefont
  {Laubscher}, \citenamefont {Sau},\ and\ \citenamefont
  {Das~Sarma}}]{Laubscher24}%
  \BibitemOpen
  \bibfield  {author} {\bibinfo {author} {\bibfnamefont {K.}~\bibnamefont
  {Laubscher}}, \bibinfo {author} {\bibfnamefont {J.~D.}\ \bibnamefont {Sau}},\
  and\ \bibinfo {author} {\bibfnamefont {S.}~\bibnamefont {Das~Sarma}},\
  }\bibfield  {title} {\bibinfo {title} {Majorana zero modes in gate-defined
  germanium hole nanowires},\ }\href
  {https://doi.org/10.1103/PhysRevB.109.035433} {\bibfield  {journal} {\bibinfo
   {journal} {Phys. Rev. B}\ }\textbf {\bibinfo {volume} {109}},\ \bibinfo
  {pages} {035433} (\bibinfo {year} {2024})}\BibitemShut {NoStop}%
\bibitem [{\citenamefont {{Sato}}\ \emph {et~al.}(2009)\citenamefont {{Sato}},
  \citenamefont {{Takahashi}},\ and\ \citenamefont {{Fujimoto}}}]{Sato09b}%
  \BibitemOpen
  \bibfield  {author} {\bibinfo {author} {\bibfnamefont {M.}~\bibnamefont
  {{Sato}}}, \bibinfo {author} {\bibfnamefont {Y.}~\bibnamefont
  {{Takahashi}}},\ and\ \bibinfo {author} {\bibfnamefont {S.}~\bibnamefont
  {{Fujimoto}}},\ }\bibfield  {title} {\bibinfo {title} {Non-{Abelian}
  topological order in {$s$}-wave superfluids of ultracold fermionic atoms},\
  }\href {https://doi.org/10.1103/PhysRevLett.103.020401} {\bibfield  {journal}
  {\bibinfo  {journal} {Phys. Rev. Lett.}\ }\textbf {\bibinfo {volume} {103}},\
  \bibinfo {eid} {020401} (\bibinfo {year} {2009})},\ \Eprint
  {https://arxiv.org/abs/0901.4693} {arXiv:0901.4693} \BibitemShut {NoStop}%
\bibitem [{\citenamefont {Zhang}\ \emph {et~al.}(2008)\citenamefont {Zhang},
  \citenamefont {Tewari}, \citenamefont {Lutchyn},\ and\ \citenamefont
  {Das~Sarma}}]{Zhang08}%
  \BibitemOpen
  \bibfield  {author} {\bibinfo {author} {\bibfnamefont {C.}~\bibnamefont
  {Zhang}}, \bibinfo {author} {\bibfnamefont {S.}~\bibnamefont {Tewari}},
  \bibinfo {author} {\bibfnamefont {R.~M.}\ \bibnamefont {Lutchyn}},\ and\
  \bibinfo {author} {\bibfnamefont {S.}~\bibnamefont {Das~Sarma}},\ }\bibfield
  {title} {\bibinfo {title} {${p}_{x}+i{p}_{y}$ superfluid from $s$-wave
  interactions of fermionic cold atoms},\ }\href
  {https://doi.org/10.1103/PhysRevLett.101.160401} {\bibfield  {journal}
  {\bibinfo  {journal} {Phys. Rev. Lett.}\ }\textbf {\bibinfo {volume} {101}},\
  \bibinfo {pages} {160401} (\bibinfo {year} {2008})}\BibitemShut {NoStop}%
\bibitem [{\citenamefont {Willett}\ \emph {et~al.}(1987)\citenamefont
  {Willett}, \citenamefont {Eisenstein}, \citenamefont {St\"ormer},
  \citenamefont {Tsui}, \citenamefont {Gossard},\ and\ \citenamefont
  {English}}]{Willett87}%
  \BibitemOpen
  \bibfield  {author} {\bibinfo {author} {\bibfnamefont {R.}~\bibnamefont
  {Willett}}, \bibinfo {author} {\bibfnamefont {J.~P.}\ \bibnamefont
  {Eisenstein}}, \bibinfo {author} {\bibfnamefont {H.~L.}\ \bibnamefont
  {St\"ormer}}, \bibinfo {author} {\bibfnamefont {D.~C.}\ \bibnamefont {Tsui}},
  \bibinfo {author} {\bibfnamefont {A.~C.}\ \bibnamefont {Gossard}},\ and\
  \bibinfo {author} {\bibfnamefont {J.~H.}\ \bibnamefont {English}},\
  }\bibfield  {title} {\bibinfo {title} {Observation of an even-denominator
  quantum number in the fractional quantum {Hall} effect},\ }\href
  {https://doi.org/10.1103/PhysRevLett.59.1776} {\bibfield  {journal} {\bibinfo
   {journal} {Phys. Rev. Lett.}\ }\textbf {\bibinfo {volume} {59}},\ \bibinfo
  {pages} {1776} (\bibinfo {year} {1987})}\BibitemShut {NoStop}%
\bibitem [{\citenamefont {Pan}\ \emph {et~al.}(1999)\citenamefont {Pan},
  \citenamefont {Xia}, \citenamefont {Shvarts}, \citenamefont {Adams},
  \citenamefont {Stormer}, \citenamefont {Tsui}, \citenamefont {Pfieffer},
  \citenamefont {Baldwin},\ and\ \citenamefont {West}}]{Pan99b}%
  \BibitemOpen
  \bibfield  {author} {\bibinfo {author} {\bibfnamefont {W.}~\bibnamefont
  {Pan}}, \bibinfo {author} {\bibfnamefont {J.-S.}\ \bibnamefont {Xia}},
  \bibinfo {author} {\bibfnamefont {V.}~\bibnamefont {Shvarts}}, \bibinfo
  {author} {\bibfnamefont {D.~E.}\ \bibnamefont {Adams}}, \bibinfo {author}
  {\bibfnamefont {H.~L.}\ \bibnamefont {Stormer}}, \bibinfo {author}
  {\bibfnamefont {D.~C.}\ \bibnamefont {Tsui}}, \bibinfo {author}
  {\bibfnamefont {L.~N.}\ \bibnamefont {Pfieffer}}, \bibinfo {author}
  {\bibfnamefont {K.~W.}\ \bibnamefont {Baldwin}},\ and\ \bibinfo {author}
  {\bibfnamefont {K.~W.}\ \bibnamefont {West}},\ }\bibfield  {title} {\bibinfo
  {title} {Exact quantization of the even-denominator fractional quantum {Hall}
  state at $\nu=5/2$ {L}andau level filling factor},\ }\href
  {https://doi.org/10.1103/PhysRevLett.83.3530} {\bibfield  {journal} {\bibinfo
   {journal} {Phys. Rev. Lett.}\ }\textbf {\bibinfo {volume} {83}},\ \bibinfo
  {pages} {3530} (\bibinfo {year} {1999})},\ \Eprint
  {https://arxiv.org/abs/cond-mat/9907356} {arXiv:cond-mat/9907356}
  \BibitemShut {NoStop}%
\bibitem [{\citenamefont {Kumar}\ \emph {et~al.}(2010)\citenamefont {Kumar},
  \citenamefont {Cs\'athy}, \citenamefont {Manfra}, \citenamefont {Pfeiffer},\
  and\ \citenamefont {West}}]{Kumar10}%
  \BibitemOpen
  \bibfield  {author} {\bibinfo {author} {\bibfnamefont {A.}~\bibnamefont
  {Kumar}}, \bibinfo {author} {\bibfnamefont {G.~A.}\ \bibnamefont {Cs\'athy}},
  \bibinfo {author} {\bibfnamefont {M.~J.}\ \bibnamefont {Manfra}}, \bibinfo
  {author} {\bibfnamefont {L.~N.}\ \bibnamefont {Pfeiffer}},\ and\ \bibinfo
  {author} {\bibfnamefont {K.~W.}\ \bibnamefont {West}},\ }\bibfield  {title}
  {\bibinfo {title} {Nonconventional odd-denominator fractional quantum {Hall}
  states in the second {Landau} level},\ }\href
  {https://doi.org/10.1103/PhysRevLett.105.246808} {\bibfield  {journal}
  {\bibinfo  {journal} {Phys. Rev. Lett.}\ }\textbf {\bibinfo {volume} {105}},\
  \bibinfo {pages} {246808} (\bibinfo {year} {2010})}\BibitemShut {NoStop}%
\bibitem [{\citenamefont {Zibrov}\ \emph {et~al.}(2017)\citenamefont {Zibrov},
  \citenamefont {Kometter}, \citenamefont {Zhou}, \citenamefont {Spanton},
  \citenamefont {Taniguchi}, \citenamefont {Watanabe}, \citenamefont
  {Zaletel},\ and\ \citenamefont {Young}}]{Zibrov17}%
  \BibitemOpen
  \bibfield  {author} {\bibinfo {author} {\bibfnamefont {A.~A.}\ \bibnamefont
  {Zibrov}}, \bibinfo {author} {\bibfnamefont {C.}~\bibnamefont {Kometter}},
  \bibinfo {author} {\bibfnamefont {H.}~\bibnamefont {Zhou}}, \bibinfo {author}
  {\bibfnamefont {E.~M.}\ \bibnamefont {Spanton}}, \bibinfo {author}
  {\bibfnamefont {T.}~\bibnamefont {Taniguchi}}, \bibinfo {author}
  {\bibfnamefont {K.}~\bibnamefont {Watanabe}}, \bibinfo {author}
  {\bibfnamefont {M.~P.}\ \bibnamefont {Zaletel}},\ and\ \bibinfo {author}
  {\bibfnamefont {A.~F.}\ \bibnamefont {Young}},\ }\bibfield  {title} {\bibinfo
  {title} {Tunable interacting composite fermion phases in a half-filled
  bilayer-graphene {Landau} level},\ }\href
  {https://doi.org/10.1038/nature23893} {\bibfield  {journal} {\bibinfo
  {journal} {Nature}\ }\textbf {\bibinfo {volume} {549}},\ \bibinfo {pages}
  {360} (\bibinfo {year} {2017})}\BibitemShut {NoStop}%
\bibitem [{\citenamefont {de~C.~Chamon}\ \emph {et~al.}(1997)\citenamefont
  {de~C.~Chamon}, \citenamefont {Freed}, \citenamefont {Kivelson},
  \citenamefont {Sondhi},\ and\ \citenamefont {Wen}}]{Chamon97}%
  \BibitemOpen
  \bibfield  {author} {\bibinfo {author} {\bibfnamefont {C.}~\bibnamefont
  {de~C.~Chamon}}, \bibinfo {author} {\bibfnamefont {D.~E.}\ \bibnamefont
  {Freed}}, \bibinfo {author} {\bibfnamefont {S.~A.}\ \bibnamefont {Kivelson}},
  \bibinfo {author} {\bibfnamefont {S.~L.}\ \bibnamefont {Sondhi}},\ and\
  \bibinfo {author} {\bibfnamefont {X.~G.}\ \bibnamefont {Wen}},\ }\bibfield
  {title} {\bibinfo {title} {Two point-contact interferometer for quantum
  {Hall} systems},\ }\href {https://doi.org/10.1103/PhysRevB.55.2331}
  {\bibfield  {journal} {\bibinfo  {journal} {Phys. Rev. B}\ }\textbf {\bibinfo
  {volume} {55}},\ \bibinfo {pages} {2331} (\bibinfo {year}
  {1997})}\BibitemShut {NoStop}%
\bibitem [{\citenamefont {Fradkin}\ \emph {et~al.}(1998)\citenamefont
  {Fradkin}, \citenamefont {Nayak}, \citenamefont {Tsvelik},\ and\
  \citenamefont {Wilczek}}]{Fradkin98}%
  \BibitemOpen
  \bibfield  {author} {\bibinfo {author} {\bibfnamefont {E.}~\bibnamefont
  {Fradkin}}, \bibinfo {author} {\bibfnamefont {C.}~\bibnamefont {Nayak}},
  \bibinfo {author} {\bibfnamefont {A.}~\bibnamefont {Tsvelik}},\ and\ \bibinfo
  {author} {\bibfnamefont {F.}~\bibnamefont {Wilczek}},\ }\bibfield  {title}
  {\bibinfo {title} {A {Chern-Simons} effective field theory for the {P}faffian
  quantum {Hall} state},\ }\href
  {https://doi.org/10.1016/S0550-3213(98)00111-4} {\bibfield  {journal}
  {\bibinfo  {journal} {Nucl. Phys. B}\ }\textbf {\bibinfo {volume} {516}},\
  \bibinfo {pages} {704} (\bibinfo {year} {1998})},\ \Eprint
  {https://arxiv.org/abs/arXiv:cond-mat/9711087} {arXiv:cond-mat/9711087}
  \BibitemShut {NoStop}%
\bibitem [{\citenamefont {Bonderson}\ \emph {et~al.}(2006)\citenamefont
  {Bonderson}, \citenamefont {Kitaev},\ and\ \citenamefont
  {Shtengel}}]{Bonderson06a}%
  \BibitemOpen
  \bibfield  {author} {\bibinfo {author} {\bibfnamefont {P.}~\bibnamefont
  {Bonderson}}, \bibinfo {author} {\bibfnamefont {A.}~\bibnamefont {Kitaev}},\
  and\ \bibinfo {author} {\bibfnamefont {K.}~\bibnamefont {Shtengel}},\
  }\bibfield  {title} {\bibinfo {title} {Detecting non-{Abelian} statistics in
  the $\nu=5/2$ fractional quantum {Hall} state},\ }\href
  {https://doi.org/10.1103/PhysRevLett.96.016803} {\bibfield  {journal}
  {\bibinfo  {journal} {Phys. Rev. Lett.}\ }\textbf {\bibinfo {volume} {96}},\
  \bibinfo {pages} {016803} (\bibinfo {year} {2006})},\ \Eprint
  {https://arxiv.org/abs/cond-mat/0508616} {arXiv:cond-mat/0508616}
  \BibitemShut {NoStop}%
\bibitem [{\citenamefont {Stern}\ and\ \citenamefont
  {Halperin}(2006)}]{Stern06}%
  \BibitemOpen
  \bibfield  {author} {\bibinfo {author} {\bibfnamefont {A.}~\bibnamefont
  {Stern}}\ and\ \bibinfo {author} {\bibfnamefont {B.~I.}\ \bibnamefont
  {Halperin}},\ }\bibfield  {title} {\bibinfo {title} {Proposed experiments to
  probe the non-{Abelian} $\nu=5/2$ quantum {Hall} state},\ }\href
  {https://doi.org/10.1103/PhysRevLett.96.016802} {\bibfield  {journal}
  {\bibinfo  {journal} {Phys. Rev. Lett.}\ }\textbf {\bibinfo {volume} {96}},\
  \bibinfo {pages} {016802} (\bibinfo {year} {2006})},\ \Eprint
  {https://arxiv.org/abs/arXiv:cond-mat/0508447} {arXiv:cond-mat/0508447}
  \BibitemShut {NoStop}%
\bibitem [{\citenamefont {Willett}\ \emph {et~al.}(2010)\citenamefont
  {Willett}, \citenamefont {Pfeiffer},\ and\ \citenamefont {West}}]{Willett10}%
  \BibitemOpen
  \bibfield  {author} {\bibinfo {author} {\bibfnamefont {R.~L.}\ \bibnamefont
  {Willett}}, \bibinfo {author} {\bibfnamefont {L.~N.}\ \bibnamefont
  {Pfeiffer}},\ and\ \bibinfo {author} {\bibfnamefont {K.~W.}\ \bibnamefont
  {West}},\ }\bibfield  {title} {\bibinfo {title} {Alternation and interchange
  of e/4 and e/2 period interference oscillations consistent with filling
  factor 5/2 non-{A}belian quasiparticles},\ }\href
  {https://doi.org/10.1103/PhysRevB.82.205301} {\bibfield  {journal} {\bibinfo
  {journal} {Phys. Rev. B}\ }\textbf {\bibinfo {volume} {82}},\ \bibinfo
  {pages} {205301} (\bibinfo {year} {2010})},\ \Eprint
  {https://arxiv.org/abs/0911.0345} {arXiv:0911.0345} \BibitemShut {NoStop}%
\bibitem [{\citenamefont {Nakamura}\ \emph {et~al.}(2020)\citenamefont
  {Nakamura}, \citenamefont {Liang}, \citenamefont {Gardner},\ and\
  \citenamefont {Manfra}}]{Nakamura20}%
  \BibitemOpen
  \bibfield  {author} {\bibinfo {author} {\bibfnamefont {J.}~\bibnamefont
  {Nakamura}}, \bibinfo {author} {\bibfnamefont {S.}~\bibnamefont {Liang}},
  \bibinfo {author} {\bibfnamefont {G.~C.}\ \bibnamefont {Gardner}},\ and\
  \bibinfo {author} {\bibfnamefont {M.~J.}\ \bibnamefont {Manfra}},\ }\bibfield
   {title} {\bibinfo {title} {Direct observation of anyonic braiding
  statistics},\ }\href {https://doi.org/10.1038/s41567-020-1019-1} {\bibfield
  {journal} {\bibinfo  {journal} {Nat. Phys.}\ }\textbf {\bibinfo {volume}
  {16}},\ \bibinfo {pages} {931} (\bibinfo {year} {2020})}\BibitemShut
  {NoStop}%
\bibitem [{Note6()}]{Note6}%
  \BibitemOpen
  \bibinfo {note} {When the overall island parity can change, the Pauli
  operators depend on the overall MZM parity.}\BibitemShut {Stop}%
\bibitem [{\citenamefont {Karzig}\ \emph
  {et~al.}(2017{\natexlab{b}})\citenamefont {Karzig}, \citenamefont {Knapp},
  \citenamefont {Lutchyn}, \citenamefont {Bonderson}, \citenamefont {Hastings},
  \citenamefont {Nayak}, \citenamefont {Alicea}, \citenamefont {Flensberg},
  \citenamefont {Plugge}, \citenamefont {Oreg} \emph
  {et~al.}}]{karzig2017scalable}%
  \BibitemOpen
  \bibfield  {author} {\bibinfo {author} {\bibfnamefont {T.}~\bibnamefont
  {Karzig}}, \bibinfo {author} {\bibfnamefont {C.}~\bibnamefont {Knapp}},
  \bibinfo {author} {\bibfnamefont {R.~M.}\ \bibnamefont {Lutchyn}}, \bibinfo
  {author} {\bibfnamefont {P.}~\bibnamefont {Bonderson}}, \bibinfo {author}
  {\bibfnamefont {M.~B.}\ \bibnamefont {Hastings}}, \bibinfo {author}
  {\bibfnamefont {C.}~\bibnamefont {Nayak}}, \bibinfo {author} {\bibfnamefont
  {J.}~\bibnamefont {Alicea}}, \bibinfo {author} {\bibfnamefont
  {K.}~\bibnamefont {Flensberg}}, \bibinfo {author} {\bibfnamefont
  {S.}~\bibnamefont {Plugge}}, \bibinfo {author} {\bibfnamefont
  {Y.}~\bibnamefont {Oreg}}, \emph {et~al.},\ }\bibfield  {title} {\bibinfo
  {title} {Scalable designs for quasiparticle-poisoning-protected topological
  quantum computation with majorana zero modes},\ }\href@noop {} {\bibfield
  {journal} {\bibinfo  {journal} {Physical Review B}\ }\textbf {\bibinfo
  {volume} {95}},\ \bibinfo {pages} {235305} (\bibinfo {year}
  {2017}{\natexlab{b}})}\BibitemShut {NoStop}%
\bibitem [{Note7()}]{Note7}%
  \BibitemOpen
  \bibinfo {note} {\protect \Cref {eqn:pa} assumes that measurement corresponds
  to distinguishing two Gaussians of equal width $\sigma $ for an SNR
  definition of the distance between Gaussians divided by $2\sigma
  $.}\BibitemShut {Stop}%
\bibitem [{Note8()}]{Note8}%
  \BibitemOpen
  \bibinfo {note} {Additionally, non-equilibrium quasiparticles may contribute
  to the same error process with different scaling than thermally excited
  quasiparticles, however for the expected qubit volumes these are estimated to
  be subleading. This is in contrast to Ref.~\protect \rev@citealp {Aghaee24},
  where a lifetime of $\SI {2}{\milli \second }$ was observed in a
  configuration where all but one segment of the qubit device are tuned into a
  trivial superconducting regime. As such a configuration does not support a
  qubit, many of the error processes discussed here do not apply, and
  non-equilibrium quasiparticles remain as the dominant source of parity
  flips.}\BibitemShut {Stop}%
\bibitem [{Note9()}]{Note9}%
  \BibitemOpen
  \bibinfo {note} {In the MBQB demonstration, it is never necessary for qubits
  to idle. Idle qubits are additionally subject to coherent rotations along the
  axis of the residual MZM coupling; these rotations result in an error since
  the measurement-based approach cannot be formulated in a rotating
  frame.}\BibitemShut {Stop}%
\bibitem [{Note10()}]{Note10}%
  \BibitemOpen
  \bibinfo {note} {In practice, the band gap $E_g$ of the semiconductor sets a
  limit $\protect \mathcal {O}(e^{-E_g/k_B T})$.}\BibitemShut {Stop}%
\bibitem [{\citenamefont {Nielsen}(2003)}]{Nielsen2003}%
  \BibitemOpen
  \bibfield  {author} {\bibinfo {author} {\bibfnamefont {M.~A.}\ \bibnamefont
  {Nielsen}},\ }\bibfield  {title} {\bibinfo {title} {Quantum computation by
  measurement and quantum memory},\ }\href
  {https://doi.org/10.1016/s0375-9601(02)01803-0} {\bibfield  {journal}
  {\bibinfo  {journal} {Physics Letters A}\ }\textbf {\bibinfo {volume}
  {308}},\ \bibinfo {pages} {96–100} (\bibinfo {year} {2003})},\ \Eprint
  {https://arxiv.org/abs/quant-ph/0108020} {arXiv:quant-ph/0108020}
  \BibitemShut {NoStop}%
\bibitem [{\citenamefont {Perdrix}(2005)}]{Perdrix2005}%
  \BibitemOpen
  \bibfield  {author} {\bibinfo {author} {\bibfnamefont {S.}~\bibnamefont
  {Perdrix}},\ }\bibfield  {title} {\bibinfo {title} {State transfer instead of
  teleportation in measurement-based quantum computation},\ }\href
  {https://doi.org/10.1142/s0219749905000785} {\bibfield  {journal} {\bibinfo
  {journal} {International Journal of Quantum Information}\ }\textbf {\bibinfo
  {volume} {03}},\ \bibinfo {pages} {219–223} (\bibinfo {year}
  {2005})}\BibitemShut {NoStop}%
\bibitem [{\citenamefont {Bonderson}\ \emph
  {et~al.}(2008{\natexlab{a}})\citenamefont {Bonderson}, \citenamefont
  {Freedman},\ and\ \citenamefont {Nayak}}]{Bonderson08a}%
  \BibitemOpen
  \bibfield  {author} {\bibinfo {author} {\bibfnamefont {P.}~\bibnamefont
  {Bonderson}}, \bibinfo {author} {\bibfnamefont {M.}~\bibnamefont
  {Freedman}},\ and\ \bibinfo {author} {\bibfnamefont {C.}~\bibnamefont
  {Nayak}},\ }\bibfield  {title} {\bibinfo {title} {Measurement-only
  topological quantum computation},\ }\href
  {https://doi.org/10.1103/PhysRevLett.101.010501} {\bibfield  {journal}
  {\bibinfo  {journal} {Phys. Rev. Lett.}\ }\textbf {\bibinfo {volume} {101}},\
  \bibinfo {pages} {010501} (\bibinfo {year} {2008}{\natexlab{a}})},\ \Eprint
  {https://arxiv.org/abs/0802.0279} {arXiv:0802.0279} \BibitemShut {NoStop}%
\bibitem [{\citenamefont {Sau}\ \emph {et~al.}(2011)\citenamefont {Sau},
  \citenamefont {Clarke},\ and\ \citenamefont {Tewari}}]{Sau11a}%
  \BibitemOpen
  \bibfield  {author} {\bibinfo {author} {\bibfnamefont {J.~D.}\ \bibnamefont
  {Sau}}, \bibinfo {author} {\bibfnamefont {D.~J.}\ \bibnamefont {Clarke}},\
  and\ \bibinfo {author} {\bibfnamefont {S.}~\bibnamefont {Tewari}},\
  }\bibfield  {title} {\bibinfo {title} {Controlling non-{Abelian} statistics
  of {Majorana} fermions in semiconductor nanowires},\ }\href
  {https://doi.org/10.1103/PhysRevB.84.094505} {\bibfield  {journal} {\bibinfo
  {journal} {Phys. Rev. B}\ }\textbf {\bibinfo {volume} {84}},\ \bibinfo
  {pages} {094505} (\bibinfo {year} {2011})},\ \Eprint
  {https://arxiv.org/abs/arXiv:1012.0561} {arXiv:1012.0561} \BibitemShut
  {NoStop}%
\bibitem [{\citenamefont {{Scheurer}}\ and\ \citenamefont
  {{Shnirman}}(2013)}]{Scheurer13}%
  \BibitemOpen
  \bibfield  {author} {\bibinfo {author} {\bibfnamefont {M.~S.}\ \bibnamefont
  {{Scheurer}}}\ and\ \bibinfo {author} {\bibfnamefont {A.}~\bibnamefont
  {{Shnirman}}},\ }\bibfield  {title} {\bibinfo {title} {{Nonadiabatic
  processes in {Majorana} qubit systems}},\ }\href
  {https://doi.org/10.1103/PhysRevB.88.064515} {\bibfield  {journal} {\bibinfo
  {journal} {Phys. Rev. B}\ }\textbf {\bibinfo {volume} {88}},\ \bibinfo {eid}
  {064515} (\bibinfo {year} {2013})},\ \Eprint
  {https://arxiv.org/abs/1305.4923} {arXiv:1305.4923} \BibitemShut {NoStop}%
\bibitem [{\citenamefont {Hell}\ \emph {et~al.}(2016)\citenamefont {Hell},
  \citenamefont {Danon}, \citenamefont {Flensberg},\ and\ \citenamefont
  {Leijnse}}]{Hell16}%
  \BibitemOpen
  \bibfield  {author} {\bibinfo {author} {\bibfnamefont {M.}~\bibnamefont
  {Hell}}, \bibinfo {author} {\bibfnamefont {J.}~\bibnamefont {Danon}},
  \bibinfo {author} {\bibfnamefont {K.}~\bibnamefont {Flensberg}},\ and\
  \bibinfo {author} {\bibfnamefont {M.}~\bibnamefont {Leijnse}},\ }\bibfield
  {title} {\bibinfo {title} {Time scales for majorana manipulation using
  coulomb blockade in gate-controlled superconducting nanowires},\ }\bibfield
  {journal} {\bibinfo  {journal} {Physical Review B}\ }\textbf {\bibinfo
  {volume} {94}},\ \href {https://doi.org/10.1103/physrevb.94.035424}
  {10.1103/physrevb.94.035424} (\bibinfo {year} {2016})\BibitemShut {NoStop}%
\bibitem [{\citenamefont {{Knapp}}\ \emph {et~al.}(2016)\citenamefont
  {{Knapp}}, \citenamefont {{Zaletel}}, \citenamefont {{Liu}}, \citenamefont
  {{Cheng}}, \citenamefont {{Bonderson}},\ and\ \citenamefont
  {{Nayak}}}]{Knapp16}%
  \BibitemOpen
  \bibfield  {author} {\bibinfo {author} {\bibfnamefont {C.}~\bibnamefont
  {{Knapp}}}, \bibinfo {author} {\bibfnamefont {M.}~\bibnamefont {{Zaletel}}},
  \bibinfo {author} {\bibfnamefont {D.~E.}\ \bibnamefont {{Liu}}}, \bibinfo
  {author} {\bibfnamefont {M.}~\bibnamefont {{Cheng}}}, \bibinfo {author}
  {\bibfnamefont {P.}~\bibnamefont {{Bonderson}}},\ and\ \bibinfo {author}
  {\bibfnamefont {C.}~\bibnamefont {{Nayak}}},\ }\bibfield  {title} {\bibinfo
  {title} {{The Nature and Correction of Diabatic Errors in Anyon Braiding}},\
  }\href {https://doi.org/10.1103/PhysRevX.6.041003} {\bibfield  {journal}
  {\bibinfo  {journal} {Phys. Rev. X}\ }\textbf {\bibinfo {volume} {6}},\
  \bibinfo {pages} {041003} (\bibinfo {year} {2016})},\ \Eprint
  {https://arxiv.org/abs/1601.05790} {arXiv:1601.05790} \BibitemShut {NoStop}%
\bibitem [{Note11()}]{Note11}%
  \BibitemOpen
  \bibinfo {note} {It is worth noting that braiding statistics have been
  experimentally emulated in qubit systems~\cite
  {Song2018,Satzinger2021,Stenger2021,Google2023,Xu2024,Iqbal2023a,Iqbal2023b}.
  However, these experiments differ from topologically protected braiding
  transformations as discussed in this paper in that the topological state is
  not prepared as the ground state of the system's Hamiltonian, but through a
  quantum circuit without active error correction. Therefore, there is no
  topological protection (such as that endowed by the topological gap) of these
  operations.}\BibitemShut {Stop}%
\bibitem [{\citenamefont {Tran}\ \emph {et~al.}(2020)\citenamefont {Tran},
  \citenamefont {Bocharov}, \citenamefont {Bauer},\ and\ \citenamefont
  {Bonderson}}]{Tran20}%
  \BibitemOpen
  \bibfield  {author} {\bibinfo {author} {\bibfnamefont {A.}~\bibnamefont
  {Tran}}, \bibinfo {author} {\bibfnamefont {A.}~\bibnamefont {Bocharov}},
  \bibinfo {author} {\bibfnamefont {B.}~\bibnamefont {Bauer}},\ and\ \bibinfo
  {author} {\bibfnamefont {P.}~\bibnamefont {Bonderson}},\ }\bibfield  {title}
  {\bibinfo {title} {Optimizing clifford gate generation for measurement-only
  topological quantum computation with majorana zero modes},\ }\href@noop {}
  {\bibfield  {journal} {\bibinfo  {journal} {SciPost Physics}\ }\textbf
  {\bibinfo {volume} {8}},\ \bibinfo {pages} {091} (\bibinfo {year}
  {2020})}\BibitemShut {NoStop}%
\bibitem [{\citenamefont {Beverland}\ \emph {et~al.}(2022)\citenamefont
  {Beverland}, \citenamefont {Murali}, \citenamefont {Troyer}, \citenamefont
  {Svore}, \citenamefont {Hoefler}, \citenamefont {Kliuchnikov}, \citenamefont
  {Low}, \citenamefont {Soeken}, \citenamefont {Sundaram},\ and\ \citenamefont
  {Vaschillo}}]{beverland2022assessing}%
  \BibitemOpen
  \bibfield  {author} {\bibinfo {author} {\bibfnamefont {M.~E.}\ \bibnamefont
  {Beverland}}, \bibinfo {author} {\bibfnamefont {P.}~\bibnamefont {Murali}},
  \bibinfo {author} {\bibfnamefont {M.}~\bibnamefont {Troyer}}, \bibinfo
  {author} {\bibfnamefont {K.~M.}\ \bibnamefont {Svore}}, \bibinfo {author}
  {\bibfnamefont {T.}~\bibnamefont {Hoefler}}, \bibinfo {author} {\bibfnamefont
  {V.}~\bibnamefont {Kliuchnikov}}, \bibinfo {author} {\bibfnamefont {G.~H.}\
  \bibnamefont {Low}}, \bibinfo {author} {\bibfnamefont {M.}~\bibnamefont
  {Soeken}}, \bibinfo {author} {\bibfnamefont {A.}~\bibnamefont {Sundaram}},\
  and\ \bibinfo {author} {\bibfnamefont {A.}~\bibnamefont {Vaschillo}},\
  }\href@noop {} {\bibinfo {title} {Assessing requirements to scale to
  practical quantum advantage}} (\bibinfo {year} {2022}),\ \Eprint
  {https://arxiv.org/abs/2211.07629} {arXiv:2211.07629 [quant-ph]} \BibitemShut
  {NoStop}%
\bibitem [{Note12()}]{Note12}%
  \BibitemOpen
  \bibinfo {note} {Although these Pauli operations are commonly referred to as
  corrections, they are very different in nature from error corrections
  inferred by a decoder associated with an error correction code. In this
  context, they are simply operations that depend on previous non-deterministic
  measurement outcomes, like the corrections that arise in quantum
  teleportation.}\BibitemShut {Stop}%
\bibitem [{\citenamefont {Knill}(2005)}]{Knill05}%
  \BibitemOpen
  \bibfield  {author} {\bibinfo {author} {\bibfnamefont {E.}~\bibnamefont
  {Knill}},\ }\bibfield  {title} {\bibinfo {title} {Quantum computing with
  realistically noisy devices},\ }\href {https://doi.org/10.1038/nature03350}
  {\bibfield  {journal} {\bibinfo  {journal} {Nature}\ }\textbf {\bibinfo
  {volume} {434}},\ \bibinfo {pages} {39–44} (\bibinfo {year}
  {2005})}\BibitemShut {NoStop}%
\bibitem [{Note13()}]{Note13}%
  \BibitemOpen
  \bibinfo {note} {Here $H$ denotes the Hadamard gate, while $S$ denotes the
  phase gate.}\BibitemShut {Stop}%
\bibitem [{\citenamefont {Nielsen}\ \emph {et~al.}(2021)\citenamefont
  {Nielsen}, \citenamefont {Gamble}, \citenamefont {Rudinger}, \citenamefont
  {Scholten}, \citenamefont {Young},\ and\ \citenamefont
  {Blume-Kohout}}]{Nielsen21}%
  \BibitemOpen
  \bibfield  {author} {\bibinfo {author} {\bibfnamefont {E.}~\bibnamefont
  {Nielsen}}, \bibinfo {author} {\bibfnamefont {J.~K.}\ \bibnamefont {Gamble}},
  \bibinfo {author} {\bibfnamefont {K.}~\bibnamefont {Rudinger}}, \bibinfo
  {author} {\bibfnamefont {T.}~\bibnamefont {Scholten}}, \bibinfo {author}
  {\bibfnamefont {K.}~\bibnamefont {Young}},\ and\ \bibinfo {author}
  {\bibfnamefont {R.}~\bibnamefont {Blume-Kohout}},\ }\bibfield  {title}
  {\bibinfo {title} {Gate {S}et {T}omography},\ }\href
  {https://doi.org/10.22331/q-2021-10-05-557} {\bibfield  {journal} {\bibinfo
  {journal} {{Quantum}}\ }\textbf {\bibinfo {volume} {5}},\ \bibinfo {pages}
  {557} (\bibinfo {year} {2021})}\BibitemShut {NoStop}%
\bibitem [{Note14()}]{Note14}%
  \BibitemOpen
  \bibinfo {note} {Similar to how partial gate-set tomography is performed for
  X and Z measurements, the sequences must be prefixed with a randomized state
  preparation sequence, which we elide here for brevity. See \protect \Cref
  {app:MBQB-details} for details.}\BibitemShut {Stop}%
\bibitem [{Note15()}]{Note15}%
  \BibitemOpen
  \bibinfo {note} {Residual MZM coupling through the wire can also result in
  coherent rotations when the measurement basis commutes with the MZM coupling
  basis. For example, MZM overlap along the same wire results in a single-qubit
  $Z$ rotation during a $ZZ$ measurement, which does not affect that
  measurement outcome but can affect a subsequent non-commuting
  measurement.}\BibitemShut {Stop}%
\bibitem [{Note16()}]{Note16}%
  \BibitemOpen
  \bibinfo {note} {Process tomography and gateset tomography accomplish similar
  tasks (the reconstruction of a noisy quantum evolution), but process
  tomography as usually described~\cite {Nielsen00} assumes knowledge of the
  state preparation and measurements used to reconstruct the evolution. As we
  are describing numerical simulations, we have knowledge of preparation and
  measurement details, so we can rely on the simpler process tomography
  procedure instead of gateset tomography.}\BibitemShut {Stop}%
\bibitem [{\citenamefont {Nielsen}(2002)}]{Nielsen2002}%
  \BibitemOpen
  \bibfield  {author} {\bibinfo {author} {\bibfnamefont {M.~A.}\ \bibnamefont
  {Nielsen}},\ }\bibfield  {title} {\bibinfo {title} {A simple formula for the
  average gate fidelity of a quantum dynamical operation},\ }\href
  {https://doi.org/https://doi.org/10.1016/S0375-9601(02)01272-0} {\bibfield
  {journal} {\bibinfo  {journal} {Physics Letters A}\ }\textbf {\bibinfo
  {volume} {303}},\ \bibinfo {pages} {249} (\bibinfo {year}
  {2002})}\BibitemShut {NoStop}%
\bibitem [{\citenamefont {Shor}(1996)}]{shor96}%
  \BibitemOpen
  \bibfield  {author} {\bibinfo {author} {\bibfnamefont {P.~W.}\ \bibnamefont
  {Shor}},\ }\bibfield  {title} {\bibinfo {title} {Fault-tolerant quantum
  computation},\ }in\ \href {https://doi.org/10.1109/SFCS.1996.548464} {\emph
  {\bibinfo {booktitle} {Proc. 37th Symp. on Foundations of Computer Science
  (FOCS)}}}\ (\bibinfo {year} {1996})\ p.~\bibinfo {pages} {96},\ \Eprint
  {https://arxiv.org/abs/quant-ph/9605011} {arXiv:quant-ph/9605011}
  \BibitemShut {NoStop}%
\bibitem [{\citenamefont {Steane}(1996)}]{Steane1996}%
  \BibitemOpen
  \bibfield  {author} {\bibinfo {author} {\bibfnamefont {A.~M.}\ \bibnamefont
  {Steane}},\ }\bibfield  {title} {\bibinfo {title} {Simple quantum
  error-correcting codes},\ }\href {https://doi.org/10.1103/PhysRevA.54.4741}
  {\bibfield  {journal} {\bibinfo  {journal} {Phys. Rev. A}\ }\textbf {\bibinfo
  {volume} {54}},\ \bibinfo {pages} {4741} (\bibinfo {year}
  {1996})}\BibitemShut {NoStop}%
\bibitem [{\citenamefont {Aharonov}\ and\ \citenamefont
  {Ben-Or}(1997)}]{Dorit1997}%
  \BibitemOpen
  \bibfield  {author} {\bibinfo {author} {\bibfnamefont {D.}~\bibnamefont
  {Aharonov}}\ and\ \bibinfo {author} {\bibfnamefont {M.}~\bibnamefont
  {Ben-Or}},\ }\bibfield  {title} {\bibinfo {title} {Fault-tolerant quantum
  computation with constant error},\ }in\ \href
  {https://doi.org/10.1145/258533.258579} {\emph {\bibinfo {booktitle}
  {Proceedings of the Twenty-Ninth Annual ACM Symposium on Theory of
  Computing}}},\ \bibinfo {series and number} {STOC '97}\ (\bibinfo
  {publisher} {Association for Computing Machinery},\ \bibinfo {address} {New
  York, NY, USA},\ \bibinfo {year} {1997})\ p.\ \bibinfo {pages}
  {176–188}\BibitemShut {NoStop}%
\bibitem [{\citenamefont {Knill}\ \emph {et~al.}(1998)\citenamefont {Knill},
  \citenamefont {Laflamme},\ and\ \citenamefont {Zurek}}]{Knill1998}%
  \BibitemOpen
  \bibfield  {author} {\bibinfo {author} {\bibfnamefont {E.}~\bibnamefont
  {Knill}}, \bibinfo {author} {\bibfnamefont {R.}~\bibnamefont {Laflamme}},\
  and\ \bibinfo {author} {\bibfnamefont {W.~H.}\ \bibnamefont {Zurek}},\
  }\bibfield  {title} {\bibinfo {title} {Resilient quantum computation},\
  }\href {https://doi.org/10.1126/science.279.5349.342} {\bibfield  {journal}
  {\bibinfo  {journal} {Science}\ }\textbf {\bibinfo {volume} {279}},\ \bibinfo
  {pages} {342} (\bibinfo {year} {1998})}\BibitemShut {NoStop}%
\bibitem [{\citenamefont {Kitaev}(2003)}]{kitaev2003fault}%
  \BibitemOpen
  \bibfield  {author} {\bibinfo {author} {\bibfnamefont {A.~Y.}\ \bibnamefont
  {Kitaev}},\ }\bibfield  {title} {\bibinfo {title} {Fault-tolerant quantum
  computation by anyons},\ }\href@noop {} {\bibfield  {journal} {\bibinfo
  {journal} {Annals of Physics}\ }\textbf {\bibinfo {volume} {303}},\ \bibinfo
  {pages} {2} (\bibinfo {year} {2003})}\BibitemShut {NoStop}%
\bibitem [{\citenamefont {Terhal}\ and\ \citenamefont
  {Burkard}(2005)}]{Terhal2005}%
  \BibitemOpen
  \bibfield  {author} {\bibinfo {author} {\bibfnamefont {B.~M.}\ \bibnamefont
  {Terhal}}\ and\ \bibinfo {author} {\bibfnamefont {G.}~\bibnamefont
  {Burkard}},\ }\bibfield  {title} {\bibinfo {title} {Fault-tolerant quantum
  computation for local non-markovian noise},\ }\href
  {https://doi.org/10.1103/PhysRevA.71.012336} {\bibfield  {journal} {\bibinfo
  {journal} {Phys. Rev. A}\ }\textbf {\bibinfo {volume} {71}},\ \bibinfo
  {pages} {012336} (\bibinfo {year} {2005})}\BibitemShut {NoStop}%
\bibitem [{\citenamefont {Aliferis}\ \emph {et~al.}(2006)\citenamefont
  {Aliferis}, \citenamefont {Gottesman},\ and\ \citenamefont
  {Preskill}}]{aliferisgottesmanpreskill05}%
  \BibitemOpen
  \bibfield  {author} {\bibinfo {author} {\bibfnamefont {P.}~\bibnamefont
  {Aliferis}}, \bibinfo {author} {\bibfnamefont {D.}~\bibnamefont
  {Gottesman}},\ and\ \bibinfo {author} {\bibfnamefont {J.}~\bibnamefont
  {Preskill}},\ }\bibfield  {title} {\bibinfo {title} {Quantum accuracy
  threshold for concatenated distance-3 codes},\ }\href
  {https://doi.org/10.26421/QIC6.2-1} {\bibfield  {journal} {\bibinfo
  {journal} {Quant. Inf. Comput.}\ }\textbf {\bibinfo {volume} {6}},\ \bibinfo
  {pages} {97} (\bibinfo {year} {2006})},\ \Eprint
  {https://arxiv.org/abs/quant-ph/0504218} {arXiv:quant-ph/0504218}
  \BibitemShut {NoStop}%
\bibitem [{\citenamefont {Willsch}\ \emph {et~al.}(2018)\citenamefont
  {Willsch}, \citenamefont {Willsch}, \citenamefont {Jin}, \citenamefont
  {De~Raedt},\ and\ \citenamefont {Michielsen}}]{willsch_2018}%
  \BibitemOpen
  \bibfield  {author} {\bibinfo {author} {\bibfnamefont {D.}~\bibnamefont
  {Willsch}}, \bibinfo {author} {\bibfnamefont {M.}~\bibnamefont {Willsch}},
  \bibinfo {author} {\bibfnamefont {F.}~\bibnamefont {Jin}}, \bibinfo {author}
  {\bibfnamefont {H.}~\bibnamefont {De~Raedt}},\ and\ \bibinfo {author}
  {\bibfnamefont {K.}~\bibnamefont {Michielsen}},\ }\bibfield  {title}
  {\bibinfo {title} {Testing quantum fault tolerance on small systems},\
  }\bibfield  {journal} {\bibinfo  {journal} {Physical Review A}\ }\textbf
  {\bibinfo {volume} {98}},\ \href {https://doi.org/10.1103/physreva.98.052348}
  {10.1103/physreva.98.052348} (\bibinfo {year} {2018}),\ \Eprint
  {https://arxiv.org/abs/1805.05227} {arXiv:1805.05227 [quant-ph]} \BibitemShut
  {NoStop}%
\bibitem [{\citenamefont {Vuillot}(2018)}]{vuillot_2018}%
  \BibitemOpen
  \bibfield  {author} {\bibinfo {author} {\bibfnamefont {C.}~\bibnamefont
  {Vuillot}},\ }\bibfield  {title} {\bibinfo {title} {Is error detection
  helpful on {IBM} 5{Q} chips?},\ }\bibfield  {journal} {\bibinfo  {journal}
  {Quantum Information and Computation}\ }\textbf {\bibinfo {volume} {18}},\
  \href {https://doi.org/10.26421/qic18.11-12} {10.26421/qic18.11-12} (\bibinfo
  {year} {2018}),\ \Eprint {https://arxiv.org/abs/1705.08957} {arXiv:1705.08957
  [quant-ph]} \BibitemShut {NoStop}%
\bibitem [{\citenamefont {Linke}\ \emph {et~al.}(2017)\citenamefont {Linke},
  \citenamefont {Gutierrez}, \citenamefont {Landsman}, \citenamefont {Figgatt},
  \citenamefont {Debnath}, \citenamefont {Brown},\ and\ \citenamefont
  {Monroe}}]{linke17edexp}%
  \BibitemOpen
  \bibfield  {author} {\bibinfo {author} {\bibfnamefont {N.~M.}\ \bibnamefont
  {Linke}}, \bibinfo {author} {\bibfnamefont {M.}~\bibnamefont {Gutierrez}},
  \bibinfo {author} {\bibfnamefont {K.~A.}\ \bibnamefont {Landsman}}, \bibinfo
  {author} {\bibfnamefont {C.}~\bibnamefont {Figgatt}}, \bibinfo {author}
  {\bibfnamefont {S.}~\bibnamefont {Debnath}}, \bibinfo {author} {\bibfnamefont
  {K.~R.}\ \bibnamefont {Brown}},\ and\ \bibinfo {author} {\bibfnamefont
  {C.}~\bibnamefont {Monroe}},\ }\bibfield  {title} {\bibinfo {title}
  {Fault-tolerant quantum error detection},\ }\href
  {https://doi.org/10.1126/sciadv.1701074} {\bibfield  {journal} {\bibinfo
  {journal} {Science Advances}\ }\textbf {\bibinfo {volume} {3}},\ \bibinfo
  {pages} {1701074} (\bibinfo {year} {2017})},\ \Eprint
  {https://arxiv.org/abs/1611.06946} {arXiv:1611.06946 [quant-ph]} \BibitemShut
  {NoStop}%
\bibitem [{\citenamefont {Cane}\ \emph {et~al.}(2021)\citenamefont {Cane},
  \citenamefont {Chandra}, \citenamefont {Ng},\ and\ \citenamefont
  {Hanzo}}]{cane2021}%
  \BibitemOpen
  \bibfield  {author} {\bibinfo {author} {\bibfnamefont {R.}~\bibnamefont
  {Cane}}, \bibinfo {author} {\bibfnamefont {D.}~\bibnamefont {Chandra}},
  \bibinfo {author} {\bibfnamefont {S.~X.}\ \bibnamefont {Ng}},\ and\ \bibinfo
  {author} {\bibfnamefont {L.}~\bibnamefont {Hanzo}},\ }\bibfield  {title}
  {\bibinfo {title} {Experimental characterization of fault-tolerant circuits
  in small-scale quantum processors},\ }\href
  {https://doi.org/10.1109/ACCESS.2021.3133483} {\bibfield  {journal} {\bibinfo
   {journal} {{IEEE} Access}\ }\textbf {\bibinfo {volume} {9}},\ \bibinfo
  {pages} {162996} (\bibinfo {year} {2021})},\ \Eprint
  {https://arxiv.org/abs/2112.04076} {arXiv:2112.04076 [quant-ph]} \BibitemShut
  {NoStop}%
\bibitem [{\citenamefont {Hong}\ \emph {et~al.}(2024)\citenamefont {Hong},
  \citenamefont {{Durso-Sabina}}, \citenamefont {Hayes},\ and\ \citenamefont
  {Lucas}}]{hong24cat4onh2}%
  \BibitemOpen
  \bibfield  {author} {\bibinfo {author} {\bibfnamefont {Y.}~\bibnamefont
  {Hong}}, \bibinfo {author} {\bibfnamefont {E.}~\bibnamefont
  {{Durso-Sabina}}}, \bibinfo {author} {\bibfnamefont {D.}~\bibnamefont
  {Hayes}},\ and\ \bibinfo {author} {\bibfnamefont {A.}~\bibnamefont {Lucas}},\
  }\bibfield  {title} {\bibinfo {title} {Entangling four logical qubits beyond
  break-even in a nonlocal code},\ }\href
  {https://doi.org/10.1103/PhysRevLett.133.180601} {\bibfield  {journal}
  {\bibinfo  {journal} {Phys. Rev. Lett.}\ }\textbf {\bibinfo {volume} {133}},\
  \bibinfo {pages} {180601} (\bibinfo {year} {2024})},\ \Eprint
  {https://arxiv.org/abs/2406.02666} {arXiv:2406.02666 [quant-ph]} \BibitemShut
  {NoStop}%
\bibitem [{\citenamefont {Bluvstein}\ \emph {et~al.}(2024)\citenamefont
  {Bluvstein}, \citenamefont {Evered}, \citenamefont {Geim}, \citenamefont
  {Li}, \citenamefont {Zhou}, \citenamefont {Manovitz}, \citenamefont {Ebadi},
  \citenamefont {Cain}, \citenamefont {Kalinowski}, \citenamefont {Hangleiter},
  \citenamefont {Ataides}, \citenamefont {Maskara}, \citenamefont {Cong},
  \citenamefont {Gao}, \citenamefont {Rodriguez}, \citenamefont {Karolyshyn},
  \citenamefont {Semeghini}, \citenamefont {Gullans}, \citenamefont {Greiner},
  \citenamefont {Vuletic},\ and\ \citenamefont
  {Lukin}}]{bluvsteinharvard23neutralatoms}%
  \BibitemOpen
  \bibfield  {author} {\bibinfo {author} {\bibfnamefont {D.}~\bibnamefont
  {Bluvstein}}, \bibinfo {author} {\bibfnamefont {S.~J.}\ \bibnamefont
  {Evered}}, \bibinfo {author} {\bibfnamefont {A.~A.}\ \bibnamefont {Geim}},
  \bibinfo {author} {\bibfnamefont {S.~H.}\ \bibnamefont {Li}}, \bibinfo
  {author} {\bibfnamefont {H.}~\bibnamefont {Zhou}}, \bibinfo {author}
  {\bibfnamefont {T.}~\bibnamefont {Manovitz}}, \bibinfo {author}
  {\bibfnamefont {S.}~\bibnamefont {Ebadi}}, \bibinfo {author} {\bibfnamefont
  {M.}~\bibnamefont {Cain}}, \bibinfo {author} {\bibfnamefont {M.}~\bibnamefont
  {Kalinowski}}, \bibinfo {author} {\bibfnamefont {D.}~\bibnamefont
  {Hangleiter}}, \bibinfo {author} {\bibfnamefont {J.~P.~B.}\ \bibnamefont
  {Ataides}}, \bibinfo {author} {\bibfnamefont {N.}~\bibnamefont {Maskara}},
  \bibinfo {author} {\bibfnamefont {I.}~\bibnamefont {Cong}}, \bibinfo {author}
  {\bibfnamefont {X.}~\bibnamefont {Gao}}, \bibinfo {author} {\bibfnamefont
  {P.~S.}\ \bibnamefont {Rodriguez}}, \bibinfo {author} {\bibfnamefont
  {T.}~\bibnamefont {Karolyshyn}}, \bibinfo {author} {\bibfnamefont
  {G.}~\bibnamefont {Semeghini}}, \bibinfo {author} {\bibfnamefont {M.~J.}\
  \bibnamefont {Gullans}}, \bibinfo {author} {\bibfnamefont {M.}~\bibnamefont
  {Greiner}}, \bibinfo {author} {\bibfnamefont {V.}~\bibnamefont {Vuletic}},\
  and\ \bibinfo {author} {\bibfnamefont {M.~D.}\ \bibnamefont {Lukin}},\
  }\bibfield  {title} {\bibinfo {title} {Logical quantum processor based on
  reconfigurable atom arrays},\ }\href
  {https://doi.org/10.1038/s41586-023-06927-3} {\bibfield  {journal} {\bibinfo
  {journal} {Nature}\ }\textbf {\bibinfo {volume} {626}},\ \bibinfo {pages}
  {58} (\bibinfo {year} {2024})},\ \Eprint {https://arxiv.org/abs/2312.03982}
  {arXiv:2312.03982 [quant-ph]} \BibitemShut {NoStop}%
\bibitem [{\citenamefont {Self}\ \emph {et~al.}(2024)\citenamefont {Self},
  \citenamefont {Benedetti},\ and\ \citenamefont {Amaro}}]{self22icebergcode}%
  \BibitemOpen
  \bibfield  {author} {\bibinfo {author} {\bibfnamefont {C.~N.}\ \bibnamefont
  {Self}}, \bibinfo {author} {\bibfnamefont {M.}~\bibnamefont {Benedetti}},\
  and\ \bibinfo {author} {\bibfnamefont {D.}~\bibnamefont {Amaro}},\ }\bibfield
   {title} {\bibinfo {title} {Protecting expressive circuits with a quantum
  error detection code},\ }\href {https://doi.org/10.1038/s41567-023-02282-2}
  {\bibfield  {journal} {\bibinfo  {journal} {Nat. Phys.}\ }\textbf {\bibinfo
  {volume} {20}},\ \bibinfo {pages} {219} (\bibinfo {year} {2024})},\ \Eprint
  {https://arxiv.org/abs/2211.06703} {arXiv:2211.06703 [quant-ph]} \BibitemShut
  {NoStop}%
\bibitem [{\citenamefont {Yamamoto}\ \emph {et~al.}(2024)\citenamefont
  {Yamamoto}, \citenamefont {Duffield}, \citenamefont {Kikuchi},\ and\
  \citenamefont {Ramo}}]{yamamoto23phaseestimattionerrordetection}%
  \BibitemOpen
  \bibfield  {author} {\bibinfo {author} {\bibfnamefont {K.}~\bibnamefont
  {Yamamoto}}, \bibinfo {author} {\bibfnamefont {S.}~\bibnamefont {Duffield}},
  \bibinfo {author} {\bibfnamefont {Y.}~\bibnamefont {Kikuchi}},\ and\ \bibinfo
  {author} {\bibfnamefont {D.~M.}\ \bibnamefont {Ramo}},\ }\bibfield  {title}
  {\bibinfo {title} {Demonstrating {B}ayesian quantum phase estimation with
  quantum error detection},\ }\href
  {https://doi.org/10.1103/PhysRevResearch.6.013221} {\bibfield  {journal}
  {\bibinfo  {journal} {Phys. Rev. Research}\ }\textbf {\bibinfo {volume}
  {6}},\ \bibinfo {pages} {013221} (\bibinfo {year} {2024})},\ \Eprint
  {https://arxiv.org/abs/2306.16608} {arXiv:2306.16608 [quant-ph]} \BibitemShut
  {NoStop}%
\bibitem [{\citenamefont {{Ryan-Anderson}}\ \emph {et~al.}(2021)\citenamefont
  {{Ryan-Anderson}}, \citenamefont {Bohnet}, \citenamefont {Lee}, \citenamefont
  {Gresh}, \citenamefont {Hankin}, \citenamefont {Gaebler}, \citenamefont
  {Francois}, \citenamefont {Chernoguzov}, \citenamefont {Lucchetti},
  \citenamefont {Brown}, \citenamefont {Gatterman}, \citenamefont {Halit},
  \citenamefont {Gilmore}, \citenamefont {Gerber}, \citenamefont {Neyenhuis},
  \citenamefont {Hayes},\ and\ \citenamefont {Stutz}}]{honeywell21steane}%
  \BibitemOpen
  \bibfield  {author} {\bibinfo {author} {\bibfnamefont {C.}~\bibnamefont
  {{Ryan-Anderson}}}, \bibinfo {author} {\bibfnamefont {J.~G.}\ \bibnamefont
  {Bohnet}}, \bibinfo {author} {\bibfnamefont {K.}~\bibnamefont {Lee}},
  \bibinfo {author} {\bibfnamefont {D.}~\bibnamefont {Gresh}}, \bibinfo
  {author} {\bibfnamefont {A.}~\bibnamefont {Hankin}}, \bibinfo {author}
  {\bibfnamefont {J.~P.}\ \bibnamefont {Gaebler}}, \bibinfo {author}
  {\bibfnamefont {D.}~\bibnamefont {Francois}}, \bibinfo {author}
  {\bibfnamefont {A.}~\bibnamefont {Chernoguzov}}, \bibinfo {author}
  {\bibfnamefont {D.}~\bibnamefont {Lucchetti}}, \bibinfo {author}
  {\bibfnamefont {N.~C.}\ \bibnamefont {Brown}}, \bibinfo {author}
  {\bibfnamefont {T.~M.}\ \bibnamefont {Gatterman}}, \bibinfo {author}
  {\bibfnamefont {S.~K.}\ \bibnamefont {Halit}}, \bibinfo {author}
  {\bibfnamefont {K.}~\bibnamefont {Gilmore}}, \bibinfo {author} {\bibfnamefont
  {J.}~\bibnamefont {Gerber}}, \bibinfo {author} {\bibfnamefont
  {B.}~\bibnamefont {Neyenhuis}}, \bibinfo {author} {\bibfnamefont
  {D.}~\bibnamefont {Hayes}},\ and\ \bibinfo {author} {\bibfnamefont {R.~P.}\
  \bibnamefont {Stutz}},\ }\bibfield  {title} {\bibinfo {title} {Realization of
  real-time fault-tolerant quantum error correction},\ }\href
  {https://doi.org/10.1103/PhysRevX.11.041058} {\bibfield  {journal} {\bibinfo
  {journal} {Phys. Rev. X}\ }\textbf {\bibinfo {volume} {11}},\ \bibinfo
  {pages} {041058} (\bibinfo {year} {2021})},\ \Eprint
  {https://arxiv.org/abs/2107.07505} {arXiv:2107.07505 [quant-ph]} \BibitemShut
  {NoStop}%
\bibitem [{\citenamefont {Postler}\ \emph {et~al.}(2024)\citenamefont
  {Postler}, \citenamefont {Butt}, \citenamefont {Pogorelov}, \citenamefont
  {Marciniak}, \citenamefont {Heu{\ss}en}, \citenamefont {Blatt}, \citenamefont
  {Schindler}, \citenamefont {Rispler}, \citenamefont {M{\"u}ller},\ and\
  \citenamefont {Monz}}]{postler23steaneec}%
  \BibitemOpen
  \bibfield  {author} {\bibinfo {author} {\bibfnamefont {L.}~\bibnamefont
  {Postler}}, \bibinfo {author} {\bibfnamefont {F.}~\bibnamefont {Butt}},
  \bibinfo {author} {\bibfnamefont {I.}~\bibnamefont {Pogorelov}}, \bibinfo
  {author} {\bibfnamefont {C.~D.}\ \bibnamefont {Marciniak}}, \bibinfo {author}
  {\bibfnamefont {S.}~\bibnamefont {Heu{\ss}en}}, \bibinfo {author}
  {\bibfnamefont {R.}~\bibnamefont {Blatt}}, \bibinfo {author} {\bibfnamefont
  {P.}~\bibnamefont {Schindler}}, \bibinfo {author} {\bibfnamefont
  {M.}~\bibnamefont {Rispler}}, \bibinfo {author} {\bibfnamefont
  {M.}~\bibnamefont {M{\"u}ller}},\ and\ \bibinfo {author} {\bibfnamefont
  {T.}~\bibnamefont {Monz}},\ }\bibfield  {title} {\bibinfo {title}
  {Demonstration of fault-tolerant {S}teane quantum error correction},\ }\href
  {https://doi.org/10.1103/PRXQuantum.5.030326} {\bibfield  {journal} {\bibinfo
   {journal} {PRX Quantum}\ }\textbf {\bibinfo {volume} {5}},\ \bibinfo {pages}
  {030326} (\bibinfo {year} {2024})},\ \Eprint
  {https://arxiv.org/abs/2312.09745} {arXiv:2312.09745 [quant-ph]} \BibitemShut
  {NoStop}%
\bibitem [{\citenamefont {{Paetznick}}\ \emph {et~al.}(2024)\citenamefont
  {{Paetznick}}, \citenamefont {da~{Silva}}, \citenamefont {{Ryan-Anderson}}
  \emph {et~al.}}]{silva24microsoft12qubitcode}%
  \BibitemOpen
  \bibfield  {author} {\bibinfo {author} {\bibfnamefont {A.}~\bibnamefont
  {{Paetznick}}}, \bibinfo {author} {\bibfnamefont {M.~P.}\ \bibnamefont
  {da~{Silva}}}, \bibinfo {author} {\bibfnamefont {C.}~\bibnamefont
  {{Ryan-Anderson}}}, \emph {et~al.},\ }\href@noop {} {\bibinfo {title}
  {Demonstration of logical qubits and repeated error correction with
  better-than-physical error rates}} (\bibinfo {year} {2024}),\ \Eprint
  {https://arxiv.org/abs/2404.02280} {arXiv:2404.02280 [quant-ph]} \BibitemShut
  {NoStop}%
\bibitem [{\citenamefont {Reichardt}\ \emph {et~al.}(2024)\citenamefont
  {Reichardt}, \citenamefont {Aasen}, \citenamefont {Chao}, \citenamefont
  {Chernoguzov} \emph {et~al.}}]{reichardtmicrosoft24tesseract}%
  \BibitemOpen
  \bibfield  {author} {\bibinfo {author} {\bibfnamefont {B.~W.}\ \bibnamefont
  {Reichardt}}, \bibinfo {author} {\bibfnamefont {D.}~\bibnamefont {Aasen}},
  \bibinfo {author} {\bibfnamefont {R.}~\bibnamefont {Chao}}, \bibinfo {author}
  {\bibfnamefont {A.}~\bibnamefont {Chernoguzov}}, \emph {et~al.},\ }\href@noop
  {} {\bibinfo {title} {Demonstration of quantum computation and error
  correction with a tesseract code}} (\bibinfo {year} {2024}),\ \Eprint
  {https://arxiv.org/abs/2409.04628} {arXiv:2409.04628 [quant-ph]} \BibitemShut
  {NoStop}%
\bibitem [{\citenamefont {Harper}\ and\ \citenamefont
  {Flammia}(2019)}]{harper19edd}%
  \BibitemOpen
  \bibfield  {author} {\bibinfo {author} {\bibfnamefont {R.}~\bibnamefont
  {Harper}}\ and\ \bibinfo {author} {\bibfnamefont {S.~T.}\ \bibnamefont
  {Flammia}},\ }\bibfield  {title} {\bibinfo {title} {Fault-tolerant logical
  gates in the {IBM} {Q}uantum {E}xperience},\ }\href
  {https://doi.org/10.1103/PhysRevLett.122.080504} {\bibfield  {journal}
  {\bibinfo  {journal} {Phys. Rev. Lett.}\ }\textbf {\bibinfo {volume} {122}},\
  \bibinfo {pages} {080504} (\bibinfo {year} {2019})},\ \Eprint
  {https://arxiv.org/abs/1806.02359} {arXiv:1806.02359 [quant-ph]} \BibitemShut
  {NoStop}%
\bibitem [{\citenamefont {Krinner}\ \emph {et~al.}(2022)\citenamefont
  {Krinner}, \citenamefont {Lacroix}, \citenamefont {Remm}, \citenamefont
  {Paolo}, \citenamefont {Genois}, \citenamefont {Leroux}, \citenamefont
  {Hellings}, \citenamefont {Lazar}, \citenamefont {Swiadek}, \citenamefont
  {Herrmann}, \citenamefont {Norris}, \citenamefont {Andersen}, \citenamefont
  {M{\"u}ller}, \citenamefont {Blais}, \citenamefont {Eichler},\ and\
  \citenamefont {Wallraff}}]{krinner21repeatedsurfaceec}%
  \BibitemOpen
  \bibfield  {author} {\bibinfo {author} {\bibfnamefont {S.}~\bibnamefont
  {Krinner}}, \bibinfo {author} {\bibfnamefont {N.}~\bibnamefont {Lacroix}},
  \bibinfo {author} {\bibfnamefont {A.}~\bibnamefont {Remm}}, \bibinfo {author}
  {\bibfnamefont {A.~D.}\ \bibnamefont {Paolo}}, \bibinfo {author}
  {\bibfnamefont {E.}~\bibnamefont {Genois}}, \bibinfo {author} {\bibfnamefont
  {C.}~\bibnamefont {Leroux}}, \bibinfo {author} {\bibfnamefont
  {C.}~\bibnamefont {Hellings}}, \bibinfo {author} {\bibfnamefont
  {S.}~\bibnamefont {Lazar}}, \bibinfo {author} {\bibfnamefont
  {F.}~\bibnamefont {Swiadek}}, \bibinfo {author} {\bibfnamefont
  {J.}~\bibnamefont {Herrmann}}, \bibinfo {author} {\bibfnamefont {G.~J.}\
  \bibnamefont {Norris}}, \bibinfo {author} {\bibfnamefont {C.~K.}\
  \bibnamefont {Andersen}}, \bibinfo {author} {\bibfnamefont {M.}~\bibnamefont
  {M{\"u}ller}}, \bibinfo {author} {\bibfnamefont {A.}~\bibnamefont {Blais}},
  \bibinfo {author} {\bibfnamefont {C.}~\bibnamefont {Eichler}},\ and\ \bibinfo
  {author} {\bibfnamefont {A.}~\bibnamefont {Wallraff}},\ }\bibfield  {title}
  {\bibinfo {title} {Realizing repeated quantum error correction in a
  distance-three surface code},\ }\href
  {https://doi.org/10.1038/s41586-022-04566-8} {\bibfield  {journal} {\bibinfo
  {journal} {Nature}\ }\textbf {\bibinfo {volume} {605}},\ \bibinfo {pages}
  {669} (\bibinfo {year} {2022})},\ \Eprint {https://arxiv.org/abs/2112.03708}
  {arXiv:2112.03708 [quant-ph]} \BibitemShut {NoStop}%
\bibitem [{\citenamefont {{Google Quantum AI}}(2023)}]{google23surface}%
  \BibitemOpen
  \bibfield  {author} {\bibinfo {author} {\bibnamefont {{Google Quantum AI}}},\
  }\bibfield  {title} {\bibinfo {title} {Suppressing quantum errors by scaling
  a surface code logical qubit},\ }\href
  {https://doi.org/10.1038/s41586-022-05434-1} {\bibfield  {journal} {\bibinfo
  {journal} {Nature}\ }\textbf {\bibinfo {volume} {614}},\ \bibinfo {pages}
  {676} (\bibinfo {year} {2023})},\ \Eprint {https://arxiv.org/abs/2207.06431}
  {arXiv:2207.06431 [quant-ph]} \BibitemShut {NoStop}%
\bibitem [{\citenamefont {Putterman}\ \emph {et~al.}(2024)\citenamefont
  {Putterman}, \citenamefont {Noh}, \citenamefont {Hann}, \citenamefont
  {MacCabe}, \citenamefont {Aghaeimeibodi} \emph {et~al.}}]{amazon24catec}%
  \BibitemOpen
  \bibfield  {author} {\bibinfo {author} {\bibfnamefont {H.}~\bibnamefont
  {Putterman}}, \bibinfo {author} {\bibfnamefont {K.}~\bibnamefont {Noh}},
  \bibinfo {author} {\bibfnamefont {C.~T.}\ \bibnamefont {Hann}}, \bibinfo
  {author} {\bibfnamefont {G.~S.}\ \bibnamefont {MacCabe}}, \bibinfo {author}
  {\bibfnamefont {S.}~\bibnamefont {Aghaeimeibodi}}, \emph {et~al.},\
  }\href@noop {} {\bibinfo {title} {Hardware-efficient quantum error correction
  using concatenated bosonic qubits}} (\bibinfo {year} {2024}),\ \Eprint
  {https://arxiv.org/abs/2409.13025} {arXiv:2409.13025 [quant-ph]} \BibitemShut
  {NoStop}%
\bibitem [{\citenamefont {Ofek}\ \emph {et~al.}(2016)\citenamefont {Ofek},
  \citenamefont {Petrenko}, \citenamefont {Heeres}, \citenamefont {Reinhold},
  \citenamefont {Leghtas}, \citenamefont {Vlastakis}, \citenamefont {Liu},
  \citenamefont {Frunzio}, \citenamefont {Girvin}, \citenamefont {Jiang} \emph
  {et~al.}}]{ofek2016extending}%
  \BibitemOpen
  \bibfield  {author} {\bibinfo {author} {\bibfnamefont {N.}~\bibnamefont
  {Ofek}}, \bibinfo {author} {\bibfnamefont {A.}~\bibnamefont {Petrenko}},
  \bibinfo {author} {\bibfnamefont {R.}~\bibnamefont {Heeres}}, \bibinfo
  {author} {\bibfnamefont {P.}~\bibnamefont {Reinhold}}, \bibinfo {author}
  {\bibfnamefont {Z.}~\bibnamefont {Leghtas}}, \bibinfo {author} {\bibfnamefont
  {B.}~\bibnamefont {Vlastakis}}, \bibinfo {author} {\bibfnamefont
  {Y.}~\bibnamefont {Liu}}, \bibinfo {author} {\bibfnamefont {L.}~\bibnamefont
  {Frunzio}}, \bibinfo {author} {\bibfnamefont {S.~M.}\ \bibnamefont {Girvin}},
  \bibinfo {author} {\bibfnamefont {L.}~\bibnamefont {Jiang}}, \emph {et~al.},\
  }\bibfield  {title} {\bibinfo {title} {Extending the lifetime of a quantum
  bit with error correction in superconducting circuits},\ }\href
  {https://doi.org/10.1038/nature18949} {\bibfield  {journal} {\bibinfo
  {journal} {Nature}\ }\textbf {\bibinfo {volume} {536}},\ \bibinfo {pages}
  {441} (\bibinfo {year} {2016})}\BibitemShut {NoStop}%
\bibitem [{\citenamefont {Sivak}\ \emph {et~al.}(2023)\citenamefont {Sivak},
  \citenamefont {Eickbusch}, \citenamefont {Royer}, \citenamefont {Singh},
  \citenamefont {Tsioutsios}, \citenamefont {Ganjam}, \citenamefont {Miano},
  \citenamefont {Brock}, \citenamefont {Ding}, \citenamefont {Frunzio} \emph
  {et~al.}}]{sivak2023real}%
  \BibitemOpen
  \bibfield  {author} {\bibinfo {author} {\bibfnamefont {V.}~\bibnamefont
  {Sivak}}, \bibinfo {author} {\bibfnamefont {A.}~\bibnamefont {Eickbusch}},
  \bibinfo {author} {\bibfnamefont {B.}~\bibnamefont {Royer}}, \bibinfo
  {author} {\bibfnamefont {S.}~\bibnamefont {Singh}}, \bibinfo {author}
  {\bibfnamefont {I.}~\bibnamefont {Tsioutsios}}, \bibinfo {author}
  {\bibfnamefont {S.}~\bibnamefont {Ganjam}}, \bibinfo {author} {\bibfnamefont
  {A.}~\bibnamefont {Miano}}, \bibinfo {author} {\bibfnamefont
  {B.}~\bibnamefont {Brock}}, \bibinfo {author} {\bibfnamefont
  {A.}~\bibnamefont {Ding}}, \bibinfo {author} {\bibfnamefont {L.}~\bibnamefont
  {Frunzio}}, \emph {et~al.},\ }\bibfield  {title} {\bibinfo {title} {Real-time
  quantum error correction beyond break-even},\ }\href
  {https://doi.org/10.1038/s41586-023-05782-6} {\bibfield  {journal} {\bibinfo
  {journal} {Nature}\ }\textbf {\bibinfo {volume} {616}},\ \bibinfo {pages}
  {50} (\bibinfo {year} {2023})}\BibitemShut {NoStop}%
\bibitem [{\citenamefont {Ni}\ \emph {et~al.}(2023)\citenamefont {Ni},
  \citenamefont {Li}, \citenamefont {Deng}, \citenamefont {Cai}, \citenamefont
  {Zhang}, \citenamefont {Wang}, \citenamefont {Yang}, \citenamefont {Yu},
  \citenamefont {Yan}, \citenamefont {Liu} \emph {et~al.}}]{ni2023beating}%
  \BibitemOpen
  \bibfield  {author} {\bibinfo {author} {\bibfnamefont {Z.}~\bibnamefont
  {Ni}}, \bibinfo {author} {\bibfnamefont {S.}~\bibnamefont {Li}}, \bibinfo
  {author} {\bibfnamefont {X.}~\bibnamefont {Deng}}, \bibinfo {author}
  {\bibfnamefont {Y.}~\bibnamefont {Cai}}, \bibinfo {author} {\bibfnamefont
  {L.}~\bibnamefont {Zhang}}, \bibinfo {author} {\bibfnamefont
  {W.}~\bibnamefont {Wang}}, \bibinfo {author} {\bibfnamefont {Z.-B.}\
  \bibnamefont {Yang}}, \bibinfo {author} {\bibfnamefont {H.}~\bibnamefont
  {Yu}}, \bibinfo {author} {\bibfnamefont {F.}~\bibnamefont {Yan}}, \bibinfo
  {author} {\bibfnamefont {S.}~\bibnamefont {Liu}}, \emph {et~al.},\ }\bibfield
   {title} {\bibinfo {title} {Beating the break-even point with a
  discrete-variable-encoded logical qubit},\ }\href
  {https://doi.org/10.1038/s41586-023-05784-4} {\bibfield  {journal} {\bibinfo
  {journal} {Nature}\ }\textbf {\bibinfo {volume} {616}},\ \bibinfo {pages}
  {56} (\bibinfo {year} {2023})}\BibitemShut {NoStop}%
\bibitem [{\citenamefont {{Google Quantum
  AI}}(2024{\natexlab{a}})}]{google24surfacecode}%
  \BibitemOpen
  \bibfield  {author} {\bibinfo {author} {\bibnamefont {{Google Quantum AI}}},\
  }\bibfield  {title} {\bibinfo {title} {Quantum error correction below the
  surface code threshold},\ }\bibfield  {journal} {\bibinfo  {journal}
  {Nature}\ }\href {https://doi.org/10.1038/s41586-024-08449-y}
  {10.1038/s41586-024-08449-y} (\bibinfo {year}
  {2024}{\natexlab{a}})\BibitemShut {NoStop}%
\bibitem [{\citenamefont {{Google Quantum
  AI}}(2024{\natexlab{b}})}]{eickbusch2024demonstrating}%
  \BibitemOpen
  \bibfield  {author} {\bibinfo {author} {\bibnamefont {{Google Quantum AI}}},\
  }\href {https://arxiv.org/abs/2412.14360} {\bibinfo {title} {Demonstrating
  dynamic surface codes}} (\bibinfo {year} {2024}{\natexlab{b}}),\ \Eprint
  {https://arxiv.org/abs/2412.14360} {arXiv:2412.14360 [quant-ph]} \BibitemShut
  {NoStop}%
\bibitem [{\citenamefont {Lacroix}\ \emph {et~al.}(2024)\citenamefont
  {Lacroix}, \citenamefont {Bourassa}, \citenamefont {Heras}, \citenamefont
  {Zhang}, \citenamefont {Bausch}, \citenamefont {Senior}, \citenamefont
  {Edlich}, \citenamefont {Shutty}, \citenamefont {Sivak}, \citenamefont
  {Bengtsson} \emph {et~al.}}]{lacroix2024scaling}%
  \BibitemOpen
  \bibfield  {author} {\bibinfo {author} {\bibfnamefont {N.}~\bibnamefont
  {Lacroix}}, \bibinfo {author} {\bibfnamefont {A.}~\bibnamefont {Bourassa}},
  \bibinfo {author} {\bibfnamefont {F.~J.}\ \bibnamefont {Heras}}, \bibinfo
  {author} {\bibfnamefont {L.~M.}\ \bibnamefont {Zhang}}, \bibinfo {author}
  {\bibfnamefont {J.}~\bibnamefont {Bausch}}, \bibinfo {author} {\bibfnamefont
  {A.~W.}\ \bibnamefont {Senior}}, \bibinfo {author} {\bibfnamefont
  {T.}~\bibnamefont {Edlich}}, \bibinfo {author} {\bibfnamefont
  {N.}~\bibnamefont {Shutty}}, \bibinfo {author} {\bibfnamefont
  {V.}~\bibnamefont {Sivak}}, \bibinfo {author} {\bibfnamefont
  {A.}~\bibnamefont {Bengtsson}}, \emph {et~al.},\ }\href@noop {} {\bibinfo
  {title} {Scaling and logic in the color code on a superconducting quantum
  processor}} (\bibinfo {year} {2024}),\ \Eprint
  {https://arxiv.org/abs/2412.14256} {arXiv:2412.14256 [quant-ph]} \BibitemShut
  {NoStop}%
\bibitem [{\citenamefont {Haah}\ and\ \citenamefont
  {Hastings}(2022)}]{Haah2022}%
  \BibitemOpen
  \bibfield  {author} {\bibinfo {author} {\bibfnamefont {J.}~\bibnamefont
  {Haah}}\ and\ \bibinfo {author} {\bibfnamefont {M.~B.}\ \bibnamefont
  {Hastings}},\ }\bibfield  {title} {\bibinfo {title} {Boundaries for the
  honeycomb code},\ }\href {https://doi.org/10.22331/q-2022-04-21-693}
  {\bibfield  {journal} {\bibinfo  {journal} {Quantum}\ }\textbf {\bibinfo
  {volume} {6}},\ \bibinfo {pages} {693} (\bibinfo {year} {2022})}\BibitemShut
  {NoStop}%
\bibitem [{\citenamefont {Gidney}\ \emph {et~al.}(2021)\citenamefont {Gidney},
  \citenamefont {Newman}, \citenamefont {Fowler},\ and\ \citenamefont
  {Broughton}}]{Gidney2021b}%
  \BibitemOpen
  \bibfield  {author} {\bibinfo {author} {\bibfnamefont {C.}~\bibnamefont
  {Gidney}}, \bibinfo {author} {\bibfnamefont {M.}~\bibnamefont {Newman}},
  \bibinfo {author} {\bibfnamefont {A.}~\bibnamefont {Fowler}},\ and\ \bibinfo
  {author} {\bibfnamefont {M.}~\bibnamefont {Broughton}},\ }\bibfield  {title}
  {\bibinfo {title} {A fault-tolerant honeycomb memory},\ }\href
  {https://doi.org/10.22331/q-2021-12-20-605} {\bibfield  {journal} {\bibinfo
  {journal} {Quantum}\ }\textbf {\bibinfo {volume} {5}},\ \bibinfo {pages}
  {605} (\bibinfo {year} {2021})}\BibitemShut {NoStop}%
\bibitem [{\citenamefont {Gidney}\ \emph {et~al.}(2022)\citenamefont {Gidney},
  \citenamefont {Newman},\ and\ \citenamefont {McEwen}}]{Gidney2022}%
  \BibitemOpen
  \bibfield  {author} {\bibinfo {author} {\bibfnamefont {C.}~\bibnamefont
  {Gidney}}, \bibinfo {author} {\bibfnamefont {M.}~\bibnamefont {Newman}},\
  and\ \bibinfo {author} {\bibfnamefont {M.}~\bibnamefont {McEwen}},\
  }\bibfield  {title} {\bibinfo {title} {Benchmarking the planar honeycomb
  code},\ }\href {https://doi.org/10.22331/q-2022-09-21-813} {\bibfield
  {journal} {\bibinfo  {journal} {Quantum}\ }\textbf {\bibinfo {volume} {6}},\
  \bibinfo {pages} {813} (\bibinfo {year} {2022})}\BibitemShut {NoStop}%
\bibitem [{\citenamefont {Horsman}\ \emph {et~al.}(2012)\citenamefont
  {Horsman}, \citenamefont {Fowler}, \citenamefont {Devitt},\ and\
  \citenamefont {Van~Meter}}]{Horsman12}%
  \BibitemOpen
  \bibfield  {author} {\bibinfo {author} {\bibfnamefont {D.}~\bibnamefont
  {Horsman}}, \bibinfo {author} {\bibfnamefont {A.~G.}\ \bibnamefont {Fowler}},
  \bibinfo {author} {\bibfnamefont {S.}~\bibnamefont {Devitt}},\ and\ \bibinfo
  {author} {\bibfnamefont {R.}~\bibnamefont {Van~Meter}},\ }\bibfield  {title}
  {\bibinfo {title} {Surface code quantum computing by lattice surgery},\
  }\href {https://doi.org/10.1088/1367-2630/14/12/123011} {\bibfield  {journal}
  {\bibinfo  {journal} {New Journal of Physics}\ }\textbf {\bibinfo {volume}
  {14}},\ \bibinfo {pages} {123011} (\bibinfo {year} {2012})}\BibitemShut
  {NoStop}%
\bibitem [{Note17()}]{Note17}%
  \BibitemOpen
  \bibinfo {note} {For the scheme described below, the ladder code can only
  correct $X$ errors.}\BibitemShut {Stop}%
\bibitem [{\citenamefont {Bacon}(2006)}]{bacon05operator}%
  \BibitemOpen
  \bibfield  {author} {\bibinfo {author} {\bibfnamefont {D.}~\bibnamefont
  {Bacon}},\ }\bibfield  {title} {\bibinfo {title} {Operator quantum error
  correcting subsystems for self-correcting quantum memories},\ }\href
  {https://doi.org/10.1103/PhysRevA.73.012340} {\bibfield  {journal} {\bibinfo
  {journal} {Phys. Rev. A}\ }\textbf {\bibinfo {volume} {73}},\ \bibinfo
  {pages} {012340} (\bibinfo {year} {2006})},\ \Eprint
  {https://arxiv.org/abs/quant-ph/0506023} {arXiv:quant-ph/0506023}
  \BibitemShut {NoStop}%
\bibitem [{\citenamefont {Shor}(1995)}]{shor1995scheme}%
  \BibitemOpen
  \bibfield  {author} {\bibinfo {author} {\bibfnamefont {P.~W.}\ \bibnamefont
  {Shor}},\ }\bibfield  {title} {\bibinfo {title} {Scheme for reducing
  decoherence in quantum computer memory},\ }\href@noop {} {\bibfield
  {journal} {\bibinfo  {journal} {Physical review A}\ }\textbf {\bibinfo
  {volume} {52}},\ \bibinfo {pages} {R2493} (\bibinfo {year}
  {1995})}\BibitemShut {NoStop}%
\bibitem [{\citenamefont {Aasen}\ \emph {et~al.}(2023)\citenamefont {Aasen},
  \citenamefont {Haah}, \citenamefont {Bonderson}, \citenamefont {Wang},\ and\
  \citenamefont {Hastings}}]{aasen23}%
  \BibitemOpen
  \bibfield  {author} {\bibinfo {author} {\bibfnamefont {D.}~\bibnamefont
  {Aasen}}, \bibinfo {author} {\bibfnamefont {J.}~\bibnamefont {Haah}},
  \bibinfo {author} {\bibfnamefont {P.}~\bibnamefont {Bonderson}}, \bibinfo
  {author} {\bibfnamefont {Z.}~\bibnamefont {Wang}},\ and\ \bibinfo {author}
  {\bibfnamefont {M.}~\bibnamefont {Hastings}},\ }\href
  {https://arxiv.org/abs/2307.03715} {\bibinfo {title} {Fault-tolerant
  hastings-haah codes in the presence of dead qubits}} (\bibinfo {year}
  {2023}),\ \Eprint {https://arxiv.org/abs/2307.03715} {arXiv:2307.03715
  [quant-ph]} \BibitemShut {NoStop}%
\bibitem [{\citenamefont {Gidney}\ and\ \citenamefont
  {Eker{\aa}}(2021)}]{Gidney21}%
  \BibitemOpen
  \bibfield  {author} {\bibinfo {author} {\bibfnamefont {C.}~\bibnamefont
  {Gidney}}\ and\ \bibinfo {author} {\bibfnamefont {M.}~\bibnamefont
  {Eker{\aa}}},\ }\bibfield  {title} {\bibinfo {title} {{How to factor 2048 bit
  RSA integers in 8 hours using 20 million noisy qubits}},\ }\href
  {https://doi.org/10.22331/q-2021-04-15-433} {\bibfield  {journal} {\bibinfo
  {journal} {Quantum}\ }\textbf {\bibinfo {volume} {5}},\ \bibinfo {pages}
  {433} (\bibinfo {year} {2021})}\BibitemShut {NoStop}%
\bibitem [{\citenamefont {Aasen}\ \emph
  {et~al.}(2025{\natexlab{a}})\citenamefont {Aasen} \emph
  {et~al.}}]{Aasen25Zenodo}%
  \BibitemOpen
  \bibfield  {author} {\bibinfo {author} {\bibfnamefont {D.}~\bibnamefont
  {Aasen}} \emph {et~al.},\ }\href {https://doi.org/10.5281/zenodo.15041776}
  {\bibinfo {title} {Roadmap to fault tolerant quantum computation using
  topological qubit arrays}} (\bibinfo {year} {2025}{\natexlab{a}})\BibitemShut
  {NoStop}%
\bibitem [{\citenamefont {Davies}\ and\ \citenamefont
  {Lewis}(1970)}]{davies1970instruments}%
  \BibitemOpen
  \bibfield  {author} {\bibinfo {author} {\bibfnamefont {E.~B.}\ \bibnamefont
  {Davies}}\ and\ \bibinfo {author} {\bibfnamefont {J.~T.}\ \bibnamefont
  {Lewis}},\ }\bibfield  {title} {\bibinfo {title} {An operational approach to
  quantum probability},\ }\href {https://doi.org/10.1007/BF01647093} {\bibfield
   {journal} {\bibinfo  {journal} {Communications in Mathematical Physics}\
  }\textbf {\bibinfo {volume} {17}},\ \bibinfo {pages} {239} (\bibinfo {year}
  {1970})}\BibitemShut {NoStop}%
\bibitem [{\citenamefont {Wiseman}\ and\ \citenamefont
  {Milburn}(2010)}]{wiseman2010measuerment}%
  \BibitemOpen
  \bibfield  {author} {\bibinfo {author} {\bibfnamefont {H.}~\bibnamefont
  {Wiseman}}\ and\ \bibinfo {author} {\bibfnamefont {G.}~\bibnamefont
  {Milburn}},\ }\href {https://books.google.com/books?id=ZNjvHaH8qA4C} {\emph
  {\bibinfo {title} {Quantum Measurement and Control}}}\ (\bibinfo  {publisher}
  {Cambridge University Press},\ \bibinfo {year} {2010})\BibitemShut {NoStop}%
\bibitem [{Note18()}]{Note18}%
  \BibitemOpen
  \bibinfo {note} {We write $\Pr ({\protect \mathcal {M}^{P}_{s}})$ instead of
  $\Pr ({\protect \mathcal {M}^{P}_{s}}; \rho )$ when the relevant quantum
  state is clear from context.}\BibitemShut {Stop}%
\bibitem [{\citenamefont {de~Bruijn}(1946)}]{debruijn1946}%
  \BibitemOpen
  \bibfield  {author} {\bibinfo {author} {\bibfnamefont {N.~G.}\ \bibnamefont
  {de~Bruijn}},\ }\bibfield  {title} {\bibinfo {title} {A combinatorial
  problem},\ }\href@noop {} {\bibfield  {journal} {\bibinfo  {journal} {Nederl.
  Akad. Wetensch., Proc.}\ }\textbf {\bibinfo {volume} {49}},\ \bibinfo {pages}
  {758} (\bibinfo {year} {1946})}\BibitemShut {NoStop}%
\bibitem [{\citenamefont {Boutin}\ \emph {et~al.}(2025)\citenamefont {Boutin},
  \citenamefont {Karzig}, \citenamefont {Dandachi}, \citenamefont {Mishmash},
  \citenamefont {Gukelberger}, \citenamefont {Lutchyn},\ and\ \citenamefont
  {Bauer}}]{Boutin25}%
  \BibitemOpen
  \bibfield  {author} {\bibinfo {author} {\bibfnamefont {S.}~\bibnamefont
  {Boutin}}, \bibinfo {author} {\bibfnamefont {T.}~\bibnamefont {Karzig}},
  \bibinfo {author} {\bibfnamefont {T.~E.}\ \bibnamefont {Dandachi}}, \bibinfo
  {author} {\bibfnamefont {R.~V.}\ \bibnamefont {Mishmash}}, \bibinfo {author}
  {\bibfnamefont {J.}~\bibnamefont {Gukelberger}}, \bibinfo {author}
  {\bibfnamefont {R.~M.}\ \bibnamefont {Lutchyn}},\ and\ \bibinfo {author}
  {\bibfnamefont {B.}~\bibnamefont {Bauer}},\ }\href
  {https://arxiv.org/abs/2502.12960} {\bibinfo {title} {Predictive simulations
  of the dynamical response of mesoscopic devices}} (\bibinfo {year} {2025}),\
  \Eprint {https://arxiv.org/abs/2502.12960} {arXiv:2502.12960
  [cond-mat.mes-hall]} \BibitemShut {NoStop}%
\bibitem [{\citenamefont {Aasen}\ \emph
  {et~al.}(2025{\natexlab{b}})\citenamefont {Aasen}, \citenamefont {Bauer},
  \citenamefont {Bonderson},\ and\ \citenamefont {Knapp}}]{ICEP}%
  \BibitemOpen
  \bibfield  {author} {\bibinfo {author} {\bibfnamefont {D.}~\bibnamefont
  {Aasen}}, \bibinfo {author} {\bibfnamefont {B.}~\bibnamefont {Bauer}},
  \bibinfo {author} {\bibfnamefont {P.}~\bibnamefont {Bonderson}},\ and\
  \bibinfo {author} {\bibfnamefont {C.}~\bibnamefont {Knapp}},\ }\bibfield
  {title} {\bibinfo {title} {Electron poisoning of majorana-based qubits},\
  }\href@noop {} {\bibfield  {journal} {\bibinfo  {journal} {In preparation}\ }
  (\bibinfo {year} {2025}{\natexlab{b}})}\BibitemShut {NoStop}%
\bibitem [{\citenamefont {Bonderson}\ \emph
  {et~al.}(2008{\natexlab{b}})\citenamefont {Bonderson}, \citenamefont
  {Freedman},\ and\ \citenamefont {Nayak}}]{bonderson2008measurement}%
  \BibitemOpen
  \bibfield  {author} {\bibinfo {author} {\bibfnamefont {P.}~\bibnamefont
  {Bonderson}}, \bibinfo {author} {\bibfnamefont {M.}~\bibnamefont
  {Freedman}},\ and\ \bibinfo {author} {\bibfnamefont {C.}~\bibnamefont
  {Nayak}},\ }\bibfield  {title} {\bibinfo {title} {Measurement-only
  topological quantum computation},\ }\href@noop {} {\bibfield  {journal}
  {\bibinfo  {journal} {Physical Review Letters}\ }\textbf {\bibinfo {volume}
  {101}},\ \bibinfo {pages} {010501} (\bibinfo {year}
  {2008}{\natexlab{b}})}\BibitemShut {NoStop}%
\bibitem [{\citenamefont {Kern}\ \emph {et~al.}(2005)\citenamefont {Kern},
  \citenamefont {Alber},\ and\ \citenamefont {Shepelyansky}}]{Kern05}%
  \BibitemOpen
  \bibfield  {author} {\bibinfo {author} {\bibfnamefont {O.}~\bibnamefont
  {Kern}}, \bibinfo {author} {\bibfnamefont {G.}~\bibnamefont {Alber}},\ and\
  \bibinfo {author} {\bibfnamefont {D.~L.}\ \bibnamefont {Shepelyansky}},\
  }\bibfield  {title} {\bibinfo {title} {Quantum error correction of coherent
  errors by randomization},\ }\href
  {https://doi.org/10.1140/epjd/e2004-00196-9} {\bibfield  {journal} {\bibinfo
  {journal} {The European Physical Journal D}\ }\textbf {\bibinfo {volume}
  {32}},\ \bibinfo {pages} {153–156} (\bibinfo {year} {2005})}\BibitemShut
  {NoStop}%
\bibitem [{\citenamefont {Eastin}\ and\ \citenamefont
  {Knill}(2009)}]{Eastin2009}%
  \BibitemOpen
  \bibfield  {author} {\bibinfo {author} {\bibfnamefont {B.}~\bibnamefont
  {Eastin}}\ and\ \bibinfo {author} {\bibfnamefont {E.}~\bibnamefont {Knill}},\
  }\bibfield  {title} {\bibinfo {title} {{Restrictions on Transversal Encoded
  Quantum Gate Sets}},\ }\href {https://doi.org/10.1103/PhysRevLett.102.110502}
  {\bibfield  {journal} {\bibinfo  {journal} {Phys. Rev. Lett.}\ }\textbf
  {\bibinfo {volume} {102}},\ \bibinfo {pages} {110502} (\bibinfo {year}
  {2009})}\BibitemShut {NoStop}%
\bibitem [{\citenamefont {Litinski}(2019)}]{Litinski2019}%
  \BibitemOpen
  \bibfield  {author} {\bibinfo {author} {\bibfnamefont {D.}~\bibnamefont
  {Litinski}},\ }\bibfield  {title} {\bibinfo {title} {Magic state
  distillation: Not as costly as you think},\ }\href
  {https://doi.org/10.22331/q-2019-12-02-205} {\bibfield  {journal} {\bibinfo
  {journal} {Quantum}\ }\textbf {\bibinfo {volume} {3}},\ \bibinfo {pages}
  {205} (\bibinfo {year} {2019})}\BibitemShut {NoStop}%
\bibitem [{\citenamefont {Bravyi}(2006)}]{Bravyi2006}%
  \BibitemOpen
  \bibfield  {author} {\bibinfo {author} {\bibfnamefont {S.}~\bibnamefont
  {Bravyi}},\ }\bibfield  {title} {\bibinfo {title} {Universal quantum
  computation with the $\ensuremath{\nu}=5/2$ fractional quantum hall state},\
  }\href {https://doi.org/10.1103/PhysRevA.73.042313} {\bibfield  {journal}
  {\bibinfo  {journal} {Phys. Rev. A}\ }\textbf {\bibinfo {volume} {73}},\
  \bibinfo {pages} {042313} (\bibinfo {year} {2006})}\BibitemShut {NoStop}%
\bibitem [{\citenamefont {Freedman}\ \emph {et~al.}(2006)\citenamefont
  {Freedman}, \citenamefont {Nayak},\ and\ \citenamefont
  {Walker}}]{Freedman2006}%
  \BibitemOpen
  \bibfield  {author} {\bibinfo {author} {\bibfnamefont {M.}~\bibnamefont
  {Freedman}}, \bibinfo {author} {\bibfnamefont {C.}~\bibnamefont {Nayak}},\
  and\ \bibinfo {author} {\bibfnamefont {K.}~\bibnamefont {Walker}},\
  }\bibfield  {title} {\bibinfo {title} {Towards universal topological quantum
  computation in the $\ensuremath{\nu}=\frac{5}{2}$ fractional quantum hall
  state},\ }\href {https://doi.org/10.1103/PhysRevB.73.245307} {\bibfield
  {journal} {\bibinfo  {journal} {Phys. Rev. B}\ }\textbf {\bibinfo {volume}
  {73}},\ \bibinfo {pages} {245307} (\bibinfo {year} {2006})}\BibitemShut
  {NoStop}%
\bibitem [{\citenamefont {Sau}\ \emph {et~al.}(2010)\citenamefont {Sau},
  \citenamefont {Tewari},\ and\ \citenamefont {Das~Sarma}}]{Sau10c}%
  \BibitemOpen
  \bibfield  {author} {\bibinfo {author} {\bibfnamefont {J.~D.}\ \bibnamefont
  {Sau}}, \bibinfo {author} {\bibfnamefont {S.}~\bibnamefont {Tewari}},\ and\
  \bibinfo {author} {\bibfnamefont {S.}~\bibnamefont {Das~Sarma}},\ }\bibfield
  {title} {\bibinfo {title} {Universal quantum computation in a semiconductor
  quantum wire network},\ }\href {https://doi.org/10.1103/PhysRevA.82.052322}
  {\bibfield  {journal} {\bibinfo  {journal} {Phys. Rev. A}\ }\textbf {\bibinfo
  {volume} {82}},\ \bibinfo {pages} {052322} (\bibinfo {year} {2010})},\
  \Eprint {https://arxiv.org/abs/1007.4204} {arXiv:1007.4204} \BibitemShut
  {NoStop}%
\bibitem [{\citenamefont {{Karzig}}\ \emph {et~al.}(2016)\citenamefont
  {{Karzig}}, \citenamefont {{Oreg}}, \citenamefont {{Refael}},\ and\
  \citenamefont {{Freedman}}}]{Karzig15a}%
  \BibitemOpen
  \bibfield  {author} {\bibinfo {author} {\bibfnamefont {T.}~\bibnamefont
  {{Karzig}}}, \bibinfo {author} {\bibfnamefont {Y.}~\bibnamefont {{Oreg}}},
  \bibinfo {author} {\bibfnamefont {G.}~\bibnamefont {{Refael}}},\ and\
  \bibinfo {author} {\bibfnamefont {M.~H.}\ \bibnamefont {{Freedman}}},\
  }\bibfield  {title} {\bibinfo {title} {{Universal Geometric Path to a Robust
  {Majorana} Magic Gate}},\ }\href {https://doi.org/10.1103/PhysRevX.6.031019}
  {\bibfield  {journal} {\bibinfo  {journal} {Phys. Rev. X}\ }\textbf {\bibinfo
  {volume} {6}},\ \bibinfo {eid} {031019} (\bibinfo {year} {2016})},\ \Eprint
  {https://arxiv.org/abs/1511.05161} {arXiv:1511.05161} \BibitemShut {NoStop}%
\bibitem [{\citenamefont {Karzig}\ \emph {et~al.}(2019)\citenamefont {Karzig},
  \citenamefont {Oreg}, \citenamefont {Refael},\ and\ \citenamefont
  {Freedman}}]{Karzig19}%
  \BibitemOpen
  \bibfield  {author} {\bibinfo {author} {\bibfnamefont {T.}~\bibnamefont
  {Karzig}}, \bibinfo {author} {\bibfnamefont {Y.}~\bibnamefont {Oreg}},
  \bibinfo {author} {\bibfnamefont {G.}~\bibnamefont {Refael}},\ and\ \bibinfo
  {author} {\bibfnamefont {M.~H.}\ \bibnamefont {Freedman}},\ }\bibfield
  {title} {\bibinfo {title} {Robust majorana magic gates via measurements},\
  }\bibfield  {journal} {\bibinfo  {journal} {Physical Review B}\ }\textbf
  {\bibinfo {volume} {99}},\ \href {https://doi.org/10.1103/physrevb.99.144521}
  {10.1103/physrevb.99.144521} (\bibinfo {year} {2019})\BibitemShut {NoStop}%
\bibitem [{\citenamefont {Clarke}\ \emph {et~al.}(2016)\citenamefont {Clarke},
  \citenamefont {Sau},\ and\ \citenamefont {Das~Sarma}}]{Clarke16}%
  \BibitemOpen
  \bibfield  {author} {\bibinfo {author} {\bibfnamefont {D.~J.}\ \bibnamefont
  {Clarke}}, \bibinfo {author} {\bibfnamefont {J.~D.}\ \bibnamefont {Sau}},\
  and\ \bibinfo {author} {\bibfnamefont {S.}~\bibnamefont {Das~Sarma}},\
  }\bibfield  {title} {\bibinfo {title} {A practical phase gate for producing
  bell violations in majorana wires},\ }\bibfield  {journal} {\bibinfo
  {journal} {Physical Review X}\ }\textbf {\bibinfo {volume} {6}},\ \href
  {https://doi.org/10.1103/physrevx.6.021005} {10.1103/physrevx.6.021005}
  (\bibinfo {year} {2016})\BibitemShut {NoStop}%
\bibitem [{\citenamefont {{Bonderson}}(2007)}]{Bonderson_thesis}%
  \BibitemOpen
  \bibfield  {author} {\bibinfo {author} {\bibfnamefont {P.~H.}\ \bibnamefont
  {{Bonderson}}},\ }\emph {\bibinfo {title} {{Non-Abelian anyons and
  interferometry}}},\ \href@noop {} {Ph.D. thesis},\ \bibinfo  {school}
  {California Institute of Technology} (\bibinfo {year} {2007})\BibitemShut
  {NoStop}%
\bibitem [{\citenamefont {Kitaev}(2006)}]{Kitaev06a}%
  \BibitemOpen
  \bibfield  {author} {\bibinfo {author} {\bibfnamefont {A.}~\bibnamefont
  {Kitaev}},\ }\bibfield  {title} {\bibinfo {title} {Anyons in an exactly
  solved model and beyond},\ }\href {https://doi.org/10.1016/j.aop.2005.10.005}
  {\bibfield  {journal} {\bibinfo  {journal} {Ann. Phys.}\ }\textbf {\bibinfo
  {volume} {321}},\ \bibinfo {pages} {2} (\bibinfo {year} {2006})},\ \Eprint
  {https://arxiv.org/abs/cond-mat/0506438} {arXiv:cond-mat/0506438}
  \BibitemShut {NoStop}%
\bibitem [{\citenamefont {Zhou}\ \emph {et~al.}(2000)\citenamefont {Zhou},
  \citenamefont {Leung},\ and\ \citenamefont {Chuang}}]{Zhou00}%
  \BibitemOpen
  \bibfield  {author} {\bibinfo {author} {\bibfnamefont {X.}~\bibnamefont
  {Zhou}}, \bibinfo {author} {\bibfnamefont {D.~W.}\ \bibnamefont {Leung}},\
  and\ \bibinfo {author} {\bibfnamefont {I.~L.}\ \bibnamefont {Chuang}},\
  }\bibfield  {title} {\bibinfo {title} {Methodology for quantum logic gate
  construction},\ }\href {https://doi.org/10.1103/PhysRevA.62.052316}
  {\bibfield  {journal} {\bibinfo  {journal} {Phys. Rev. A}\ }\textbf {\bibinfo
  {volume} {62}},\ \bibinfo {pages} {052316} (\bibinfo {year}
  {2000})}\BibitemShut {NoStop}%
\bibitem [{\citenamefont {Danos}\ \emph {et~al.}(2007)\citenamefont {Danos},
  \citenamefont {Kashefi},\ and\ \citenamefont {Panangaden}}]{Danos07}%
  \BibitemOpen
  \bibfield  {author} {\bibinfo {author} {\bibfnamefont {V.}~\bibnamefont
  {Danos}}, \bibinfo {author} {\bibfnamefont {E.}~\bibnamefont {Kashefi}},\
  and\ \bibinfo {author} {\bibfnamefont {P.}~\bibnamefont {Panangaden}},\
  }\bibfield  {title} {\bibinfo {title} {The measurement calculus},\ }\href
  {https://doi.org/10.1145/1219092.1219096} {\bibfield  {journal} {\bibinfo
  {journal} {J. ACM}\ }\textbf {\bibinfo {volume} {54}},\ \bibinfo {pages}
  {8–es} (\bibinfo {year} {2007})}\BibitemShut {NoStop}%
\bibitem [{\citenamefont {Raussendorf}\ and\ \citenamefont
  {Briegel}(2001)}]{Raussendorf01}%
  \BibitemOpen
  \bibfield  {author} {\bibinfo {author} {\bibfnamefont {R.}~\bibnamefont
  {Raussendorf}}\ and\ \bibinfo {author} {\bibfnamefont {H.~J.}\ \bibnamefont
  {Briegel}},\ }\bibfield  {title} {\bibinfo {title} {A one-way quantum
  computer},\ }\href {https://doi.org/10.1103/PhysRevLett.86.5188} {\bibfield
  {journal} {\bibinfo  {journal} {Phys. Rev. Lett.}\ }\textbf {\bibinfo
  {volume} {86}},\ \bibinfo {pages} {5188} (\bibinfo {year}
  {2001})}\BibitemShut {NoStop}%
\bibitem [{\citenamefont {Bomb\'{\i}n}\ \emph {et~al.}(2023)\citenamefont
  {Bomb\'{\i}n}, \citenamefont {Dawson}, \citenamefont {Mishmash},
  \citenamefont {Nickerson}, \citenamefont {Pastawski},\ and\ \citenamefont
  {Roberts}}]{PRXQuantum.4.020303}%
  \BibitemOpen
  \bibfield  {author} {\bibinfo {author} {\bibfnamefont {H.}~\bibnamefont
  {Bomb\'{\i}n}}, \bibinfo {author} {\bibfnamefont {C.}~\bibnamefont {Dawson}},
  \bibinfo {author} {\bibfnamefont {R.~V.}\ \bibnamefont {Mishmash}}, \bibinfo
  {author} {\bibfnamefont {N.}~\bibnamefont {Nickerson}}, \bibinfo {author}
  {\bibfnamefont {F.}~\bibnamefont {Pastawski}},\ and\ \bibinfo {author}
  {\bibfnamefont {S.}~\bibnamefont {Roberts}},\ }\bibfield  {title} {\bibinfo
  {title} {Logical blocks for fault-tolerant topological quantum computation},\
  }\href {https://doi.org/10.1103/PRXQuantum.4.020303} {\bibfield  {journal}
  {\bibinfo  {journal} {PRX Quantum}\ }\textbf {\bibinfo {volume} {4}},\
  \bibinfo {pages} {020303} (\bibinfo {year} {2023})}\BibitemShut {NoStop}%
\bibitem [{\citenamefont {Kliuchnikov}\ \emph {et~al.}(2023)\citenamefont
  {Kliuchnikov}, \citenamefont {Beverland},\ and\ \citenamefont
  {Paetznick}}]{kliuchnikov2023stabilizercircuitverification}%
  \BibitemOpen
  \bibfield  {author} {\bibinfo {author} {\bibfnamefont {V.}~\bibnamefont
  {Kliuchnikov}}, \bibinfo {author} {\bibfnamefont {M.}~\bibnamefont
  {Beverland}},\ and\ \bibinfo {author} {\bibfnamefont {A.}~\bibnamefont
  {Paetznick}},\ }\href {https://arxiv.org/abs/2309.08676} {\bibinfo {title}
  {Stabilizer circuit verification}} (\bibinfo {year} {2023}),\ \Eprint
  {https://arxiv.org/abs/2309.08676} {arXiv:2309.08676 [quant-ph]} \BibitemShut
  {NoStop}%
\bibitem [{\citenamefont {Delfosse}\ and\ \citenamefont
  {Paetznick}(2023)}]{delfosse2023spacetimecodescliffordcircuits}%
  \BibitemOpen
  \bibfield  {author} {\bibinfo {author} {\bibfnamefont {N.}~\bibnamefont
  {Delfosse}}\ and\ \bibinfo {author} {\bibfnamefont {A.}~\bibnamefont
  {Paetznick}},\ }\href {https://arxiv.org/abs/2304.05943} {\bibinfo {title}
  {Spacetime codes of clifford circuits}} (\bibinfo {year} {2023}),\ \Eprint
  {https://arxiv.org/abs/2304.05943} {arXiv:2304.05943 [quant-ph]} \BibitemShut
  {NoStop}%
\bibitem [{\citenamefont {Chao}\ \emph {et~al.}(2017)\citenamefont {Chao},
  \citenamefont {Reichardt}, \citenamefont {Sutherland},\ and\ \citenamefont
  {Vidick}}]{chao2017overlapping}%
  \BibitemOpen
  \bibfield  {author} {\bibinfo {author} {\bibfnamefont {R.}~\bibnamefont
  {Chao}}, \bibinfo {author} {\bibfnamefont {B.~W.}\ \bibnamefont {Reichardt}},
  \bibinfo {author} {\bibfnamefont {C.}~\bibnamefont {Sutherland}},\ and\
  \bibinfo {author} {\bibfnamefont {T.}~\bibnamefont {Vidick}},\ }\href@noop {}
  {\bibinfo {title} {Overlapping qubits}} (\bibinfo {year} {2017}),\ \Eprint
  {https://arxiv.org/abs/1701.01062} {arXiv:1701.01062 [quant-ph]} \BibitemShut
  {NoStop}%
\bibitem [{\citenamefont {Irfan}\ \emph {et~al.}(2020)\citenamefont {Irfan},
  \citenamefont {Mayer}, \citenamefont {Ortiz},\ and\ \citenamefont
  {Knill}}]{Irfan20}%
  \BibitemOpen
  \bibfield  {author} {\bibinfo {author} {\bibfnamefont {A.~A.~M.}\
  \bibnamefont {Irfan}}, \bibinfo {author} {\bibfnamefont {K.}~\bibnamefont
  {Mayer}}, \bibinfo {author} {\bibfnamefont {G.}~\bibnamefont {Ortiz}},\ and\
  \bibinfo {author} {\bibfnamefont {E.}~\bibnamefont {Knill}},\ }\bibfield
  {title} {\bibinfo {title} {Certified quantum measurement of majorana
  fermions},\ }\href {https://doi.org/10.1103/PhysRevA.101.032106} {\bibfield
  {journal} {\bibinfo  {journal} {Phys. Rev. A}\ }\textbf {\bibinfo {volume}
  {101}},\ \bibinfo {pages} {032106} (\bibinfo {year} {2020})}\BibitemShut
  {NoStop}%
\bibitem [{\citenamefont {Song}\ \emph {et~al.}(2018)\citenamefont {Song},
  \citenamefont {Xu}, \citenamefont {Zhang}, \citenamefont {Wang},
  \citenamefont {Guo}, \citenamefont {Liu}, \citenamefont {Xu}, \citenamefont
  {Deng}, \citenamefont {Huang}, \citenamefont {Zheng} \emph
  {et~al.}}]{Song2018}%
  \BibitemOpen
  \bibfield  {author} {\bibinfo {author} {\bibfnamefont {C.}~\bibnamefont
  {Song}}, \bibinfo {author} {\bibfnamefont {D.}~\bibnamefont {Xu}}, \bibinfo
  {author} {\bibfnamefont {P.}~\bibnamefont {Zhang}}, \bibinfo {author}
  {\bibfnamefont {J.}~\bibnamefont {Wang}}, \bibinfo {author} {\bibfnamefont
  {Q.}~\bibnamefont {Guo}}, \bibinfo {author} {\bibfnamefont {W.}~\bibnamefont
  {Liu}}, \bibinfo {author} {\bibfnamefont {K.}~\bibnamefont {Xu}}, \bibinfo
  {author} {\bibfnamefont {H.}~\bibnamefont {Deng}}, \bibinfo {author}
  {\bibfnamefont {K.}~\bibnamefont {Huang}}, \bibinfo {author} {\bibfnamefont
  {D.}~\bibnamefont {Zheng}}, \emph {et~al.},\ }\bibfield  {title} {\bibinfo
  {title} {Demonstration of topological robustness of anyonic braiding
  statistics with a superconducting quantum circuit},\ }\href
  {https://doi.org/10.1103/PhysRevLett.121.030502} {\bibfield  {journal}
  {\bibinfo  {journal} {Physical Review Letters}\ }\textbf {\bibinfo {volume}
  {121}},\ \bibinfo {pages} {030502} (\bibinfo {year} {2018})}\BibitemShut
  {NoStop}%
\bibitem [{\citenamefont {Satzinger}\ \emph {et~al.}(2021)\citenamefont
  {Satzinger}, \citenamefont {Liu}, \citenamefont {Smith}, \citenamefont
  {Knapp}, \citenamefont {Newman}, \citenamefont {Jones}, \citenamefont {Chen},
  \citenamefont {Quintana}, \citenamefont {Mi}, \citenamefont {Dunsworth} \emph
  {et~al.}}]{Satzinger2021}%
  \BibitemOpen
  \bibfield  {author} {\bibinfo {author} {\bibfnamefont {K.}~\bibnamefont
  {Satzinger}}, \bibinfo {author} {\bibfnamefont {Y.-J.}\ \bibnamefont {Liu}},
  \bibinfo {author} {\bibfnamefont {A.}~\bibnamefont {Smith}}, \bibinfo
  {author} {\bibfnamefont {C.}~\bibnamefont {Knapp}}, \bibinfo {author}
  {\bibfnamefont {M.}~\bibnamefont {Newman}}, \bibinfo {author} {\bibfnamefont
  {C.}~\bibnamefont {Jones}}, \bibinfo {author} {\bibfnamefont
  {Z.}~\bibnamefont {Chen}}, \bibinfo {author} {\bibfnamefont {C.}~\bibnamefont
  {Quintana}}, \bibinfo {author} {\bibfnamefont {X.}~\bibnamefont {Mi}},
  \bibinfo {author} {\bibfnamefont {A.}~\bibnamefont {Dunsworth}}, \emph
  {et~al.},\ }\bibfield  {title} {\bibinfo {title} {Realizing topologically
  ordered states on a quantum processor},\ }\href
  {https://doi.org/10.1126/science.abi8378} {\bibfield  {journal} {\bibinfo
  {journal} {Science}\ }\textbf {\bibinfo {volume} {374}},\ \bibinfo {pages}
  {1237} (\bibinfo {year} {2021})}\BibitemShut {NoStop}%
\bibitem [{\citenamefont {Stenger}\ \emph {et~al.}(2021)\citenamefont
  {Stenger}, \citenamefont {Bronn}, \citenamefont {Egger},\ and\ \citenamefont
  {Pekker}}]{Stenger2021}%
  \BibitemOpen
  \bibfield  {author} {\bibinfo {author} {\bibfnamefont {J.~P.}\ \bibnamefont
  {Stenger}}, \bibinfo {author} {\bibfnamefont {N.~T.}\ \bibnamefont {Bronn}},
  \bibinfo {author} {\bibfnamefont {D.~J.}\ \bibnamefont {Egger}},\ and\
  \bibinfo {author} {\bibfnamefont {D.}~\bibnamefont {Pekker}},\ }\bibfield
  {title} {\bibinfo {title} {Simulating the dynamics of braiding of majorana
  zero modes using an ibm quantum computer},\ }\href
  {https://doi.org/10.1103/PhysRevResearch.3.033171} {\bibfield  {journal}
  {\bibinfo  {journal} {Physical Review Research}\ }\textbf {\bibinfo {volume}
  {3}},\ \bibinfo {pages} {033171} (\bibinfo {year} {2021})}\BibitemShut
  {NoStop}%
\bibitem [{\citenamefont {{Google Quantum AI and
  Collaborators}}(2023)}]{Google2023}%
  \BibitemOpen
  \bibfield  {author} {\bibinfo {author} {\bibnamefont {{Google Quantum AI and
  Collaborators}}},\ }\bibfield  {title} {\bibinfo {title} {{Non-Abelian
  braiding of graph vertices in a superconducting processor}},\ }\href
  {https://doi.org/10.1038/s41586-023-05954-4} {\bibfield  {journal} {\bibinfo
  {journal} {Nature}\ }\textbf {\bibinfo {volume} {618}},\ \bibinfo {pages}
  {264} (\bibinfo {year} {2023})}\BibitemShut {NoStop}%
\bibitem [{\citenamefont {Xu}\ \emph {et~al.}(2024)\citenamefont {Xu},
  \citenamefont {Sun}, \citenamefont {Wang}, \citenamefont {Li}, \citenamefont
  {Zhu}, \citenamefont {Dong}, \citenamefont {Deng}, \citenamefont {Zhang},
  \citenamefont {Chen}, \citenamefont {Wu} \emph {et~al.}}]{Xu2024}%
  \BibitemOpen
  \bibfield  {author} {\bibinfo {author} {\bibfnamefont {S.}~\bibnamefont
  {Xu}}, \bibinfo {author} {\bibfnamefont {Z.-Z.}\ \bibnamefont {Sun}},
  \bibinfo {author} {\bibfnamefont {K.}~\bibnamefont {Wang}}, \bibinfo {author}
  {\bibfnamefont {H.}~\bibnamefont {Li}}, \bibinfo {author} {\bibfnamefont
  {Z.}~\bibnamefont {Zhu}}, \bibinfo {author} {\bibfnamefont {H.}~\bibnamefont
  {Dong}}, \bibinfo {author} {\bibfnamefont {J.}~\bibnamefont {Deng}}, \bibinfo
  {author} {\bibfnamefont {X.}~\bibnamefont {Zhang}}, \bibinfo {author}
  {\bibfnamefont {J.}~\bibnamefont {Chen}}, \bibinfo {author} {\bibfnamefont
  {Y.}~\bibnamefont {Wu}}, \emph {et~al.},\ }\bibfield  {title} {\bibinfo
  {title} {Non-abelian braiding of fibonacci anyons with a superconducting
  processor},\ }\href {https://doi.org/10.1038/s41567-024-02529-6} {\bibfield
  {journal} {\bibinfo  {journal} {Nature Physics}\ }\textbf {\bibinfo {volume}
  {20}},\ \bibinfo {pages} {1469} (\bibinfo {year} {2024})}\BibitemShut
  {NoStop}%
\bibitem [{\citenamefont {Iqbal}\ \emph
  {et~al.}(2024{\natexlab{a}})\citenamefont {Iqbal}, \citenamefont
  {Tantivasadakarn}, \citenamefont {Gatterman}, \citenamefont {Gerber},
  \citenamefont {Gilmore}, \citenamefont {Gresh}, \citenamefont {Hankin},
  \citenamefont {Hewitt}, \citenamefont {Horst}, \citenamefont {Matheny},
  \citenamefont {Mengle}, \citenamefont {Neyenhuis}, \citenamefont
  {Vishwanath}, \citenamefont {Foss-Feig}, \citenamefont {Verresen},\ and\
  \citenamefont {Dreyer}}]{Iqbal2023a}%
  \BibitemOpen
  \bibfield  {author} {\bibinfo {author} {\bibfnamefont {M.}~\bibnamefont
  {Iqbal}}, \bibinfo {author} {\bibfnamefont {N.}~\bibnamefont
  {Tantivasadakarn}}, \bibinfo {author} {\bibfnamefont {T.~M.}\ \bibnamefont
  {Gatterman}}, \bibinfo {author} {\bibfnamefont {J.~A.}\ \bibnamefont
  {Gerber}}, \bibinfo {author} {\bibfnamefont {K.}~\bibnamefont {Gilmore}},
  \bibinfo {author} {\bibfnamefont {D.}~\bibnamefont {Gresh}}, \bibinfo
  {author} {\bibfnamefont {A.}~\bibnamefont {Hankin}}, \bibinfo {author}
  {\bibfnamefont {N.}~\bibnamefont {Hewitt}}, \bibinfo {author} {\bibfnamefont
  {C.~V.}\ \bibnamefont {Horst}}, \bibinfo {author} {\bibfnamefont
  {M.}~\bibnamefont {Matheny}}, \bibinfo {author} {\bibfnamefont
  {T.}~\bibnamefont {Mengle}}, \bibinfo {author} {\bibfnamefont
  {B.}~\bibnamefont {Neyenhuis}}, \bibinfo {author} {\bibfnamefont
  {A.}~\bibnamefont {Vishwanath}}, \bibinfo {author} {\bibfnamefont
  {M.}~\bibnamefont {Foss-Feig}}, \bibinfo {author} {\bibfnamefont
  {R.}~\bibnamefont {Verresen}},\ and\ \bibinfo {author} {\bibfnamefont
  {H.}~\bibnamefont {Dreyer}},\ }\bibfield  {title} {\bibinfo {title}
  {Topological order from measurements and feed-forward on a trapped ion
  quantum computer},\ }\bibfield  {journal} {\bibinfo  {journal}
  {Communications Physics}\ }\textbf {\bibinfo {volume} {7}},\ \href
  {https://doi.org/10.1038/s42005-024-01698-3} {10.1038/s42005-024-01698-3}
  (\bibinfo {year} {2024}{\natexlab{a}})\BibitemShut {NoStop}%
\bibitem [{\citenamefont {Iqbal}\ \emph
  {et~al.}(2024{\natexlab{b}})\citenamefont {Iqbal}, \citenamefont
  {Tantivasadakarn}, \citenamefont {Verresen}, \citenamefont {Campbell},
  \citenamefont {Dreiling}, \citenamefont {Figgatt}, \citenamefont {Gaebler},
  \citenamefont {Johansen}, \citenamefont {Mills}, \citenamefont {Moses},
  \citenamefont {Pino}, \citenamefont {Ransford}, \citenamefont {Rowe},
  \citenamefont {Siegfried}, \citenamefont {Stutz}, \citenamefont {Foss-Feig},
  \citenamefont {Vishwanath},\ and\ \citenamefont {Dreyer}}]{Iqbal2023b}%
  \BibitemOpen
  \bibfield  {author} {\bibinfo {author} {\bibfnamefont {M.}~\bibnamefont
  {Iqbal}}, \bibinfo {author} {\bibfnamefont {N.}~\bibnamefont
  {Tantivasadakarn}}, \bibinfo {author} {\bibfnamefont {R.}~\bibnamefont
  {Verresen}}, \bibinfo {author} {\bibfnamefont {S.~L.}\ \bibnamefont
  {Campbell}}, \bibinfo {author} {\bibfnamefont {J.~M.}\ \bibnamefont
  {Dreiling}}, \bibinfo {author} {\bibfnamefont {C.}~\bibnamefont {Figgatt}},
  \bibinfo {author} {\bibfnamefont {J.~P.}\ \bibnamefont {Gaebler}}, \bibinfo
  {author} {\bibfnamefont {J.}~\bibnamefont {Johansen}}, \bibinfo {author}
  {\bibfnamefont {M.}~\bibnamefont {Mills}}, \bibinfo {author} {\bibfnamefont
  {S.~A.}\ \bibnamefont {Moses}}, \bibinfo {author} {\bibfnamefont {J.~M.}\
  \bibnamefont {Pino}}, \bibinfo {author} {\bibfnamefont {A.}~\bibnamefont
  {Ransford}}, \bibinfo {author} {\bibfnamefont {M.}~\bibnamefont {Rowe}},
  \bibinfo {author} {\bibfnamefont {P.}~\bibnamefont {Siegfried}}, \bibinfo
  {author} {\bibfnamefont {R.~P.}\ \bibnamefont {Stutz}}, \bibinfo {author}
  {\bibfnamefont {M.}~\bibnamefont {Foss-Feig}}, \bibinfo {author}
  {\bibfnamefont {A.}~\bibnamefont {Vishwanath}},\ and\ \bibinfo {author}
  {\bibfnamefont {H.}~\bibnamefont {Dreyer}},\ }\bibfield  {title} {\bibinfo
  {title} {Non-abelian topological order and anyons on a trapped-ion
  processor},\ }\href {https://doi.org/10.1038/s41586-023-06934-4} {\bibfield
  {journal} {\bibinfo  {journal} {Nature}\ }\textbf {\bibinfo {volume} {626}},\
  \bibinfo {pages} {505–511} (\bibinfo {year}
  {2024}{\natexlab{b}})}\BibitemShut {NoStop}%
\end{thebibliography}%
\end{document}